\documentclass[11pt,a4paper]{article}

\def\data{{\cal{D}}}
\def\auxdata{{\cal{A}}}
\def\Ndata{{{\cal{N}}_{\cal{D}}}}
\def\reference{{\cal{R}}}
\def\Nreference{{\rm{N}}_{\cal{R}}}

\def\NRef{{\rm{N}}({\rm{R}})}
\def\NRefnu{{\rm{N}}(\RH)}
\def\NRefCV{{\rm{N}}(\CVR)}
\def\NH{{\rm{N}}(\HH)}
\newcommand{\PRef}[1]{P(#1|\RH)}
\newcommand{\ndRef}[1]{n(#1|\RH)}
\newcommand{\ndRefCV}[1]{n(#1|\CVR)} 

\newcommand{\ndH}[1]{n(#1|\HH)}
\def\Lik{\mathcal{L}}
\newcommand\nui{{\boldsymbol\nu}}
\newcommand\RH{{\rm{R}}_{\boldsymbol\nu}}
\newcommand\CVR{{\rm{R}}_{\boldsymbol0}}
\newcommand\w{{\rm{\bf{w}}}}
\newcommand\HH{{\rm{H}}_{{\rm{\bf{w}}},{\boldsymbol\nu}}}
\usepackage{bold-extra}

\usepackage{graphicx,enumitem}

\graphicspath{ {./Figures/} }

\usepackage{authblk}
\usepackage{comment}

\pdfoutput=1
\usepackage{color}
\usepackage{todonotes}
\usepackage{subcaption}
\usepackage{booktabs}
\usepackage{array}
\usepackage{multirow,rotating}
\usepackage{amsmath,amssymb,bm,cite,graphicx,wasysym,mathrsfs,dsfont}
\usepackage{slashed,relsize}
\usepackage[colorlinks=true,linkcolor=blue,anchorcolor=black,citecolor=blue,filecolor=cyan,menucolor=red,runcolor=filecolor,urlcolor=blue,bookmarks=true,bookmarksnumbered=true]{hyperref}

\usepackage[font=small,labelfont=bf]{caption}

\def\beq{\begin{equation}\displaystyle}
\def\eeq{\end{equation}}
\def\bea{\begin{eqnarray}\displaystyle} 
\def\eea{\end{eqnarray}}

\def\({\left(}
\def\){\right)}
\def\bry{\begin{array}}
\def\ery{\end{array}}

\usepackage{soul}

\definecolor{readableRTD}{rgb}{0.7,0.1,0.2}

\definecolor{colorRTD}{rgb}{.2,.2,.7}

\title{\bf Learning New Physics from an \\ Imperfect Machine \vspace{0.4cm}} 
\date{}

\author[1]{Raffaele Tito D'Agnolo}

\author[2,3]{Gaia Grosso}

\author[2]{Maurizio Pierini}

\author[3,4]{Andrea Wulzer}

\author[3]{Marco Zanetti}

\affil[1]{\emph{Institut de Physique Th\'eorique, Universit\'e Paris Saclay, CEA, F-91191 Gif-sur-Yvette, France}}
\affil[2]{\emph{Experimental Physics Department, CERN, Geneva, Switzerland}}
\affil[3]{\emph{Dipartimento di Fisica e Astronomia, Universit\'a di Padova, and INFN, Sezione di Padova, Italy}}
\affil[4]{\emph{Institut de Th\'eorie des Ph\'enomenes Physiques, EPFL, Lausanne, Switzerland}}

\begin{document}
\baselineskip=14pt

\maketitle

\begin{abstract}
We show how to deal with uncertainties on the Standard Model predictions in an agnostic new physics search strategy that exploits artificial neural networks. Our approach builds directly on the specific Maximum Likelihood ratio treatment of uncertainties as nuisance parameters for hypothesis testing that is routinely employed in high-energy physics. After presenting the conceptual foundations of our method, we first illustrate all aspects of its implementation and extensively study its performances on a toy one-dimensional problem. We then show how to implement it in a multivariate setup by studying the impact of two typical sources of experimental uncertainties in two-body final states at the LHC.
\end{abstract}

\thispagestyle{empty}

\newpage

\begingroup
\tableofcontents
\endgroup 

\setcounter{equation}{0}
\setcounter{footnote}{0}
\setcounter{page}{1} 

\newpage

\section{Introduction}\label{intro}

Experimental results in the last several decades consolidated our knowledge of fundamental physics as described by ``standard'' theoretical models such as the Standard Model (SM) of particle physics or the \mbox{$\Lambda$CDM} model of cosmology. On the other hand we lack understanding of the microscopic origin of several ingredients of these models, such as  the Dark Matter and Dark Energy densities in \mbox{$\Lambda$CDM}, the electroweak scale and the Yukawa couplings structure in the SM. These considerations, as well as the theoretical incompleteness of our current theory of gravity, guarantee the existence of new fundamental laws waiting to be discovered, but do not sharply outline a path towards their actual experimental discovery. 

One can take the incompleteness of the standard models as guidance to formulate putative ``new physics'' models or scenarios that complete the standard models in one or several aspects. Then one can organize the exploration of new fundamental laws as the search for the experimental manifestations of such models. We call these searches ``model-dependent'' as they target the signal expected in one specific model and have poor or no sensitivity to unexpected signals. The problem with this strategy is that each new physics model only offers one possible solution to the problems of the standard models. Even searching for all of them experimentally, we are not guaranteed to achieve a discovery, as the actual solution might be one that we have not yet hypothesized. This possibility should be taken seriously also in light of the lack of discovery so far in the vast program of model-dependent searches carried out at past and ongoing experiments.

The development of ``model-independent'' strategies to search for new physics emerges in this context as a priority of fundamental physics. We dub model-independent those strategies that aim at assessing the compatibility of data with the predictions of a Reference theoretical Model, to be interpreted as one of the ``standard'' models previously discussed, rather than at probing the signatures of a specific alternative model, as in traditional model-dependent searches. It should be noted on the one hand that testing one Reference hypothesis with no assumption on the set of allowed alternative hypotheses is an ill-defined statistical concept. On the other hand, it is often trivial in practice to tell the level of compatibility of the Reference Model with the  data of an experiment whose outcome consists of a single or a few measurements. The statistical distribution of the measurements is known and can be compared with the one predicted by the Reference Model. Combining a limited number of measurements does not spoil the sensitivity even if the departure from the Reference Model is present in one single measurement. However the problem becomes practically and conceptually non-trivial in modern fundamental physics experiments where the data are extremely rich and the number of possible measurements is essentially infinite. In model-dependent strategies one restricts the set of measurements to those where the specific new physics model is expected to contribute significantly, and/or one exploits the correlation between the outcome of different measurements predicted by the new physics model. Obviously this is not an option in the model-independent case.

We consider here the model-independent method that we proposed and developed in Ref.s~\cite{DAgnolo:2018cun,DAgnolo:2019vbw} for data analysis at particle colliders such as the Large Hadron Collider (LHC). In this case the data $\data=\{x_1,\ldots,x_{\Ndata}\}$ consist of $\Ndata$ independent and identically-distributed measurements of a vector of features $x$. 
The physical knowledge of the Reference Model (the SM) can be used to produce a synthetic set of Reference data $\reference=\{x_1,\ldots,x_{\Nreference}\}$, whose elements follow the probability distribution of $x$ in the Reference hypothesis ``${\rm{R}}$''. In general, $\reference$ could be a weighted event sample. The Reference Model can also predict the total number of events $\NRef$ expected in the experiment, around which the number of observations $\Ndata$ is Poisson-distributed. Model-independent search strategies aim at exploiting these elements for a test of compatibility between the hypothesis ${\rm{R}}$ and the data.\footnote{A concise overview of the fast-growing literature on model-independent LHC searches and a categorization of the different approaches is reported in Appendix~\ref{app:MI}. In particular the origin and the role played by the Reference data set $\reference$ in each type of approach is discussed there.}
In order to be useful, the test should be capable to detect ``generic'' departures of the data distribution from the Reference expectation. Moreover it should target ``small" departures in the distribution. The significance of the discrepancy can be large, but the signal can be sizable (i.e., given by a number of events that is large, relative to the Reference model expectation) only in a small (low-probability) region of the features space, or its significance emerge from correlated small differences in a large region. This is because previous experiments and theoretical considerations generically exclude the viability of new physics models that produce a radical deformation of the LHC data distribution, which are furthermore easier to detect.  

As said, the Reference sample $\reference$ consists of synthetic instances of the variable $x$ that follow the distribution predicted by the Reference Model.  It plays conceptually the same role as the background dataset in regular model-dependent searches and it can be obtained either by a first-principle Monte Carlo simulation based on the fundamental physical laws of the Reference Model, or with data-driven methods. In the latter case, one extrapolates the background from data measured in a control region, using transfer functions that are extracted from Monte Carlo simulations. In both cases, $\reference$ results from a knowledge of the Reference Model that is unavoidably imperfect. Therefore it provides only an approximate representation of the data distribution in the Reference (or background) hypothesis. Uncertainties emerge from all the ingredients of the simulations such as the value of the Reference Model input parameters, of the parton distribution functions and of the detector response, as well as from the finite accuracy of the underlying theoretical calculations. The impact of all these uncertainties must be assessed and included if needed in any LHC analysis. In this paper we define a strategy to deal with them in our framework for model-independent new physics searches. 

\subsection{Overview of the methodology}
In this work we develop a full treatment of systematic uncertainties within a model-independent search. Our treatment follows closely the canonical high-energy physics profile likelihood approach, reviewed in Ref.~\cite{Zyla:2020zbs}. Each source of imperfection in the knowledge of the Reference Model is associated with a nuisance parameter $\nu$. Its (true) value is unknown but statistically constrained by an ``auxiliary'' dataset $\auxdata$, which produces a $\nu$-dependent multiplicative term in the likelihood, $\Lik(\nui|\auxdata)$. The Reference Model prediction for the distribution of the variable $x$ depends on the nuisance parameters, which we collect in a vector $\nui$. The Reference Model is thus interpreted as a composite (parameter-dependent) statistical hypothesis $\RH$, to be identified with the null hypothesis $H_0$ of the statistical test. The alternative hypothesis $H_1$ is defined as a local (in the features space) rescaling of the Reference distribution by the exponential of a neural network function $f(x;\w)$. The $H_1$ hypothesis is clearly also a composite one. We denote it as $\HH$, where $\w$ represents the trainable parameters of the neural network. Our strategy consists of performing a hypothesis test, based on the Maximum Likelihood log-ratio test statistic~\cite{Neyman:1933wgr,Wilks:1938dza,Wald1943}, between the $\RH$ and $\HH$ hypotheses. Namely our test statistic $t$ (see eq.~(\ref{tstat})) is twice the logarithm of the ratio between the likelihood of $\HH$ given the data (times the auxiliary likelihood $\Lik(\nui|\auxdata)$), maximized over $\w$ and $\nui$, and the likelihood of $\RH$ (times $\Lik(\nui|\auxdata)$) maximized over $\nui$.

The concept is literally the same as in Ref.s~\cite{DAgnolo:2018cun,DAgnolo:2019vbw}, with the difference that the Reference hypothesis is now composite rather than simple (i.e., $\nui$-independent) and the $H_1$ hypothesis also depends on the nuisances and not only on the neural network parameters $\w$. As in Ref.s~\cite{DAgnolo:2018cun,DAgnolo:2019vbw}, the choice of a neural network model for $H_1$ is motivated by the quest for an unbiased flexible approximant that can adapt itself to generic departures of the data from the Reference distribution, in order to maximize the sensitivity of the hypothesis test to generic new physics.

The first goal of the present paper is to construct a practical algorithm that computes the Maximum Likelihood log-ratio test statistic as defined above, including the effect of nuisance parameters. The basic idea is to normalize the $\HH$ and $\RH$ likelihoods to the likelihood of the ``central-value'' Reference hypothesis $\CVR$, namely the one where the nuisance parameters are set to their central value ($\nui=0$) that maximizes the observed auxiliary likelihood. In this way we divide the calculation of the test statistic $t$ in the evaluation of two separate terms. One of them merely consists of the likelihood log-ratio between the nuisance-dependent $\RH$ likelihood maximized over $\nui$, and the likelihood of the central-value $\CVR$ hypothesis. Maximizing the background-only likelihood as a function of the nuisance parameters is a necessary step of any LHC analysis. It serves in the first place to quantify the pull of the best-fit values of the nuisances, that maximize the complete likelihood (including the likelihood of the data of interest and of the auxiliary data, $\auxdata$), relative to their central value estimates and uncertainties as obtained from the auxiliary likelihood alone. Therefore the determination of the first term in $t$ does not pose any novel challenge, and could be in principle performed with the standard strategy of employing a binned approximation of the likelihood after modeling the dependence of the cross section in each bin on the nuisances. In practice for the numerical experiments performed in this paper we will find more effective and more easy to employ an un-binned likelihood reconstructed by neural networks~\cite{Cranmer:2015bka,Baldi:2016fzo,Brehmer:2018hga,Brehmer:2019xox,Chen:2020mev,Chen:1,Chen:2}. 

The other term required for the determination of the test statistic $t$ involves the neural network and requires the maximization over the neural network parameters $\w$ (and over $\nui$). It will be obtained by neural network training (with simultaneous minimization over $\nui$), with a strategy that is a relatively straightforward generalization of the one we already employed~\cite{DAgnolo:2018cun,DAgnolo:2019vbw} in the absence of nuisance parameters. As in Ref.s~\cite{DAgnolo:2018cun,DAgnolo:2019vbw}, the training data are the observed dataset $\data$ and the Reference dataset $\reference$. The Reference data are supposed to represent the distribution in the central-value hypothesis $\CVR$, therefore they are obtained fixing each nuisance parameter to its central value. They do not contain any information on the variability of the Reference distribution due to the nuisances, which is taken into account by the first term of the test statistic. This avoids employing in the  training Reference samples with multiple values of the nuisance parameters. The algorithm is thus  not more computationally expensive than the one in the absence of nuisances. 

Like any other frequentist hypothesis test, the practical feasibility of our strategy is linked to the validity of asymptotic formulae for the distribution of the test statistic $t$ in the null hypothesis $\RH$,  $P(t|H_0)=P(t|\RH)$. In particular the asymptotic formulae are needed to ensure the independence of $P(t|\RH)$ on the nuisance parameters $\nui$~\cite{Zyla:2020zbs,Cowan:2010js}. The Wilks--Wald Theorem~\cite{Wilks:1938dza,Wald1943} predicts a $\chi^2$ distribution for $t$ in the asymptotic (infinite sample) limit, but it gives no quantitative information on how ``large'' the dataset should be, in order for $P(t|\RH)$ to be similar to a $\chi^2$. Furthermore there is obviously no universal lower threshold on the data statistics after which the asymptotic result starts applying. The threshold depends on the problem and, crucially, on the complexity of the statistical model that is being considered. For instance if a simple one-parameter linear model was used for the numerator hypothesis instead of a neural network, a statistics of a few data events might suffice to reach the asymptotic limit accurately. Larger and larger datasets will be needed if the expressivity of the model is increased using neural networks of increasing complexity. One can of course also adopt the opposite viewpoint, which is more convenient in our case where the statistics of the data is fixed, and consider the upper threshold for the model complexity below which the asymptotic limit is reached and the distribution of $t$ starts following the $\chi^2$ distribution. 

We need the asymptotic formula to hold in order to eliminate or mitigate the dependence of $P(t|\RH)$ on $\nui$. On the other hand, we would like our model to be as complex and expressive as possible in order to be sensitive to the largest possible variety of putative new physics effects. Therefore the optimal complexity for the neural network model is right at the threshold of loosing the $\chi^2$ compatibility. In Ref.~\cite{DAgnolo:2019vbw} we already advocated this $\chi^2$ compatibility criterion for the selection of the neural network model, with the motivation that the $t$ distribution not following the asymptotic formula signals that $t$ is sensitive to low-statistics regions of the dataset, a fact which in turn can be interpreted as ``overfitting'' in our context. This heuristic motivation remains, but it is accompanied by the stronger technical argument associated with the feasibility of the hypothesis test including nuisance parameters.

\subsection{Structure of the paper}

The rest of the paper is organized as follows. In Section~\ref{sec:found} we describe the statistical foundations of our method. Namely we show how to turn the mathematical definition of the Maximum Likelihood ratio test statistic into a practical algorithm for its evaluation along the lines described above. The implementation of the algorithm in all its aspects, including the selection of the neural network hyperparameters by the $\chi^2$ compatibility criterion, is described in Section~\ref{sec:one} for an illustrative univariate problem. In that section we will obtain a first validation of our method by studying how it reacts to toy datasets generated with values of the nuisance parameters that are different from the central values employed for the Reference training set. We will see that the term in $t$ coming from the neural network is typically large, its distribution over the toys shifts to the right and gets strongly distorted  with respect to the distribution one obtains when the toy data are instead generated with central-value nuisances. The other term in $t$, associated with the $\RH/\CVR$ likelihood ratio as previously described, engineers a non-trivial cancellation on the total value of $t$ for each individual toy. A $\chi^2$ distribution is eventually recovered for the total $t$ distribution, compatibly with the Wilks--Wald Theorem, regardless of the value of $\nui$ used in the generation of the toy data. Similar tests are performed in Section~\ref{sec:two} in a slightly more realistic problem with five features (kinematical variables) that represent a dataset that one might encounter in the study of the production of two particles at the LHC. Two common sources of uncertainties are included, and their impact on the sensitivity of our strategy to benchmark putative signals is quantified. 
We report our Conclusions in Section~\ref{sec:conc}. Appendix~\ref{app:MI} provides an overview of model-independent strategies in connection and comparison with ours. 

\section{Foundations}\label{sec:found}

\subsection{Hypothesis testing}

As explained in Section~\ref{intro}, our method consists of a hypothesis test between a null hypothesis $H_0=\RH$ and an alternative $H_1=\HH$. We now characterize the two hypotheses in turn, starting from the null $\RH$ Reference (i.e., the SM) hypothesis. The data collected in the region of interest for the analysis are denoted as $\data=\{x_1,\ldots,x_{\Ndata}\}$ and consist of $\Ndata$ instances of a multi-dimensional variable $x$. For instance, the region of interest for the analysis could be defined as the subset of the entire experimental dataset where a given experimental signature (e.g., two high-$p_{\rm T}$ muons reconstructed within a certain detector acceptance) has been observed. 
The features $x$ would then consist of the reconstructed momenta of these particles. The region of interest might be further restricted by selection cuts that define the region $X$ of the phase space ($x\in X$) to which the particle momenta belong. Each instance of $x$ in $\data$ is thrown with a probability distribution that we denote as $\PRef{x\,}$ in the Reference hypothesis $\RH$. The total number of instances of $x$, $\Ndata$, is Poisson-distributed with a mean $\NRefnu$ that equals the total cross section in the region $X$ times the integrated luminosity. The likelihood of the $\RH$ hypothesis, given the observation of the dataset $\data$, is thus provided by the extended likelihood
\beq
\Lik(\RH|\data)=\frac{\NRefnu^\Ndata}{\Ndata !}e^{-\NRefnu}\prod\limits_{x\in\data}\PRef{x}=\frac{e^{-\NRefnu}}{\Ndata !}\prod\limits_{x\in\data}\ndRef{x}\,.
\eeq
In the previous equation we defined for shortness
\beq
\ndRef{x}=\NRefnu\PRef{x}\,.
\eeq
We will denote $n(x|H)$, in different hypotheses $H$, the ``distribution'' of the variable $x$. 

The Reference hypothesis distribution for $x$ depends on a set of nuisance parameters $\nui$. They model all the imperfections in the knowledge of the Reference Model, ranging from theoretical uncertainties like those in the determination of the parton distribution functions, to the calibration of the detector response. The nuisance parameters are (often, see below) statistically constrained by ``auxiliary'' measurements performed using data sets independent of $\data$, that we collectively denote as $\auxdata$. The $\RH$ hypothesis provides a $\nui$-dependent prediction also for the statistical distribution of the auxiliary measurements. The total likelihood of $\RH$, given the observation of both the data of interest and of the auxiliary data, thus reads
\beq
\Lik(\RH|\data,\auxdata)=\Lik(\RH|\data)\cdot \Lik(\nui|\auxdata)\,,
\eeq
where we denoted, for brevity, $\Lik(\RH|\auxdata)$ as $\Lik(\nui|\auxdata)$. It should be noted that this simple picture of the nuisances constraint term in the likelihood emerging from auxiliary measurements only holds for uncertainties of purely statistical origin. Genuinely systematic uncertainties such as theoretical errors associated to missing higher-orders in calculations are heuristically treated in the same manner, even if a rigorous statistical interpretation of this type of uncertainties is currently not available to our knowledge.

We now turn to the alternative hypothesis $H_1=\HH$. This hypothesis should include potential departures in the distribution of the variable $x$ from the Reference (i.e., SM) expectation. As anticipated in Section~\ref{intro}, 
 we parametrize these departures as a local rescaling of the Reference distribution by the exponential of a single-output neural network. Following the approach of Ref.s~\cite{DAgnolo:2018cun,DAgnolo:2019vbw} we postulate
\beq\label{nnh}
\ndH{x}=e^{f(x;\w)}\ndRef{x}\,,
\eeq
where $f$ is the neural network and $\w$ denotes its trainable parameters. The neural network architecture and hyper-parameters are problem-dependent. The general criteria for their optimization are discussed in Section~\ref{af} and illustrated in Sections~\ref{sec3:model} and~\ref{sec:MS5f} in greater detail.

We further postulate that new physics is absent in the auxiliary data. Namely that the distribution of the auxiliary data in the $\HH$ hypothesis is the same one as in hypothesis $\RH$
\beq\label{auxd}
\Lik(\HH|\auxdata)=\Lik(\RH|\auxdata)=\Lik(\nui|\auxdata)\,.
\eeq
Therefore the total likelihood of $\HH$ is 
\beq
\Lik(\HH|\data,\auxdata)=\Lik(\HH|\data)\cdot \Lik(\nui|\auxdata)\,,
\eeq
where $\Lik(\HH|\data)$ is the extended likelihood
\beq
\Lik(\HH|\data)=\frac{e^{-\NH}}{\Ndata !}\prod\limits_{x\in\data}\ndH{x}\,,
\eeq
with $\ndH{x}$ as in eq.~(\ref{nnh}). The total number of expected events $\NH$ is the integral of $\ndH{x}$ over the features space. A discussion of the implications of postulating the absence of new physics in the auxiliary data as in eq.~(\ref{auxd}), and of related aspects, is postponed to Section~\ref{sec:npcr}.

The test statistic variable we aim at computing and employing for the hypothesis test is the Maximum Likelihood log ratio~\cite{Zyla:2020zbs,Neyman:1933wgr,Cowan:2010js}
\beq\label{tstat}
t(\data,\auxdata)=2\,\log\frac{\max\limits_{\w,\nui}\left[\Lik(\HH|\data,\auxdata)\right]}{\max\limits_{\nui}\left[\Lik(\RH|\data,\auxdata)\right]}\,.
\eeq
Notice that this definition of the test statistic, and in turn its properties~\cite{Wilks:1938dza,Wald1943}, assumes that the composite hypothesis in the denominator ($H_0$) is contained in the numerator hypothesis ($H_1$). This holds in our case since the neural network function in eq.~(\ref{nnh}) is equal to zero when all its weights and biases $\w$ vanish. Therefore $(\HH)|_{\w=0}=\RH$. Also notice that the test statistic variable $t$ depends on all the data that are employed in the analysis. In particular it depends on the auxiliary data $\auxdata$ as well as on the data of interest $\data$. We now address the problem of evaluating $t$, once the data are made available either from the actual experiment or artificially by generating toy datasets.

\subsection{The central-value Reference hypothesis}\label{sec:cvr}

In order to proceed, we consider the special point in the space of nuisance parameters that corresponds to their central-value determination as obtained from the auxiliary data alone. If we call $\auxdata_0$ the observed auxiliary dataset, namely the one that is observed in the actual experiment, the central values of the nuisance parameters are those maximizing the auxiliary likelihood function $\Lik(\nui|\auxdata_0)$. It is always possible to choose the coordinates in the nuisance parameters space such that the central values of all the parameters sit at $\nu=0$. So we have, by definition
\beq\label{cvn}
\max\limits_{\nui}\left[\Lik(\nui|\auxdata_0)\right]=\Lik({\boldsymbol0}|\auxdata_0)\,.
\eeq
We stress again that $\auxdata_0$ represents one single outcome of the auxiliary measurements (the one observed in the actual experiment), unlike $\auxdata$ (and $\data$) that describe all the possible experimental outcomes. Therefore $\auxdata_0$, and in turn the central value of the nuisance parameters that we have set to $\nui={\boldsymbol0}$, is not a statistical variable and therefore it will not fluctuate when we will generate toy experiments, unlike $\auxdata$ and $\data$.

The central-value Reference hypothesis $\CVR$ predicts a distribution for the variable $x$, $\ndRefCV{x}$, that can be regarded as the ``best guess'' we can make for the actual SM distribution of $x$ before analyzing the dataset of interest $\data$. Correspondingly, $\nui={\boldsymbol0}$ is the best prior guess for the value of the nuisances. The likelihood of $\CVR$, given by
\beq
\Lik(\CVR|\data,\auxdata)=\Lik(\CVR|\data) \cdot \Lik({\boldsymbol0}|\auxdata) =\frac{e^{-\NRefCV}}{\Ndata !}\prod\limits_{x\in\data}\ndRefCV{x} \cdot \Lik({\boldsymbol0}|\auxdata)\,,
\eeq
is thus conveniently used to ``normalize'' the likelihoods at the numerator and denominator in eq.~(\ref{tstat}). Namely we multiply and divide the argument of the log by $\Lik(\CVR|\data,\auxdata)$ and we obtain
\beq\label{tstat1}
t(\data,\auxdata)=\tau(\data,\auxdata)-\Delta(\data,\auxdata)\,,
\eeq
where $\tau$ involves the maximization over the neural network parameters $\w$ and over $\nui$
\beq\label{tau}
\tau(\data,\auxdata)=2\,\max\limits_{\w,\nui}\log\left[\frac{\Lik(\HH|\data)}{\Lik(\CVR|\data)}\cdot\frac{\Lik(\nui|\auxdata)}{\Lik({\boldsymbol0}|\auxdata)}\right]\,,
\eeq
while the ``correction'' term $\Delta$ does not contain the neural network and involves exclusively the Reference hypothesis
\beq\label{delta}
\Delta(\data,\auxdata)=2\,\max\limits_{\nui}\log\left[\frac{\Lik(\RH|\data)}{\Lik(\CVR|\data)}\cdot\frac{\Lik(\nui|\auxdata)}{{\Lik({\boldsymbol0}|\auxdata)}}\right]\,.
\eeq
Both $\tau$ and $\Delta$ are positive-definite. Since they contribute with opposite sign, the test statistic $t$ will emerge from a cancellation between these two terms. The cancellation is more and more severe the more the data happen to favor a value of $\nui$ that is far from the central value. In Section~\ref{af} we will describe the nature and the origin of this cancellation in connection with the asymptotic formulae for the distribution of $t$. Below we outline our strategy for computing $\tau$ and $\Delta$, starting from the latter term.

\subsection{Learning the effect of nuisance parameters}\label{sec:nuplearn}

The correction term $\Delta$ in eq.~(\ref{delta}) is the log-ratio between the likelihood of the Reference hypothesis evaluated with best-fit values of the nuisance parameters, and the one with central-value nuisance parameters. This object is of interest for any statistical analysis to be performed on the dataset $\data$, as it provides a first indication of the data compatibility with the Reference hypothesis. In particular a sizable departure of the best-fit nuisance parameters from the central values should be monitored as an indication of a mis-modeling of the Reference hypothesis or possibly of a new physics effect. 

In order to introduce our strategy for the evaluation of $\Delta$, it is convenient to first recall the standard approach, employed in most LHC analyses, based on a binned Poisson likelihood approximation of $\Lik(\RH|\data)$. In this approach, the dataset gets binned and the observed counting in each bin is compared with the corresponding $\nui$-dependent cross section prediction. The predictions are obtained by computing each cross section for multiple values of the nuisance parameters and interpolating with a polynomial (or with the exponential of a polynomial, to enforce cross section positivity) around the central value $\nui=0$. A simple polynomial is sufficient to model the dependence of the cross section on the nuisances if their effect is small. The polynomial interpolation produces analytic expressions for the cross sections as a function of $\nui$, which are fed into the Poisson likelihood. The maximization over $\nui$ in eq.~(\ref{delta}) is then performed with standard computer packages.

In principle we could proceed to the evaluation of $\Delta$ exactly as described above. However we found it simpler and more effective to employ an un-binned $\Lik(\RH|\data)$ likelihood, obtained by reconstructing the ratio between the $\ndRef{x}$ and $\ndRefCV{x}$ distributions locally in the feature space. This is achieved by a rather straightforward adaptation of likelihood-reconstruction techniques based on neural networks developed in the literature~\cite{Cranmer:2015bka,Baldi:2016fzo,Brehmer:2018hga,Brehmer:2019xox,Chen:2020mev,Chen:1, Chen:2}. In particular, our implementation (briefly summarized below) closely follows Refs.~\cite{Chen:2020mev, Chen:1, Chen:2} to which we refer the reader for a more in-depth exposition. As for the regular binned approach, the basic idea is to employ a polynomial approximation for the dependence of the distribution on the nuisances. The polynomial coefficients, functions of the input $x$, are expressed as suitably trained neural networks. For instance, in the case of a single nuisance parameter $\nu$ we would write
\beq\label{nupar}
r(x;\nui)\equiv \frac{\ndRef{x}}{\ndRefCV{x}}= \exp\left[\nu\,\delta_1(x)+\frac12\nu^2\,\delta_2(x)+\ldots\right]\,,
\eeq
with the Taylor series expansion in the exponent truncated at some finite order. Clearly the truncation is justified only if the effect of the nuisance is a relatively small correction to the central-value distribution. More precisely, nuisance effects must be small when $\nu$ is in a ``plausibility'' range around $0$, compatibly with the shape of the auxiliary likelihood $\Lik(\nui|\auxdata)$. For instance, if the auxiliary likelihood is Gaussian with standard deviation $\sigma_\nu$, we should worry about the validity of the approximation in eq.~(\ref{nupar}) only for $\nu$ within few times $\pm\sigma_\nu$. Larger values are not relevant for the maximization in eq.~(\ref{delta}) because they are suppressed by $\Lik(\nui|\auxdata)$. Notice that in eq.~(\ref{nupar}) we might have opted for a polynomial approximation of the ratio $r$ rather than of its logarithm. However the latter choice guarantees the positivity of $r$ even when the numerical minimization algorithm is led to explore regions where $\nu$ is large. Furthermore working with $\log \,r(x;\nui)$ is more convenient for our purposes, as we will readily see. The polynomial expansion in eq.~(\ref{nupar}) can be straightforwardly generalized to deal with several nuisance parameters, including if needed mixed quadratic terms to capture the correlated effects of two different parameters.

Approximations $\widehat\delta(x)$ of the $\delta(x)$ coefficient functions are obtained as follows. Consider a continuous-output classifier $c(x;\nui)\in(0,1)$ defined as 
\beq
c(x;\nui)\equiv\frac1{1+{\widehat{r}}(x;\nui)}\,,
\eeq
where ${\widehat{r}}$ has the same dependence on the nuisance parameter as the true distribution ratio $r$. For instance in the case of a single nuisance parameter, and truncating  eq.~(\ref{nupar}) at the quadratic order, we have
\beq\label{rhat}
{\widehat{r}}(x;\nui)= \exp\left[\nu\,\widehat\delta_1(x)+\frac12\nu^2\,\widehat\delta_2(x)\right]\,,
\eeq
where $\widehat\delta_{1,2}(x)$ represents two suitably trained single-output neural network models.\footnote{Alternatively, the $\widehat\delta_{1,2,}(x)$ coefficient functions might be described by a single network with two outputs. The choice between the two options, as well as the choice of the neural networks hyper-parameters, obviously depends on the specific problem. Other models could also be considered for $\widehat\delta$ in alternative to neural networks.} 

The training is performed on a set of data samples ${\rm{S}}_0(\nui_i)$ that follow the distribution of $x$ in the $\RH$ hypothesis at different points $\nui=\nui_i\neq {\boldsymbol0}$ in the nuisance parameters space. Two distinct $\nui_i$ points are sufficient to learn the two coefficient functions associated to a single nuisance parameter at the quadratic order. Employing more points is possible and typically convenient for the accuracy of the coefficient functions reconstruction. Data samples produced in the central-value Reference hypothesis $\nui={\boldsymbol0}$ are also employed, one for each ${\rm{S}}_0(\nui_i)$ sample. These central-value Reference samples are denoted as ${\rm{S}}_1(\nui_i)$, in spite of the fact that they all follow the $\CVR$ hypothesis. Each event ``e'' in the samples has a weight $w_{\rm{e}}$, normalized such that the sum of the weights in each sample equals the total number of expected events in the corresponding hypothesis (i.e., ${\rm{N}}({{\rm{R}}_{{\boldsymbol\nu}_i}})$ for ${\rm{S}}_0(\nui_i)$ and $\NRefCV$ for ${\rm{S}}_1(\nui_i)$). The loss function is
\beq\label{lossnul}
L[\widehat\delta(\cdot)]=\sum\limits_{\nui_i}\left\{
\sum\limits_{{\rm{e}\in {\rm{S}}_0}(\nui_i)}w_{\rm{e}}[c(x_{\rm{e}};\nui_i)]^2+\sum\limits_{{\rm{e}\in {\rm{S}}_1}(\nui_i)}w_{\rm{e}}[1-c(x_{\rm{e}};\nui_i)]^2
\right\}\,.
\eeq
It is not difficult to show~\cite{Chen:2020mev} that the $\widehat\delta$ networks trained with the loss in eq.~(\ref{lossnul}) converge to the corresponding coefficient function $\delta$ in the limit where the samples are large, provided of course the true distribution ratio is in the form of eq.~(\ref{nupar}).

The basic strategy outlined above can be improved and refined in several aspects~\cite{Chen:1,Chen:2}, whose detailed description falls however outside the scope of the present paper. For our purposes it is sufficient to know that the coefficient functions in eq.~(\ref{nupar}) can be rather easily and accurately reconstructed. As such, the dependence on $\nui$ of the distribution ratio $r(x;\nui)$ is known analytically at each point $x$ of the features space. This solves our problem of evaluating the correction term $\Delta$ in eq.~(\ref{delta}), because $\Delta$ is
\beq\label{delta1}
\Delta(\data,\auxdata)=2\,\max\limits_{\nui}\left\{
\sum\limits_{x\in\data}\log[r(x;\nui)]-\NRefnu+\NRefCV+\log\left[\frac{\Lik(\nui|\auxdata)}{{\Lik({\boldsymbol0}|\auxdata)}}\right]
\right\}\,.
\eeq
Thanks to the fact that we adopted an exponential parametrization for $r$~(\ref{nupar}), the first term in the curly brackets is a polynomial in $\nui$. The constant term of the polynomial vanishes. The higher degree terms are the sum over $x\in\data$ of the corresponding $\delta(x)$ coefficients, approximated with the reconstructed $\widehat\delta(x)$ that are provided by the trained neural networks. The second term, $\NRefnu$, is proportional to the total cross section in the $\RH$ hypothesis. It can be approximated with a polynomial or with the exponential of the polynomial as in regular binned likelihood analyses. Finally, $\NRefCV$ is a constant and the log ratio between the $\nui$ and the ${\mathbf0}$ auxiliary likelihoods is also known in an analytical form. Actually in most cases the auxiliary likelihood is Gaussian and  $\log[\Lik(\nui|\auxdata)/\Lik({\boldsymbol0}|\auxdata)]$ is merely a quadratic polynomial. In summary, all the terms in the curly brackets of eq.~(\ref{delta1}) are known analytically. The maximization required to evaluate $\Delta$ is thus a trivial numerical operation for dedicated computer packages. 

\subsection{Maximum likelihood from minimal loss}\label{sec:mlml}

We now turn to the evaluation of the $\tau$ term defined in eq.~(\ref{tau}). This term involves the $\HH$ hypothesis, which foresees possible non-SM effects (i.e., departures from the Reference Model) in the distribution of $x$. Non-SM effects are parametrized by the neural network $f(x;\w)$ as in eq.~(\ref{nnh}). The calculation of $\tau$ involves the maximization over the neural network weights and biases, $\w$, and over the nuisance parameters $\nui$. The maximization will be performed by running a training algorithm, treating both $\w$ and $\nui$ as trainable parameters. The algorithm will exploit the knowledge of the $\delta$ coefficient functions that is provided by the $\widehat\delta$ neural networks as explained in the previous section. However the latter networks are pre-trained. Therefore their parameters are not trainable during the evaluation of $\tau$, even if they do appear in the loss function as we will readily see.

In order to turn the evaluation of $\tau$ into a training problem, the first step is to combine eq.~(\ref{nnh}) with the definition of $r$ in eq.~(\ref{nupar}), obtaining
\beq
\ndH{x}=e^{f(x;\w)}r(x;\nui)\,\ndRefCV{x}\,.
\eeq
We then rewrite $\tau$ in the form
\beq\label{tau1}
\tau(\data,\auxdata)=2\,\max\limits_{\w,\nui}\left\{
\sum\limits_{x\in\data}\left[f(x;\w)+\log(r(x;\nui))\right]-\NH+\NRefCV+\log\left[\frac{\Lik(\nui|\auxdata)}{\Lik({\boldsymbol0}|\auxdata)}\right]
\right\}.
\eeq
The first, third and fourth terms in the curly brackets are easily available. The first one depends on the neural network $f(x;\w)$, as well as on the coefficient functions $\delta$ (approximated by the neural networks $\widehat\delta$) through $r(x;\nui)$ in eq.~(\ref{nupar}). The second term is the total number of events in the $\HH$ hypothesis, given by
\beq
\NH=\int_{X}\hspace{-5pt}dx\,\ndH{x}=\int_{X}\hspace{-5pt}dx\,\ndRefCV{x}\cdot \exp\left[f(x;\w)+\log(r(x;\nui))\right]\,.
\eeq
Clearly $\NH$ is not easily available because $\ndRefCV{x}$ is not known in closed form and even if it was, computing the integral as a function of $\w$ and $\nui$ is numerically unfeasible. 

Evaluating $\NH$ requires us to employ a Reference data set $\reference=\{x_1,\ldots,x_{\Nreference}\}$. As described in Section~\ref{intro}, $\reference$ consists of synthetic instances of the variable $x$ that follow the Reference Model distribution. The $\reference$ set plus the data $\data$ constitute the sample that we will employ for training the neural network $f(x;\w)$. Notice that the $\reference$ dataset follows, by construction, the central-value distribution $\ndRefCV{x}$. It might result from a first-principle Monte Carlo simulation, or have data-driven origin. In both cases it might take the form of a weighted event sample.~\footnote{For instance, a data-driven background sample could be obtained from a MC-assisted reweighting of control region data as it is often done in SUSY searches.} 
 We choose the normalization of the weights such that
\beq\label{wnorm}
\sum\limits_{{\rm{e}\in \reference}}w_{\rm{e}}=\NRefCV\,.
\eeq
If the $\reference$ sample is ``unweighted'', all the weights are equal, and equal to $w_{\rm{e}}=\NRefCV/\Nreference$, with $\Nreference$ the Reference sample size. The Reference sample plays conceptually the same role as the central-value in regular model-dependent LHC searches. Its composition and origin is the same one of the samples ${\rm{S}}_1(\nui_i)$ employed to learn the effect of nuisance parameters with the strategy outlined in the previous section. 

With the normalization~(\ref{wnorm}), the weighted sum of a function of $x$ over the Reference sample approximates the integral of the function with integration measure $\ndRefCV{x}dx$. Therefore 
\beq\label{nhRef}
\NH\simeq \sum\limits_{{\rm{e}\in \reference}}w_{\rm{e}} \exp\left[f(x_{\rm{e}};\w)+\log(r(x_{\rm{e}};\nui))\right]\,,
\eeq
where the accuracy of the approximation improves with (square root of) the size of the Reference sample. In what follows we are going to assume an infinitely abundant Reference sample and turn the approximate equality above into a strict equality. Clearly in so doing we are ignoring the uncertainties associated with finite statistics of $\reference$. This is justified if $\Nreference\gg \NRefCV\sim \Ndata$, because in this case the statistical variability of $\tau$ is expectedly dominated by the statistical fluctuation of the data sample $\data$. All the results of the present paper are compatible with this expectation for $\Nreference$ a few times larger than $\Ndata$. 

By combining eq.s~(\ref{tau1}) and (\ref{nhRef}) (and (\ref{wnorm})) and by factoring out a minus sign to turn the maximization into a minimization, we express
\beq\label{tau2}
\tau(\data,\auxdata)=-2\,\min\limits_{\w,\nui}\left\{L\left[f(\cdot;\w),\,\nui;\,\widehat\delta(\cdot)\right]\right\}\,,
\eeq
where $L$ has the form of a loss function for a supervised training between the $\data$ and $\reference$ samples
\bea\label{loss}
&&L\left[f(\cdot;\w),\,\nui;\,\widehat\delta(\cdot)\right] = 
-\sum\limits_{x\in\data}\left[f(x;\w)+\log(r(x;\nui))\right]+ \sum\limits_{{\rm{e}\in \reference}}w_{\rm{e}} \left[
e^{f(x_{\rm{e}};\w)+\log(r(x_{\rm{e}};\nui))}-1\right]\nonumber\\
&&\hspace{104pt}-\log\left[\frac{\Lik(\nui|\auxdata)}{\Lik({\boldsymbol0}|\auxdata)}\right]\,.
\eea
The loss depends on the neural network function $f(\cdot;\w)$ and in particular on its trainable parameters $\w$. It also depends on the nuisance parameters $\nui$, through the ratio $r$ and through the auxiliary likelihood ratio term. The minimization over the nuisances is requested by eq.~(\ref{tau2}), therefore the nuisances should be treated as trainable parameters on the same footing as the neural network parameters $\w$. This is relatively straightforward to implement in standard deep learning packages, provided the loss depends on $\nui$ through analytically differentiable functions. This is the case for $r(x;\nui)$, and typically also for the auxiliary likelihood ratio. The loss also depends on the reconstructed coefficient functions $\widehat\delta$. However this dependence is purely parametric and the parameters of the $\widehat\delta$ networks are fixed at their optimal values, opportunely determined in a previous training as described in Section~\ref{sec:nuplearn}. After training, $\tau$ is obtained as minus two times the minimal loss owing to eq.~(\ref{tau2}).

Our strategy to evaluate $\tau$ is a relatively straightforward extension of the one developed in Ref.s~\cite{DAgnolo:2018cun,DAgnolo:2019vbw}. In the absence of nuisance parameters, namely in the limit where $r(x;\nui)$ is independent of $\nui$ and identically equal to one, the loss in eq.~(\ref{loss}) reduces to the one of Ref.s~\cite{DAgnolo:2018cun,DAgnolo:2019vbw}, plus the auxiliary log likelihood ratio that carries all the dependence on $\nui$ and can be minimized independently. The latter term however cancels in the test statistic $t$ when subtracting the correction term $\Delta$~(\ref{delta1}) and the results of Ref.s~\cite{DAgnolo:2018cun,DAgnolo:2019vbw} are recovered in the absence of nuisances, as it should be.

\subsection{Asymptotic formulae}\label{af}

We now discuss the actual feasibility of a frequentist hypothesis test based on our variable $t$~(\ref{tstat}). The generic problem with frequentist tests stems from the determination of the distribution of the $t$ variable in the null hypothesis, $P(t|H_0)$, out of which the $p$-value of the observed data is extracted. If the null hypothesis is a simple one, this can be obtained rigorously by running toy experiments, with a procedure that is computationally demanding but not unfeasible, especially if one does not target probing the extreme tail of the $t$  distribution. If instead the null hypothesis $H_0=\RH$ is composite as in this case, due to the nuisances, and if $P(t|\RH)$ (and in turn the $p$-value) depends on the value of $\nui$, the problem becomes unsolvable as one should in principle run toy experiments and compute $P(t|\RH)$ for each value of $\nui$. Indeed in frequentist statistics there is no notion of probability for the parameters. Consequently each value of $\nui$ defines an equally ``likely'' hypothesis in the null hypotheses set $\RH$. We can thus quantify the level of incompatibility of the data with the null hypothesis only by defining the $p$-value as the maximum $p$-value that is obtainable by a scan over the $\nui$ parameters in their entire allowed range. Since this is not feasible, the only option is to employ a suitably-designed test statistic variable, such that $P(t|\RH)$ is independent of $\nui$ to a good approximation. 

The considerations above are deeply rooted in the standard treatment of nuisance parameters. They actually constitute the very reason for the choice, in LHC analyses~\cite{Cowan:2010js}, of a specific Maximum Likelihood ratio test statistic, whose distribution is in fact independent of $\nui$ in the asymptotic limit where the number of observations is large. Specifically, $P(t|\RH)$ approaches a $\chi^2$ distribution with a number of degrees of freedom equal to the number of free parameters in the ``numerator'' hypothesis $\HH$, minus the number of parameters of the ``denominator'' hypothesis $\RH$, owing to the Wilks--Wald Theorem~\cite{Wald1943,Wilks:1938dza}. In a regular model-dependent search~\cite{Cowan:2010js}, the number of degrees of freedom of the $\chi^2$ equals the number of free parameters of the new physics model that is being searched for (i.e., the so-called ``parameters of interest''). The exact same asymptotic result applies in our case because our test statistic is also defined and rigorously computed as a Maximum Likelihood ratio. Its distribution in the null hypothesis will thus be independent of $\nui$ and approach the $\chi^2$. The number of degrees of freedom is given in this case by the number of trainable parameters $\w$ of the neural network. 

As already stressed in Section~\ref{intro}, however, asymptotic formulae such as the Wilks--Wald Theorem only hold in the limit of an infinitely large data set and therefore they offer no guarantee that $P(t|\RH)$ will resemble a $\chi^2$ (and be independent of $\nui$) in concrete analyses where the dataset has a finite size. At fixed dataset size, whether this is the case or not depends on the complexity (or expressivity) of the parameter-dependent hypothesis that is being compared with the data. When fitted by the likelihood maximization, an extremely flexible hypothesis will adapt its free parameters to reproduce (overfit) the observed data points individually. Therefore the value of $t$ that results from the maximization can be driven by low-statistics portions of the dataset and thus violate the asymptotic condition even if the total size of the dataset is large. The expressivity of our hypothesis is driven by the architecture (number of neurons and layers) of the neural network $f(x;\w)$, and by the other hyper-parameters (a weight clipping, in our implementation) that regularize the network preventing it from developing overly sharp features. We can thus enforce the validity of the asymptotic formula, i.e. ensure that $P(t|\RH)$ is close to a $\chi^2$ and independent of $\nui$, by properly selecting the neural network hyper-parameters.

For the selection of the hyper-parameters according to the $\chi^2$ compatibility criterion we proceed as in Ref.~\cite{DAgnolo:2019vbw}, where this criterion had been already introduced on a more heuristic basis, unrelated with nuisance parameters. We generate toy datasets following the central-value hypothesis $\CVR$, we compute $t$ and we compare its empirical distribution with a $\chi^2$ with as many degrees of freedom as the number of parameters of the neural network. We select the largest neural network architecture and the maximal weight clipping for which a good level of compatibility is found. Notice that whether or not a given neural network model is sufficiently ``simple'' to respect the asymptotic formula is conceptually unrelated with the presence of nuisance parameters. Furthermore our goal is to show that the presence of nuisances does not affect the distribution of $t$. Therefore when we enforce the $\chi^2$ compatibility, with the strategy outlined above, we compute $t$ as if nuisance parameters were absent. After the model is selected, based on the Wilks--Wald Theorem we expect that the distribution of $t$ will be a $\chi^2$ with the same number of degrees of freedom even in the presence of nuisance parameters. This can be verified by recomputing the distribution of $t$, including this time the effect of nuisances, on the $\CVR$ toys and on new toy samples generated according to $\RH$ with different values $\nui\neq{\mathbf{0}}$ of the nuisance parameters. Explicit implementations of this procedure, and confirmations of the validity of the asymptotic formulae, will be described in Sections~\ref{sec:one} and~\ref{sec:two}.

The Wilks--Wald Theorem also enables us to develop a qualitative understanding of the interplay between the $\tau$ and $\Delta$ terms in the determination of $t$ (eq.~(\ref{tstat1})). Both $\tau$ (eq.~(\ref{tau})) and $\Delta$ (eq.~(\ref{delta})) are Maximum Likelihood log-ratios, with the simple hypothesis $\CVR$ playing the role of the denominator hypothesis. Therefore $\tau$ and $\Delta$ are also distributed as a $\chi^2_d$ with $d$ degrees of freedom, if the data follow the $\CVR$ hypothesis itself. In the case of $\tau$, $d$ is the number of neural network parameters plus the number of nuisance parameters. The number of degrees of freedom of $\Delta$ is instead given by the number of nuisance parameters. The test statistic $t$, whose value emerges from a cancellation between $\tau$ and $\Delta$, has $d$ equal to the number of neural network parameters, as previously discussed. The cancellation is not severe in this case, because the number of nuisance parameters is typically smaller than the number of neural network parameters. Namely the values of $\tau$ and $\Delta$ for each individual toy will not be, on average, much larger that $t=\tau-\Delta$. Suppose instead that the data follow $\RH$ with some $\nui\neq{\boldsymbol0}$. This hypothesis belongs to the numerator (composite) hypothesis in the definitions of $\tau$ and $\Delta$. The Wilks--Wald Theorem predicts in this case non-central  $\chi^2$ distributions~\cite{Wilks:1938dza}, with increasingly large non-centrality parameters as we increase the distance between $\nui$ and ${\boldsymbol0}$. Therefore when we compute $P(t|\RH)$ with larger and larger $\nui$, the $\tau$ and $\Delta$ distributions shift more and more to the right and their typical value over the toys becomes large. The typical value of $t$ is instead given by the number of neural network parameters, because $t$ follows a central $\chi^2$ distribution independently of $\nui$. A sharp correlation between $\tau$ and $\Delta$ will thus engineer a delicate cancellation on toys generated with very large values of the nuisance parameters. The occurrence of the cancellation amplifies the uncertainties in the calculation of $\tau$ and $\Delta$ that emerge (dominantly) from the imperfect modeling of the $\delta(x)$ coefficient functions. Obtaining a $\chi^2$ for the distribution of $t$ is thus increasingly demanding at large $\nui$, as we will see in Sections~\ref{sec:one} and~\ref{sec:two}.

\subsection{New physics in auxiliary measurements or in control regions}\label{sec:npcr}

The step we took in eq.~(\ref{auxd}) of postulating the absence of new physics in the auxiliary data deserves further comments. In regular model-dependent searches for new physics the alternative hypothesis $H_1$ is a physical model that accounts for new phenomena in addition to the SM ones. One can thus assess whether or not these new phenomena can manifest themself in the auxiliary data. If they do not, eq.~(\ref{auxd}) is justified. The situation is different in model-independent searches. On one hand, there is no way to tell if eq.~(\ref{auxd}) holds because the new physics model is not given. On the other hand, in our framework we are always free to postulate eq.~(\ref{auxd}). In a model-dependent search eq.~(\ref{auxd}) could be wrong, in our case it is a restriction on the set of new physics models that we are testing.

Still it is interesting to discuss how the model-independent strategy that we are constructing would react to the presence of new physics effects in the auxiliary data. 
New (or mis-modeled) effects in auxiliary data could in general reduce the sensitivity of the test to new physics, however it is not obvious that this reduction will be significant. Consider the extreme case in which new physics is absent from the dataset of interest, and is present only in the auxiliary measurements. The new physics effects make the true auxiliary likelihood function different from the postulated one, $\Lik(\nui|\auxdata)$. Therefore, in the likelihood maximization, the $\Lik(\nui|\auxdata)$ term will push $\nui$ to values that are different from the true values of the nuisance parameters. This will occur both in the maximization of the $\Lik(\RH|\data,\auxdata)$ and of the $\Lik(\HH|\data,\auxdata)$ likelihoods. For these incorrect values of the nuisance parameters, $\ndRef{x}$ does not provide a good description of the distribution of the data of interest $\data$. Therefore the maximal likelihood of $\RH$ will be small, due to the mismatch between the data and the Reference distribution estimated from the ``signal-polluted'' auxiliary dataset. The $\HH$ hypothesis instead possesses enough flexibility to adapt $\ndH{x}$ according to the data of interest, thanks to the flexibility of the neural network~(\ref{nnh}). The likelihood of $\HH$ will thus possess a high maximum, in the configuration where $\nui$ maximizes the auxiliary likelihood and the neural network accounts for the discrepancy between the $x$ distribution at that value of $\nui$ and the true $x$ distribution at the true value of the nuisance parameters. This can enable our test to reveal a tension of the data with the Reference Model even in this limiting configuration, as we will see happening in Section~\ref{sec:sens} in a simple setup. New physics effects in the auxiliary data might thus not spoil the potential to achieve a discovery. On the other hand, they would complicate its interpretation.

Similar considerations hold for possible new physics contaminations in the Reference dataset $\reference$ employed for training. These contaminations emerge if $\reference$ has a data-driven origin, and if new physics affects the distribution of the data control region. Since the control region data are transferred to the region of interest by assuming the validity of the Reference Model, the net effect is a mismatch between the true distribution of $x$ in the (central-value) Reference Model and the actual distribution of the instances of $x$ in the Reference sample. As for auxiliary measurements, new physics in control regions does not necessarily spoil the sensitivity to new physics. Indeed our test is sensitive to generic departures of the observed data distribution with respect to the distribution of the Reference dataset. Departures which are due to a mis-modeling of the Reference induced by new physics in the control region, rather than to new physics in the data of interest, could still be seen. Our strategy would instead loose completely sensitivity if new physics affects the control region and the data of interest in the exact same way, because in this case there would be strictly no difference between the distribution of the data and the one of the Reference dataset. 

\section{Step-by-Step implementation}\label{sec:one}

The present Section describes the detailed implementation of our strategy and its validation in a simple case study that will serve as an explanatory example throughout the presentation of the algorithm. In particular, we consider a one-dimensional feature $x\in[0,\infty)$ with exponentially falling distribution in the Reference hypothesis. We assume that our knowledge of the Reference hypothesis is not perfect and that our lack of knowledge is described by a two-dimensional nuisance parameters vector $\nui=(\nu_{\rm\textsc{n}},\nu_{\rm\textsc{s}})$. The two parameters account, respectively, for the imperfect knowledge of the normalization of the distribution (i.e., of the total number of expected events $\NRefnu\equiv e^{\nu_{\rm\textsc{n}}}\NRefCV$) and of a multiplicative ``scale'' factor (defined by $x\hspace{-2pt}=\hspace{-2pt}x_{\rm{meas.}}\hspace{-2pt}=\hspace{-2pt}e^{\nu_{\rm\textsc{s}}}x_{\rm{true}}$) in the measurement of $x$. The Reference Model distribution of $x$ reads
\beq\label{1dDist}
\ndRef{x}=n(x|{\rm{R}}_{\nu_{\rm\textsc{n}},\nu_{\rm\textsc{s}}}) =  \NRefCV\, \exp\left[{-x\,e^{-\nu_{\rm\textsc{s}}}}-{\nu_{\rm\textsc{s}}}+{\nu_{\rm\textsc{n}}}\right]\,,
\eeq
with the total number of expected events in the central-value hypothesis, $\NRefCV$, fixed at $\NRefCV=2\,000$. As discussed in Section~\ref{sec:cvr}, the central-value Reference hypothesis $\CVR$ is defined to be at the point $(\nu_{\rm\textsc{n}},\nu_{\rm\textsc{s}})=(0,0)$ in the nuisances' parameter space. We have parametrized the normalization, $e^{\nu_{\rm\textsc{n}}}$, and the scale factor, $e^{\nu_{\rm\textsc{s}}}$, so that they are positive in the entire real plane spanned by $(\nu_{\rm\textsc{n}},\nu_{\rm\textsc{s}})$. 

We suppose that the normalization and the scale nuisances are measured independently using an auxiliary set of data $\auxdata$. The estimators of the measurements central values are denoted as $\widehat\nu_{\rm\textsc{n}}=\widehat\nu_{\rm\textsc{n}}(\auxdata)$ and $\widehat\nu_{\rm\textsc{s}}=\widehat\nu_{\rm\textsc{s}}(\auxdata)$. We assume that these estimators are unbiased and Gaussian-distributed with standard deviations $\sigma_{\rm\textsc{n}}$ and $\sigma_{\rm\textsc{s}}$. The auxiliary likelihood log-ratio thus reads
\beq\label{1dnui}
2\,\log\left[\frac{\Lik(\nui|\auxdata)}{{\Lik({\boldsymbol0}|\auxdata)}}\right]=-
\left(\frac{
\widehat\nu_{\rm\textsc{n}}-\nu_{\rm\textsc{n}}}{\sigma_{\rm\textsc{n}}}
\right)^2+
\left(\frac{
\widehat\nu_{\rm\textsc{n}}}{\sigma_{\rm\textsc{n}}}
\right)^2
-\left(\frac{
\widehat\nu_{\rm\textsc{s}}-\nu_{\rm\textsc{s}}}{\sigma_{\rm\textsc{s}}}
\right)^2+
\left(\frac{
\widehat\nu_{\rm\textsc{s}}}{\sigma_{\rm\textsc{s}}}
\right)^2\,.
\eeq
It should be noted that $\widehat\nu_{\rm\textsc{n}}$ and $\widehat\nu_{\rm\textsc{s}}$ are statistical variables, owing to their dependence on the auxiliary data $\auxdata$. Therefore we must let them fluctuate when generating toys. Namely when generating toys in the ${{\rm{R}}_{\boldsymbol{\nu^*}}}$ Reference hypothesis, $\widehat\nu_{\rm\textsc{n}}$ and $\widehat\nu_{\rm\textsc{s}}$ must be thrown from Gaussian distributions with standard deviations $\sigma_{\rm\textsc{n}}$ and $\sigma_{\rm\textsc{s}}$, centered at some chosen true values of the nuisance parameters $\nu_{\rm\textsc{n}}=\nu_{\rm\textsc{n}}^*$ and $\nu_{\rm\textsc{s}}=\nu_{\rm\textsc{s}}^*$, respectively.  

The rest of this section is structured as follows. In Section~\ref{sec3:model} we describe the selection of the neural network model and regularization parameters based on the $\chi^2$ compatibility criterion introduced in the previous section (and in Ref.~\cite{DAgnolo:2019vbw}), and in particular in Section~\ref{af}. Next, in Section~\ref{sec:lm}, we illustrate the reconstruction of the coefficient functions that model the dependence of the Reference Model distribution on the nuisance parameters, following Section~\ref{sec:nuplearn}. In Section~\ref{sec:t} we present our implementation of the calculation of the test statistic as in Section~\ref{sec:mlml}. In Section~\ref{sec:validation} we validate our strategy by verifying the asymptotic formulae of Section~\ref{af} and in turn the independence of the distribution of the test statistic on the true value of the nuisance parameters. Finally, in Section~\ref{sec:sens} we study the sensitivity to putative ``new physics'' signals that distort the distribution of $x$ relative to the Reference Model expectation in eq.~(\ref{1dDist}). It should be emphasized that this latter study has a merely illustrative purpose. All the steps that are needed to set up our strategy, from the model selection to the evaluation of the distribution of the test statistic, are performed based exclusively on knowledge of the Reference Model and not on putative new physics signals, as appropriate for a model-independent search strategy.

While presented in the context of a simple univariate problem that is rather far from a realistic LHC data analysis problem, the technical implementation of all the steps described in the present section is straightforwardly applicable to more complex situations. The application to a more realistic problem will be discussed in Section~\ref{sec:two}.

\subsection{Model selection}\label{sec3:model}

The first step towards the implementation of our strategy is to select the hyper-parameters of the neural network model ``$f(x;\w)$'', which we employ to parametrize possible new physics (or Beyond the SM, BSM) effects as in eq.~(\ref{nnh}). We restrict our attention to fully-connected feedforward neural networks, with an upper bound on the absolute value of each weight and bias. The upper limit is set by a weight clipping regularization parameter that needs to be selected. The other hyper-parameters are the number of hidden layers and of neurons per layer that define the neural network architecture.

According to the general principles outlined in Section~\ref{af}, the model selection results from two competing principles. The first one is that the model should have the highest complexity that can be handled by the available computational resources in a reasonable amount of time. This maximizes the model's capability to fit complex departures from the Reference Model expectation, making it sensitive to the largest possible variety of putative new physics signals. On the other hand, the model should be simple enough for the distribution of the associated test statistic to be in the asymptotic regime, given the finite amount of training data. This condition is enforced by monitoring the compatibility with the $\chi^2$ asymptotic formula for the test statistic distribution. 

As explained in Section~\ref{af}, the $\chi^2$ compatibility condition that underlies the selection of the neural network hyperparameters will be enforced in the limit where the nuisance parameters do not affect the distribution of the variable $x$ or, equivalently, in the limit where the auxiliary measurements of the nuisance parameters are infinitely accurate (i.e., $\sigma_{\rm\textsc{n,s}}\to0$). It is easy to see from the results of Section~\ref{sec:found}, or from Ref.s~\cite{DAgnolo:2018cun,DAgnolo:2019vbw}, that the test statistic in this limit becomes
\beq\label{tbar}
\overline{t}(\data)=2\,\max\limits_{\w}\,\log\left[\frac{\Lik(H_\w|\data)}{\Lik(R_0|\data)}\right]=-2\,\min\limits_{\w}\left\{{\bar{L}}\left[f(\cdot;\w);\right]\right\}\,.
\eeq
The minimization is performed by training the network $f$ with the loss function
\beq\label{sec3:loss}
{\bar{L}}\left[f(\cdot;\w)\right] = 
-\sum\limits_{x\in\data}\left[f(x;\w)\right]+ \sum\limits_{{{\rm{e}}\in \reference}}w_{\rm{e}} \left[e^{f(x_{\rm{e}};\w)}-1\right].
\eeq
The asymptotic distribution of $\overline{t}$ is a $\chi^2$ with a number of degrees of freedom which is equal to the number of trainable parameters of the neural network. The $\chi^2$ compatibility of a given neural network model will be monitored by generating toy instances of the dataset $\data$ in the $\CVR$ hypothesis, running the training algorithm on each of them, computing the empirical probability distribution of $\overline{t}$ and comparing it with the $\chi^2$. 

\begin{figure}[t]
\begin{center}
\includegraphics[width=16.2cm]{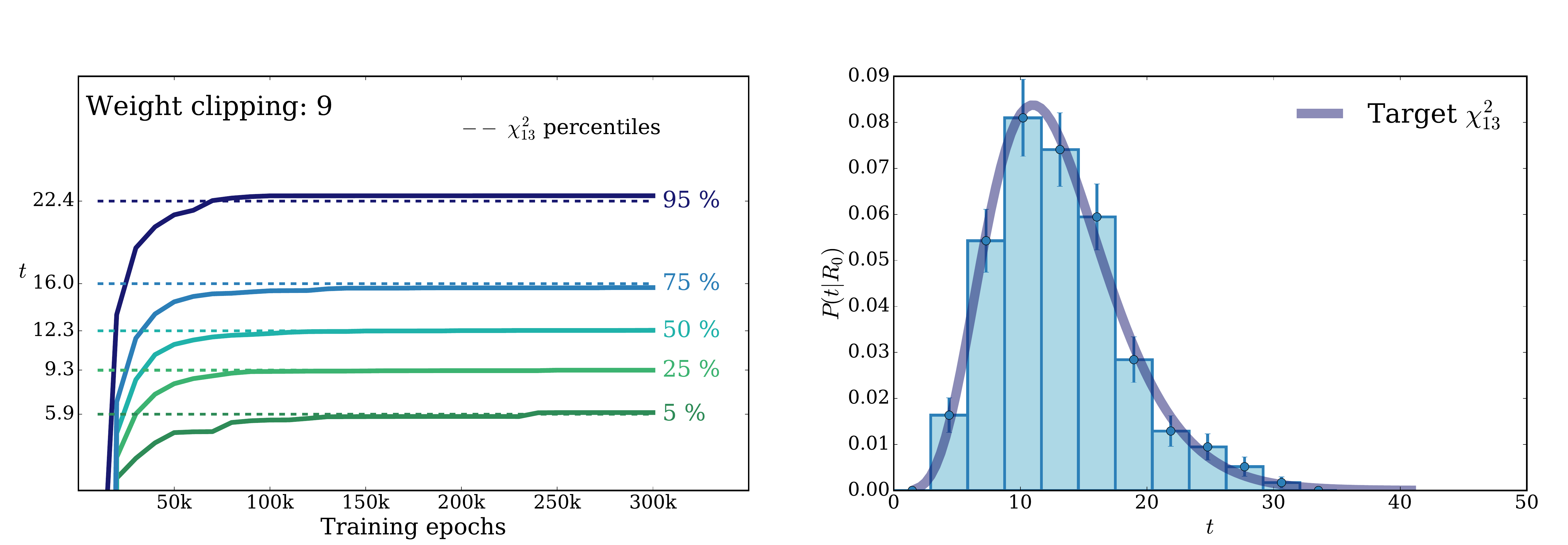}\\
\includegraphics[width=16.8cm]{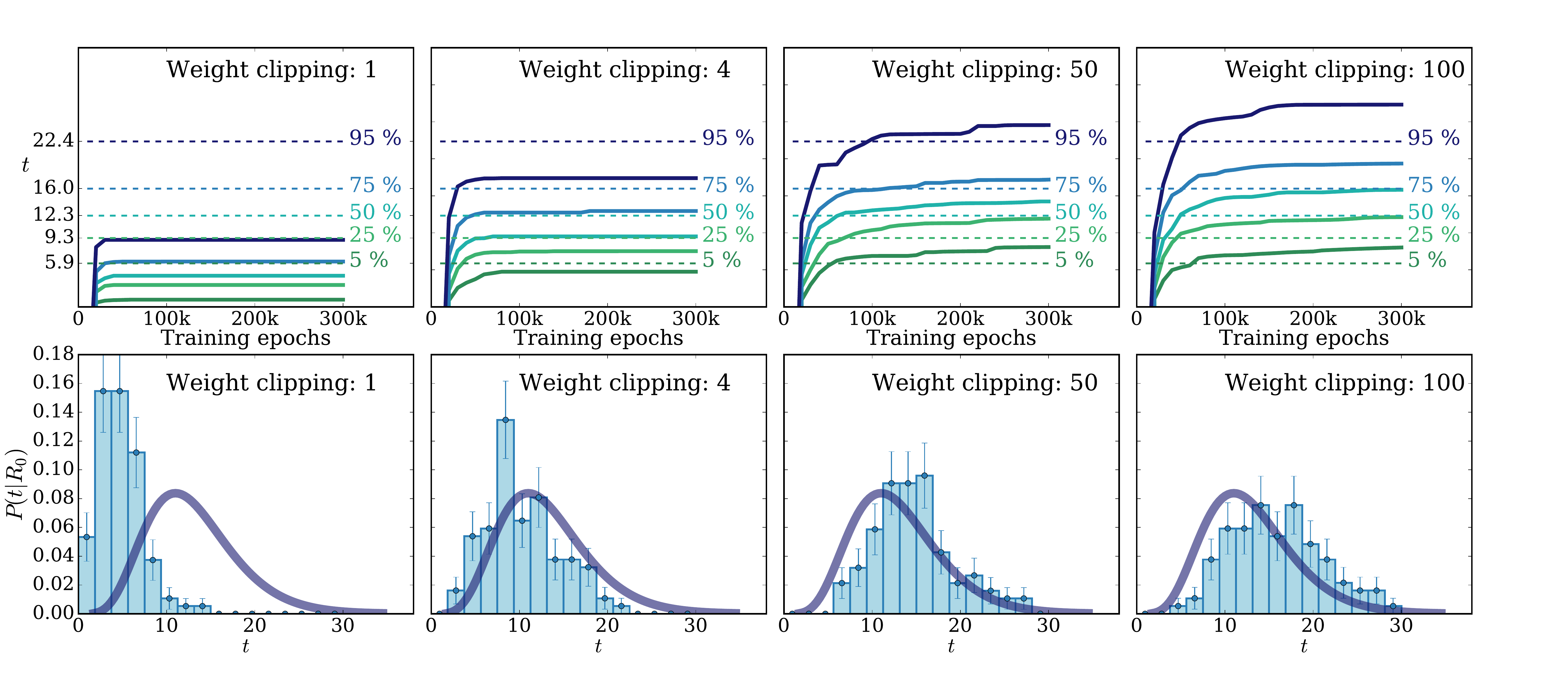}
\caption{Empirical distributions of ${\overline{t}}$ after $300\,000$ training epochs for different values of the weight clipping parameter, compared with the $\chi^2_{13}$ distribution expected in the asymptotic limit for the $(1,4,1)$ network. The evolution during training of the ${\overline{t}}$ distribution percentiles, compared with the $\chi^2_{13}$ expectation, is also shown. Only $100$ toy datasets are employed to produce the results shown in the figure, except for the ones for weight clipping equal to $9$ where all the $400$ toys are used.}
\label{WCtuning_1D}
\end{center}
\end{figure}

We first discuss how to select the weight clipping regularization parameter for a given architecture of the neural network. We consider for illustration, in the simple univariate example at hand, a network with four nodes in the hidden layer (and one-dimensional input and output). We refer to this architecture as $(1,4,1)$, for brevity. This network has a total of $13$ trainable parameters, therefore the target $\overline{t}$ distribution is a $\chi^2_{13}$ with $13$ degrees of freedom. We generated a Reference sample $\reference$, with ${{\rm{N}}_{\cal{R}}}=200\,000=100\,\NRefCV$ entries, following the $\CVR$ distribution of the variable $x$ as given by eq.~(\ref{1dDist}) for $\nu_{\rm\textsc{n,s}}=0$. The sample is unweighted, therefore the weights in the sample are all equal and $w_{\rm{e}}=\NRefCV/{{\rm{N}}_{\cal{R}}}=0.01$. We also generate $400$ toy instances of the dataset $\data$ in the same hypothesis. The number of instances of $x$ in $\data$, $\Ndata$, is thrown from a Poisson distribution with mean $\NRefCV=2\,000$ in accordance with the $\CVR$ expectation. For different values of the weight clipping parameter, ranging from $1$ to $100$, we train the neural network with the loss in eq.~(\ref{sec3:loss}) and we compute $\overline{t}(\data)$ on the toy datasets using eq.~(\ref{tbar}). The empirical $P(\overline{t}|\CVR)$ distributions obtained in this way after $300\,000$ training epochs, and some of its percentiles as a function of the number of epochs, are reported in Figure~\ref{WCtuning_1D}.

We see that for large values of the weight clipping parameter the distribution sits slightly to the right of the target $\chi^2$ with $13$ degrees of freedom. Furthermore the training is not stable and significant changes in the $\overline{t}$ percentiles (especially the $95\%$ one) occur even after $150\,000$ epochs. Very small values of the weight clipping make the distribution stable with training, but push it lower than the $\chi^2_{13}$ expectation. A good compatibility is instead obtained for intermediate values of the weight clipping parameter. We see that a weight clipping equal to $9$ reproduces the $\chi^2_{13}$ formula quite accurately.

The strategy to find the value of the weight clipping parameter that best complies with the $\chi^2$ compatibility criterion can be refined and optimized. We can start from one small and one large value of the weight clipping, for which we expect that the distribution of ${\overline{t}}$ will, respectively, undershoot and overshoot the $\chi^2$ expectation, and compute ${\overline{t}}$ by running the training algorithm on a limited number $n$ of toy datasets. The average of ${\overline{t}}$ over the $n$ toys will be below (above) the mean of the target $\chi^2$ distribution (i.e., $13$, in the case at hand) for the small (large) value of the weight clipping. We thus obtain a window of values where the optimal weight clipping sits, which can be further narrowed by applying a standard root finding algorithm on the average ${\overline{t}}$ compared with the expected mean. Clearly the average ${\overline{t}}$ will be affected by a relatively large error if $n$ is small. Therefore after a few iterations of the root finding algorithm, it will become compatible with the expected mean, preventing us from further restricting the weight clipping compatibility window. 

Rather than looking at the compatibility of the average, a more powerful compatibility test should be employed at this stage in order to pick up the optimal weight clipping value inside the window. Furthermore this test should be sensitive to the entire shape of the distribution and not only to its central value. One can consider for instance a Kolmogorov--Smirnov (KS) test and maximize, in the window, the $p$-value for the compatibility with the target $\chi^2$ of the empirical ${\overline{t}}$ distribution. \footnote{Other compatibility tests can be adopted as well. For instance, one could minimize pdf-distance metrics such as the Kullback-Leibler divergence or the Earth Mover distance.}

It is advantageous to implement the strategy described above using a rather small number $n$ of toy datasets, because training could become computationally demanding in realistic applications of our strategy. On the other hand, if $n$ is small the KS compatibility test has limited power, leaving space for considerable departures from the target $\chi^2$ of the true distribution of ${\overline{t}}$, even with the value of the weight clipping that has been selected as ``optimal''. A more accurate determination of the optimal weight clipping could however be obtaining by increasing $n$ and repeating the previous optimization step. Clearly at this stage one could restrict to the much narrower window obtained at the end of the previous step, and benefit from the previous determination of the optimal weight clipping in order to speed up the convergence. The entire procedure could be further repeated with an even larger $n$, until a certain compatibility goal is achieved. For instance, one might require a KS $p$-value larger than some threshold, at the optimal weight clipping point, with a relatively large number $n$ (say, $400$) of toy experiments. 

\begin{table}[t]
\begin{center}
\scalebox{0.75}{
\begin{tabular}{|c|cc|cc|cc|}
	\hline
	{\textbf{Weight}}
	& \multicolumn{2}{c|}{{\textbf{40 toys}}}	&  \multicolumn{2}{c|}{\textbf{100 toys}} 	&  \multicolumn{2}{c|}{\textbf{400 toys}}\\
	{\textbf{Clipping}}							& KS $p$-value 	& $\langle\overline{t}\rangle-13$		& KS $p$-value 	& $\langle\overline{t}\rangle-13$		& KS $p$-value 	& $\langle\overline{t}\rangle-13$\\
	\hline
	35		&0.10	& 1.0 $\pm$ 0.7			& $<10^{-5}$   		& 2.6 $\pm$ 0.6		&	&\\
	15		&0.09	& 2.0 $\pm$ 1.5			& 0.01 			& 1.5 $\pm$ 0.7		&	&\\
	11		&0.36	& 1.0 $\pm$ 0.8			& 0.01 			& 1.2 $\pm$ 0.5		&	&\\ 
	10		&0.86	& 0.6 $\pm$ 0.9			& 0.56 			& 0.9 $\pm$ 0.6		& 0.78		& 0.4 $\pm$ 0.3	\\ 
	9		& 0.68	& 0.5 $\pm$ 0.9			& 1.0 			& 0.0 $\pm$ 0.5		& 0.93		& 0.0 $\pm$ 0.3	\\ 
	8		& 0.44 	& 0.3 $\pm$ 0.7			& 0.53			&-0.4 $\pm$ 0.4		& 0.42 		& -0.3 $\pm$ 0.2	\\
	7		& 0.40	& -0.6 $\pm$ 0.8			& 0.21			& -0.8 $\pm$ 0.4		&0.02		&-0.7 $\pm$ 0.2	\\ 
	4		& 0.11	& -1.4 $\pm$ 0.7			& $<10^{-5}$		& -2.7 $\pm$ 0.4		&	&						\\ 
	\hline
\end{tabular}
}
\end{center}
\caption{Kolmogorov-Smirnov $p$-value and average $\overline{t}$ (minus the expected mean of $13$) for the $(1,4,1)$ network trained over samples of $40$, $100$ and $400$ toy datasets, for different values of the weight clipping regularization parameter}
\label{tab_1D_wclip_optimization}
\end{table}%

The results reported in Table~\ref{tab_1D_wclip_optimization} illustrate the weight clipping optimization strategy described above for the $(1,4,1)$ network in the univariate problem under consideration. Actually a systematic optimization strategy is not needed to deal with the simple problem at hand, because training is sufficiently fast to test many points in the weight clipping parameter space with a large number of toys. Furthermore the departures from the $\chi^2$ of the empirical $\overline{t}$ distribution are rather mild, as shown by Figure~\ref{WCtuning_1D}, in a rather wide range of weight clipping values. We will instead need the optimization strategy in order to deal with the more realistic five-features problem of Section~\ref{sec:two} where training is longer and the distribution is more sensitive to the weight clipping parameter. 

Up to now we have considered a single architecture, and found one choice of the weight clipping parameter that ensures a good level of $\chi^2$ compatibility. According to general principles of model selection, we should now switch to more complex architectures, with more neurons and/or hidden layers, aiming at selecting the most complex network that respects the asymptotic formula and that can be practically handled by the available computational resources. We saw in Ref.~\cite{DAgnolo:2019vbw} that computational considerations play an important role in the selection, however the univariate problem at hand is not sufficiently demanding to illustrate this aspect. Indeed we have found $\chi^2$-compatible networks with up to one hundred neurons, which are clearly an overkill for the univariate problem. Therefore, we will not describe the process of architecture optimization in the univariate example and postpone the discussion to a more realistic context in Section~\ref{sec:two}. The $(1,4,1)$ network, with weight clipping equal to $9$, will be employed in the rest of the present section.

\subsection{Learning nuisances}\label{sec:lm}

We now turn to the problem of learning the effect of the nuisance parameters on the distribution of the variable $x$, following the methodology described in Section~\ref{sec:nuplearn}. In the simple univariate problem at hand, we have access to the distribution in closed form (eq.~(\ref{1dDist})), and in turn to the exact analytic expression for the log distribution ratio 
\beq\label{lran}
\log{r(x;\nui)} = \log\frac{\ndRef{x}}{\ndRefCV{x}}={\nu_{\rm\textsc{n}}}+x\,(1-e^{-\nu_{\rm\textsc{s}}})-{\nu_{\rm\textsc{s}}}\,.
\eeq
The dependence on the normalization nuisance ${\nu_{\rm\textsc{n}}}$ is trivial and it can be incorporated analytically, both in the univariate problem and in realistic analyses. The dependence on the scale nuisance ${\nu_{\rm\textsc{s}}}$ is more complex, and not analytically available in realistic problems. We thus approximate it by a Taylor series as in eq.~(\ref{nupar}). Namely we define
\beq\label{lrapp}
\log\,{{\widehat{r}}(x;\nui)}={\nu_{\rm\textsc{n}}}+{\nu_{\rm\textsc{s}}}\,\widehat\delta_1(x)+\frac12 \nu_{\rm\textsc{s}}^2\, \widehat\delta_2(x) 
+\ldots\,.
\eeq
Truncations of the ${\nu_{\rm\textsc{s}}}$ series at the first and at the second order will be considered in what follows. 

We model each $\widehat\delta_a(x)$ coefficient function (with $a$ ranging from $1$ to the desired order of the series truncation in eq.~(\ref{lrapp})) with fully-connected $(1,4,1)$ neural networks with ReLU activation functions, trained with the loss function in eq.~(\ref{lossnul}). The training samples ${{\rm{S}}_1}(\nui_i)$ and ${{\rm{S}}_0}(\nui_i)$ contain $20\,000$ events each. The events in ${{\rm{S}}_1}(\nui_i)$ are thrown according to the probability distribution of $x$ in the $\CVR$ hypothesis. The ones in ${{\rm{S}}_0}(\nui_i)$ are thrown according to the $\RH$ hypothesis at selected points $\nui_i=(0,\nu_{{\rm\textsc{s}},i})$ in the nuisance parameters space. The choice of the $\nu_{{\rm\textsc{s}},i}$ values used for training has a considerable impact on the quality of the reconstruction of the $\widehat\delta_a(x)$ functions. They should be such as to expose the dependence of the distribution ratio on each monomial of the expansion. For instance, when dealing with the quadratic approximation one would employ a relatively small value of $\nu_{{\rm\textsc{s}}}$, for which the linear term dominates, in order to learn $\widehat\delta_1(x)$, and a relatively large one for the reconstruction of $\widehat\delta_2(x)$. At least one additional value of $\nu_{{\rm\textsc{s}}}$ would be needed in order to go to the cubic order. This value would be taken even larger, namely in the regime where the quadratic approximation starts becoming insufficient and the dependence of the distribution ratio on the cubic term plays a role. Employing a redundant set of $\nu_{{\rm\textsc{s}},i}$'s (for instance, $4$ points rather than $2$ at the quadratic order) is beneficial. In general it is convenient to pick up the $\nu_{{\rm\textsc{s}},i}$'s in pairs of opposite sign, symmetric around the origin. 

\begin{figure}[t]
\includegraphics[width=16.2cm]{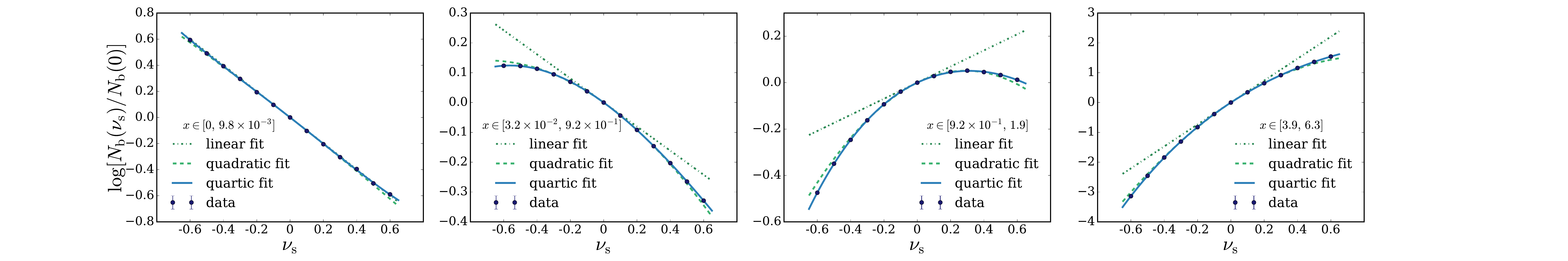}
\caption{The dependence on $\nu_{{\rm\textsc{s}}}$ of $\log{N_{\rm{b}}(\nu_{{\rm\textsc{s}}})}/{N_{\rm{b}}(0)}$ in selected bins. The dots represent the true value of the log-ratio. The linear, quadratic and quartic fits are performed using a subset of the true values points as explained in the main text.}
\label{1D_learning_nu2}
\end{figure}

The set of $\nu_{{\rm\textsc{s}},i}$'s that duly captures all the terms in the Taylor expansion can be determined by inspecting the dependence on $\nu_{{\rm\textsc{s}}}$ of the distribution integrated in bins, and identifying the points on the $\nu_{{\rm\textsc{s}}}$ axis where a change of regime (say, from linear to quadratic) is observed. This is illustrated in Figure~\ref{1D_learning_nu2}, where we plot the dependence on $\nu_{{\rm\textsc{s}}}$ of $\log{N_{\rm{b}}(\nu_{{\rm\textsc{s}}})}/{N_{\rm{b}}(0)}$, with $N_{\rm{b}}$ the integral of the distribution in selected bins of the variable $x$. The points represent the true value of the log ratio as obtained from the distribution in eq.~(\ref{1dDist}). The dot-dashed, dashed and continuous lines are the fit to these points with polynomials of order $1$, $2$ and $4$, respectively. More precisely the first-order polynomial fit only employs the points in the interval $\nu_{{\rm\textsc{s}}}\in[-0.1,0.1]$, the second-order one employs the range $\nu_{{\rm\textsc{s}}}\in[-0.3,0.3]$, while the fourth-order polynomial fit is performed on all the points. Compatibly with eq.~(\ref{lran}), we see that the behavior is almost exactly linear when $x$ is very small. Considerable departures from linearity are instead present, for bigger $x$, when $\nu_{{\rm\textsc{s}}}$ is as large as $0.3$ in absolute value. Based on these plots, for training the linear order we selected the set of values $\nu_{{\rm\textsc{s}},i}\in\{\pm0.05,\pm0.1\}$, for which the linear approximation is valid.\footnote{Obviously the linear approximation is also valid for even smaller $\nu_{{\rm\textsc{s}}}$. However, very small $\nu_{{\rm\textsc{s}},i}$'s reduces the impact of the nuisance parameters of the distribution of ${{\rm{S}}_0}(\nui_i)$ relative to the one of ${{\rm{S}}_1}(\nui_i)$, making training harder. The best option is to employ the largest $\nu_{{\rm\textsc{s}},i}$'s for which the linear approximation is still satisfactory.} The set $\nu_{{\rm\textsc{s}},i}\in\{\pm0.05,\,\pm0.3\}$ was instead employed for the quadratic order approximation. The figure also suggests that the quadratic order truncation in eq.~(\ref{lrapp}) should be sufficient to model the dependence of $\log{r(x;\nui)}$ on $\nu_{{\rm\textsc{s}}}$ in the entire phase-space of $x$, at least if we limit ourselves to the range $\nu_{{\rm\textsc{s}}}\in[-0.6,0.6]$.

\begin{figure}[t]
\centering
   \includegraphics[width=1\linewidth]{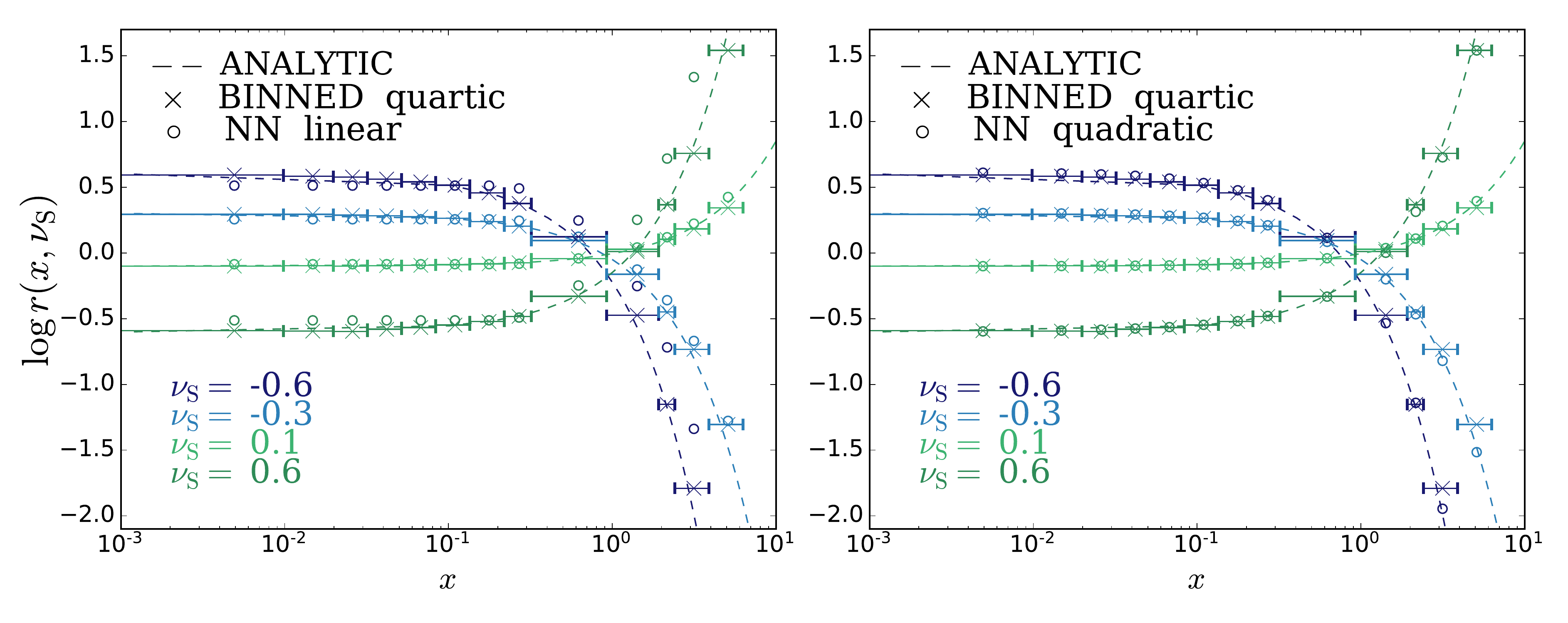}
\caption{The reconstructed distribution log-ratio (empty dots) for different values of $\nu_{{\rm\textsc{s}}}$, compared with the exact log-ratio and with the fourth-order binned approximation described in the main text. The two panels correspond to truncations of the series in eq.~(\ref{lrapp}) at linear and at the quadratic order.}
\label{1D_learning_nu}
\end{figure}

The quality of the reconstruction of the log-ratio is displayed in Figure~\ref{1D_learning_nu} for the two different polynomial orders (linear and quadratic) that we have considered for the truncation of the series in eq.~(\ref{lrapp}). The exact analytic log-ratio in eq.~(\ref{lran}) is represented as dashed lines, to be compared with the reconstructed ratio reported as empty dots. The different colors correspond to different values of $\nu_{{\rm\textsc{s}}}$. As expected, the first-order truncation is accurate only if $\nu_{{\rm\textsc{s}}}$ is small. The accuracy improves with the quadratic truncation, for which the reconstructed log-ratio is essentially identical to the exact log-ratio. It should be kept in mind that, as explained in Section~\ref{sec:nuplearn}, the $\nu_{{\rm\textsc{s}}}$ range where an accurate reconstruction is needed depends on the allowed range of variability of $\nu_{{\rm\textsc{s}}}$, namely on its standard deviation $\sigma_{\rm\textsc{s}}$. From the figure we see that the linear polynomial modeling is adequate only if $\sigma_{\rm\textsc{s}}$ is below around $0.3$, while with the quadratic one $\sigma_{\rm\textsc{s}}$ could be as large as $0.6$.\footnote{If $\sigma_{\rm\textsc{s}}$ was even larger, additional polynomial orders would be needed up to the point where the convergence of the Taylor series breaks down. After that, the only way to proceed would be to replace the Taylor series with a more rapidly convergent expansion of the log-ratio, if it exists, or to employ a parametric neural network~\cite{Baldi:2016fzo}.} The figure also reports the binned prediction for the log-ratio, as obtained from the quartic fit to $\log{N_{\rm{b}}(\nu_{{\rm\textsc{s}}})}/{N_{\rm{b}}(0)}$ previously described and displayed in Figure~\ref{1D_learning_nu2}. In realistic examples where the analytic log-ratio is not available, the binned prediction can be employed to monitor the quality of the reconstruction provided by the $\widehat\delta_a(x)$ networks. A more stringent test of the accuracy of the distribution log-ratio approximation, connected with the final validation of our strategy and its robustness to nuisances, will be discussed in Section~\ref{sec:validation}.

\subsection{Computing the test statistic}\label{sec:t}

\begin{figure}[t]
\begin{center}
\includegraphics[width=1.0\textwidth]{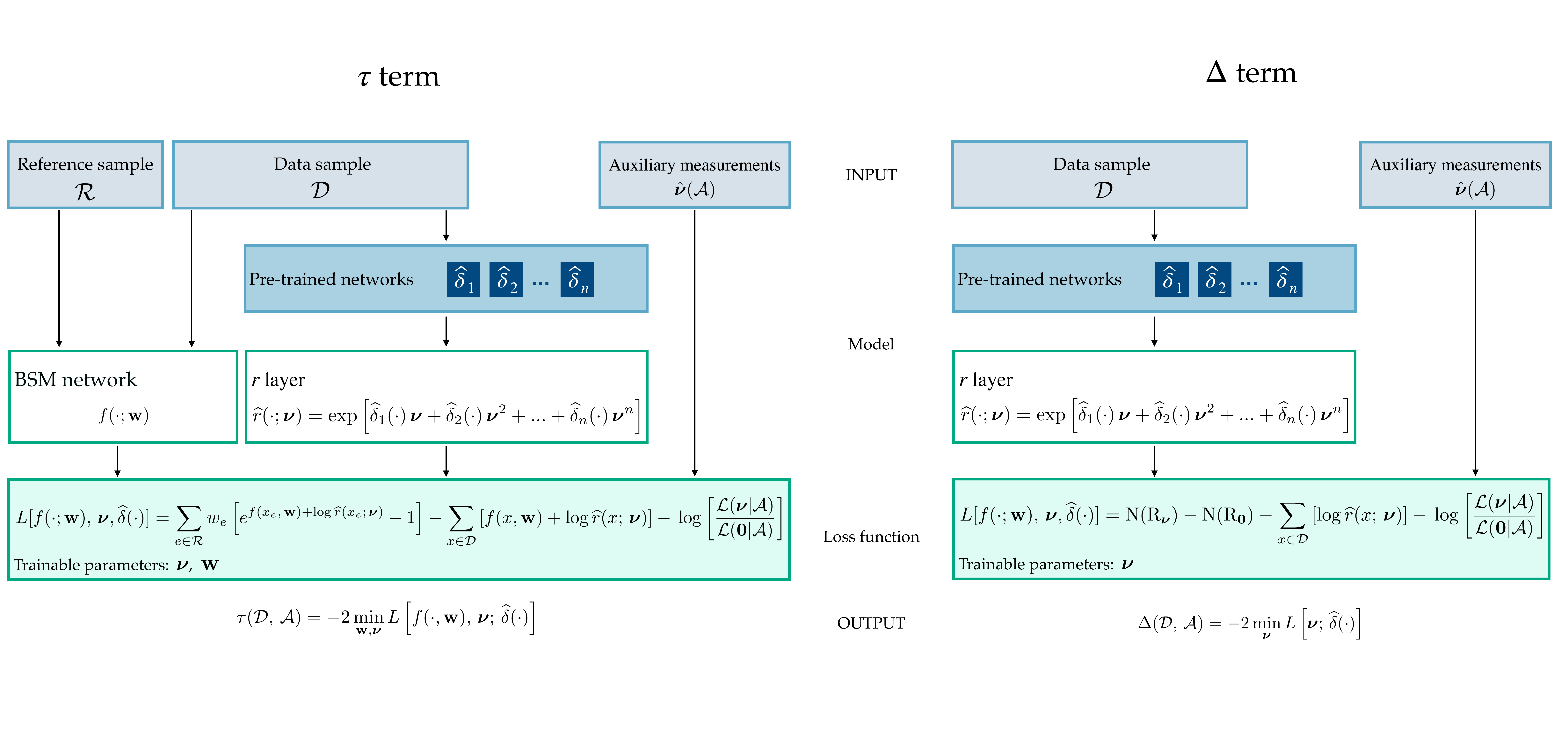}
\caption{Schematic representation of the TensorFlow implementation of our algorithm. }
\label{schematic2}
\end{center}
\end{figure}

We finally have at our disposal all the ingredients to compute the test statistic $t(\data,\auxdata)$. This consists of the $\tau$ term, subtracted by the correction $\Delta$. We now illustrate the evaluation of the two terms in turn, as implemented in the TensorFlow~\cite{tensorflow2015-whitepaper} package. The implementation is schematically represented in Figure~\ref{schematic2}, and the corresponding code is available at~\cite{Grosso_New_Physics_Learning_2021}. 

As described in Section~\ref{sec:mlml}, computing $\tau$ requires the simultaneous optimization of the parameters $\w$ of the neural network model $f(x;\w)$ (dubbed ``BSM network'' in the figure) and of the nuisance parameters $\nui$. The loss function is the one of eq.~(\ref{loss}). It depends on $\nui$ through the distribution ratio  $r$, or more precisely through its estimate ${\widehat{r}}(x;\nui)$ as in eq.~(\ref{rhat}). The estimated  ${\widehat{r}}$ ratio is implemented as a TensorFlow ``$\lambda$-layer'' (denoted as ``$r$ layer" in the figure) that takes as input the output of the $\widehat\delta$ networks and builds the required polynomial function of $\nui$. Notice that the parameters of the $\widehat\delta$ networks are ``fixed'' parameters during training, namely they are not optimized. Indeed, the $\widehat\delta$ networks have been trained at a previous stage of the implementation, as described in Section~\ref{sec:lm}. The evaluation of $\tau$ thus proceeds as shown in the left panel of Figure~\ref{schematic2}. The inputs are the Reference sample, the (observed or toy) Data and the central value of the auxiliary likelihood $\widehat\nui(\auxdata)$. Notice that $\widehat\nui(\auxdata)={\boldsymbol{0}}$ by construction in the true experiment, but it fluctuates in the toy experiments as discussed at the beginning of the present section. As in the figure, the Reference data feed only the BSM network, while the Data feed both the BSM and the $r$-layer, after passing through the pre-trained $\widehat\delta$ networks. The loss function takes as input the BSM network, the $r$-layer output and $\widehat\nui(\auxdata)$, which enters in the auxiliary term of the Likelihood. The only trainable parameters are the ones of the BSM network and of the $r$-layer, namely $\w$ and $\nui$. The loss at the end of training, times $-2$, produces the $\tau$ term.

The evaluation of the $\Delta$ term, depicted on the right panel of Figure~\ref{schematic2}, follows the strategy described in Section~\ref{sec:nuplearn}. It has been implemented in TensorFlow employing the same building blocks used for the evaluation of $\tau$, apart from the BSM network that does not participate in the evaluation of $\Delta$. The Reference dataset is similarly not employed at this step. The loss function is merely given by minus the argument of the maximum in eq.~(\ref{delta1}), so that $\Delta$ is the minimal loss at the end of training, times $-2$. For the evaluation of $\Delta$, the parameters $\nui$ of the $r$-layer are the only ones to be optimized by the training algorithm.

The TensorFlow modules described above are also employed for the preliminary steps of the algorithm described in Sections~\ref{sec3:model} and~\ref{sec:lm}. In the latter, the $\widehat\delta$ networks are trained using the loss function in eq.~(\ref{sec3:loss}) and the relevant datasets. In the former step, namely the selection of the BSM network hyper-parameters, the $r$-layer and the $\widehat\delta$ networks are not employed and the loss function is replaced with the one, in eq.~(\ref{sec3:loss}), where the effect of nuisance parameters is not taken into account.

\subsection{Validation}\label{sec:validation}

As previously emphasized, it is vital for the applicability of our strategy that the distribution $P(t|\RH)$ of the test statistic is nearly independent of $\nui$. This is ensured in line of principle by the asymptotic formulae described in Section~\ref{af}. Verifying in practice the validity of the asymptotic formulae is thus the crucial validation step, which we will perform by computing the empirical $P(t|\RH)$ distribution on toy experiments. toy datasets are generated according to the $\RH$ hypothesis, at different points $\nui={\boldsymbol{\nu^*}}=(\nu_{\rm\textsc{n}}^*,\nu_{\rm\textsc{s}}^*)$ of the nuisances' parameter space. Each toy dataset $\data$ is accompanied by one instance of the nuisance parameters estimators ${\boldsymbol{\widehat\nu}}=(\widehat\nu_{\rm\textsc{n}},\widehat\nu_{\rm\textsc{s}})$. As explained at the beginning of the present section, the estimators are thrown as Gaussians with standard deviations $\sigma_{{\rm\textsc{n}},{\rm\textsc{s}}}$ centered at $\nu_{{\rm\textsc{n}},{\rm\textsc{s}}}^*$. They appear in the auxiliary likelihood log-ratio as in eq.~(\ref{1dnui}).

\begin{figure}[t]
\begin{center}
\includegraphics[width=18cm]{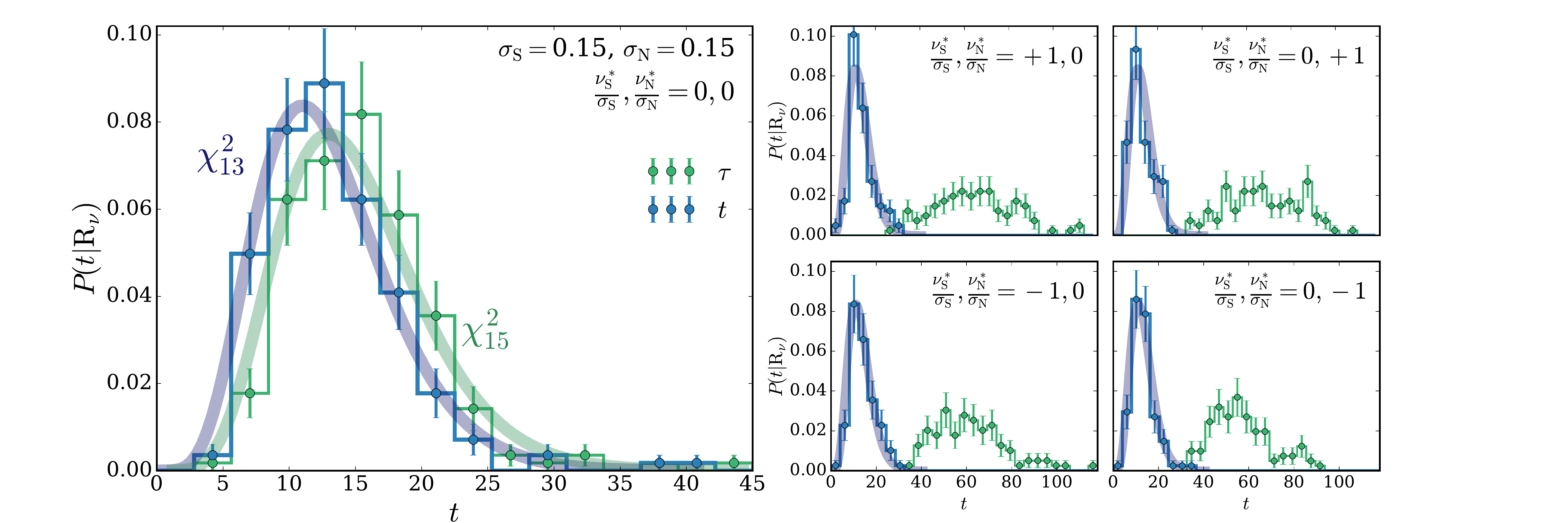}
\caption{The empirical distribution of $\tau$ (in green) and of $t$ (in blue) computed by $100$ toy experiments performed in the $\RH$ hypothesis at different points in the nuisances' parameters space. The $\chi^2_{13}$ distribution is reported in blue in all the plots. The $\chi^2_{15}$ distribution is shown in green on the left plot.}
\label{1D_validation}
\end{center}
\end{figure}

We start by setting $\sigma_{{\rm\textsc{n}}}=\sigma_{{\rm\textsc{s}}}=0.15$, and from central-value nuisance parameters $(\nu_{\rm\textsc{n}}^*,\nu_{\rm\textsc{s}}^*)=(0,0)$, obtaining the results on the left panel of Figure~\ref{1D_validation}. The plot shows the empirical $\tau$ distribution in green and, in blue, the distribution of $t=\tau-\Delta$. In spite of the fact that the toys are generated according to the central-value Reference hypothesis, which is the same hypothesis under which we enforced compatibility with the $\chi^2_{13}$ by choosing the weight clipping parameter in Section~\ref{sec3:model} (see Figure~\ref{WCtuning_1D}), the distribution of $\tau$ is slightly different from the $\chi^2_{13}$. This is not surprising because the $\chi^2$-compatibility was enforced on the variable $\overline{t}$~(\ref{tbar}), which does not account for the presence of nuisances and is different from $\tau$. The distribution of $\tau$ is instead quite close to the $\chi^2$ with a number of degrees of freedom equal to $15$, which is the number of parameters of the neural network plus the number of nuisance parameters. This is compatible with the asymptotic expectation as discussed in Section~\ref{af}. Again compatibly with the asymptotic formulae, we see in the figure that the distribution of $t=\tau-\Delta$ is instead a $\chi^2$ with $13$ degrees of freedom.

The left panel of Figure~\ref{1D_validation} provides a first confirmation of the validity of the asymptotic formula for $P(t|\RH)$, thought not a particularly striking one because the $\tau$ distribution is not vastly different from the one of $t$, meaning that the correction term $\Delta$ does not play an extremely significant role in this case. A more interesting result is obtained when setting $\nu_{\rm\textsc{n}}^*$ or $\nu_{\rm\textsc{s}}^*$ one $\sigma$ away from the central value, as shown in the four plots in the right panel of the figure. In this case, as expected from the asymptotic formulae, the $\tau$ distribution is radically different from the one of $t$. It is expected to follow a non-central $\chi^2$ with a non-centrality parameter that is controlled by the departure of the true values of the nuisances from the central values. The correction term $\Delta$ has a big impact on the distribution of $t$, bringing it back to the expected $\chi^2_{13}$. The effect is due to a strong correlation between the $\tau$ and $\Delta$ distribution over the toys, which engineers a cancellation in $t=\tau-\Delta$.

A more quantitative and systematic validation of the compatibility of $t$ with the $\chi^2_{13}$ can be obtained by computing the Kolmogorov--Smirnov test $p$-value as in Section~\ref{sec3:model}. The results are reported in Table~\ref{tab:1D_validationNN}. The ``w/o correction'' columns report the $p$-value obtained by comparing the distribution of $\tau$ (i.e., without the $\Delta$ correction term) with the $\chi^2_{13}$. The ``w/ correction'' columns report the $p$-value for the distribution of $t$, including the correction. The table contains the results obtained for $\sigma_{{\rm\textsc{n}},{\rm\textsc{s}}} =0.15$, as well as those for lower values of the nuisances' standard deviations $\sigma_{{\rm\textsc{n}},{\rm\textsc{s}}}=0.10,\,0.05$.

The above results establish the validity of the asymptotic formulae when the standard deviation of the nuisance parameters is of order $15\%$ or less. Notice that it is increasingly simple to deal with smaller standard deviations (i.e., with more precisely measured nuisances), merely because when $\nui$ is small the ratio ${{\widehat{r}}(x;\nui)}$ approaches $1$ becoming independent of $\nui$, regardless of the accuracy with which it is reconstructed by the $\widehat\delta_a(x)$ networks. Consequently the maximization over $\w$ in $\tau$~(\ref{tau2}) tends to decouple from the maximization over $\nui$ and the cancellation between $\tau$ and $\Delta$ in the determination of $t$ is guaranteed. On the contrary, larger standard deviations are more difficult to handle. Indeed, as explained in Section~\ref{af}, larger values of $\nui$ push the $\tau$ distribution away from the target $\chi^2$, forcing the correction term to engineer an increasingly delicate cancellation. This enhances the impact of all the imperfections that are present in the implementation of the algorithm, and in particular of the ones related with the quality of the reconstruction of ${{\widehat{r}}}$ that is achieved by the $\widehat\delta_a(x)$ networks. The results presented up to now (namely, Figure~\ref{1D_validation} and Table~\ref{tab:1D_validationNN}) are obtained by employing the linear-order reconstruction for $\log\,{{\widehat{r}}}$. The good observed level of compatibility with the asymptotic formula thus shows that the linear-order reconstruction is sufficiently accurate in order to deal with $\sigma_{{\rm\textsc{n}},{\rm\textsc{s}}}\leq15\%$. However the accuracy is expected to become insufficient for larger $\sigma_{{\rm\textsc{n}},{\rm\textsc{s}}}$, owing to the considerable departures of the exact $\log\,{{{r}}}$ from linearity described in Section~\ref{sec:lm}.

\begin{table}[t]
\begin{center}
\scalebox{0.78}{
\begin{tabular}{c|cc|cc|cc|}
	\multirow{3}{*}{{$\displaystyle\left(\frac{\nu_{\rm\textsc{s}}}{\sigma_{\rm\textsc{s}}},\,\frac{\nu_{\rm\textsc{n}}}{\sigma_{\rm\textsc{n}}}\right)$}} 	& \multicolumn{2}{c|}{$\sigma_{\rm\textsc{s}} = 5\%$, $\sigma_{\rm\textsc{n}}= 5\%$} 	& \multicolumn{2}{c|}{$\sigma_{\rm\textsc{s}} = 10\%$, $\sigma_{\rm\textsc{n}}= 10\%$} 	& \multicolumn{2}{c|}{$\sigma_{\rm\textsc{s}} = 15\%$, $\sigma_{\rm\textsc{n}}= 15\%$}\\
																& \multicolumn{2}{c|}{KS $p$-value}			& \multicolumn{2}{c|}{KS $p$-value}				& \multicolumn{2}{c|}{KS $p$-value}	\\
																& w/o correction 	& w/ correction					& w/o correction 	& w/ correction						& w/o correction 	& w/ correction	\\
	\hline
	(0, 0)								&  $< 10^{-5}$		&  $0.33$  &  $< 10^{-5}$		&  $0.49$		&  $< 10^{-5}$	&  $0.08$ \\
	(+1, 0)							&  $< 10^{-5}$		&  $0.41$ 	&  $< 10^{-5}$		&  $0.55$  	&  $< 10^{-5}$	&  $0.72$\\
	(0, +1)							&  $< 10^{-5}$		&  $0.80$ 	&  $< 10^{-5}$		&  $0.86$ 		&  $< 10^{-5}$	&  $0.45$\\
	(-1, 0)							&  $< 10^{-5}$		&  $0.33$ 	&  $< 10^{-5}$		&  $0.88$ 		&  $< 10^{-5}$	&  $0.37$\\
	(0, -1)							&  $< 10^{-5}$		&  $0.47$ 	&  $< 10^{-5}$		&  $0.82$  	&  $< 10^{-5}$	&  $0.36$\\
\end{tabular}
}
\end{center}
\caption{Kolmogorov--Smirnov $p$-value for the compatibility of the $\tau$ (``w/o correction'' columns) and of the $t$ (``w/ correction'' columns) distributions with the $\chi^2_{13}$. The KS test is based $100$ toy experiments performed in the $\RH$ hypothesis at different points in the nuisance parameters space.}
\label{tab:1D_validationNN}
\end{table}

\begin{figure}[t]
\begin{center}
\includegraphics[width=15cm]{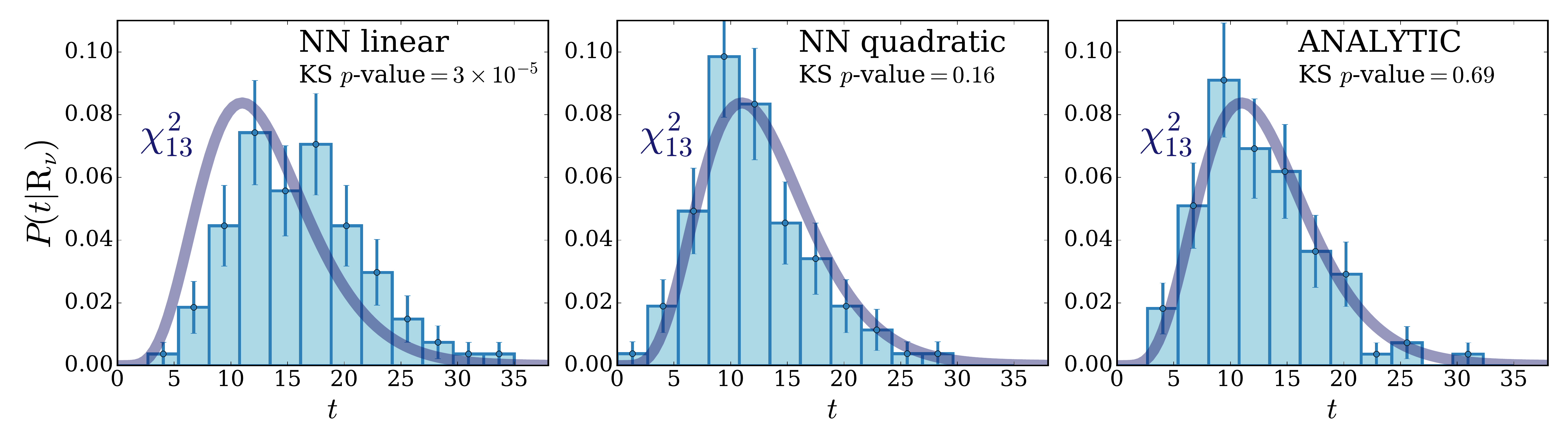}
\caption{The empirical distribution of $t$ computed with $100$ toy experiments for $(\nu_{\rm\textsc{n}}^*,\nu_{\rm\textsc{s}}^*)=(0,-0.6)$. Increasingly accurate modelings of $\log\,{{\widehat{r}}(x;\nui)}$ are employed in the three panels, namely the linear- and quadratic-order approximations and the analytic log-ratio in eq.~(\ref{lran}).}
\label{1D_learning_nuS-0.6}
\end{center}
\end{figure}

We illustrate this aspect by computing the empirical $t$ distribution for $\sigma_{{\rm\textsc{n}},{\rm\textsc{s}}}=0.6$ and setting $(\nu_{\rm\textsc{n}}^*,\nu_{\rm\textsc{s}}^*)=(0,-0.6)$.\footnote{We set $\nu_{\rm\textsc{n}}^*=0$ for this study because non-vanishing values of $\nu_{\rm\textsc{n}}^*$ are easy to deal with, since the dependence of $r$ on the normalization nuisance is known exactly.} The result reported in the left panel of Figure~\ref{1D_learning_nuS-0.6} employ the linear-order approximation of $\log\,{{\widehat{r}}}$. The ones in the middle panel are obtained with the quadratic order approximation while the exact $\log\,{{{r}}}$~(\ref{lran}) is employed in the right panel. The figure shows that the linear-order approximation is insufficient, while a good compatibility with the target $\chi^2_{13}$ is found with the quadratic approximation and with the exact log-ratio. 

\begin{figure}[t]
\begin{center}
\includegraphics[width=0.98\linewidth]{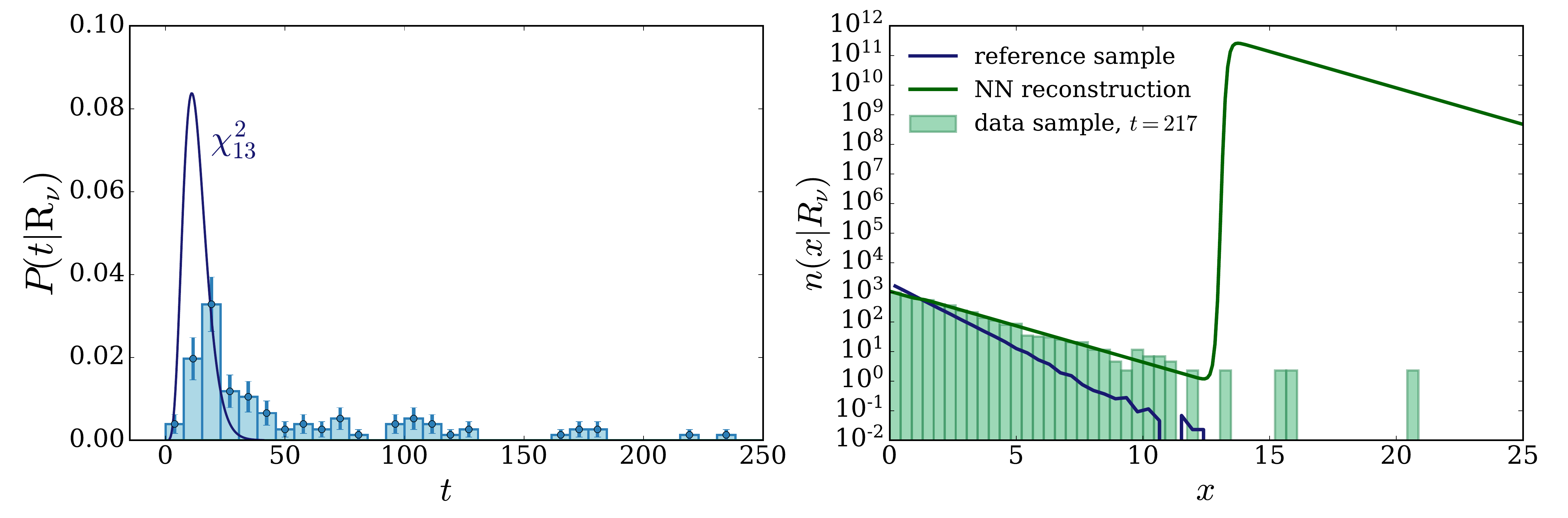}
\caption{Left panel: the empirical distribution of $t$ computed with $100$ toy experiments for $(\nu_{\rm\textsc{n}}^*,\nu_{\rm\textsc{s}}^*)=(0,0.6)$. Right panel: neural network reconstruction of the $x$ variable distribution (using eq.s~(\ref{nnh}) and~(\ref{1dDist})) of a single toy experiment for which the test statistic output is an outlier ($t\simeq217$). }
\label{1D_learning_nuS-p0.6}
\end{center}
\end{figure}

A similar test performed with $(\nu_{\rm\textsc{n}}^*,\nu_{\rm\textsc{s}}^*)=(0,+0.6)$ produced however a non-satisfactory level of compatibility as shown on the left panel of  Figure~\ref{1D_learning_nuS-p0.6}. The reason is that for positive and relatively large $\nu_{\rm\textsc{s}}^*=+0.6$, the scale factor $e^{\nu_{\rm\textsc{s}}}\simeq1.8$ is considerably larger than one and pushes the Reference Model distribution~(\ref{1dDist}) towards large $x$. Therefore, toy data generated with positive and large $\nu_{\rm\textsc{s}}^*$ can often display instances of $x$ that fall in a region that is not populated by the Reference sample. The ``new physics'' network $f$ identifies these instances as highly anomalous, since they do not have any counterpart in the Reference sample, producing outliers in the $\tau$ distribution and in turn in the one of $t$. An illustration of this behavior is displayed on the right panel of the figure. For the toy experiment under consideration, the large observed $t=217$ is due to the data points at $x\gtrsim 13$, which falls well above the largest instance of $x$ ($\simeq11$) that is present in the Reference sample. Such problematic outliers with no counterpart in the Reference sample can not occur if $\nu_{\rm\textsc{s}}^*$ is sufficiently small, such that the $n(x|{{\rm{R}}_{\boldsymbol{\nu^*}}})$ distribution is similar to the central-vale $\ndRefCV{x}$ distribution according to which the Reference sample is generated, because the Reference sample is more abundant ($100$ times, in the case at hand) than the data. But they can occur if, as for $|\nu_{\rm\textsc{s}}^*|=0.6$, the nuisance parameters are so large to modify at order one the central-value distribution and, as for $\nu_{\rm\textsc{s}}^*=+0.6$, they push it towards phase-space regions that are particularly rare in the central-value hypothesis. This potential issue should be kept in mind when dealing with poorly constrained nuisance parameters. Similar problems occur in traditional analyses, whenever the reference control sample statistics is insufficient. A typical mitigation of this effect is obtained binning the dataset with larger binwidths on distribution tails. 
For our method, which in its generic formulation does not make use of bins, possible solutions are either to restrict the variables to a region that is well-populated by the available Reference sample, or to produce a Reference sample that populates the tail of the features distribution more effectively. Further discussion on this point is postponed to Section~\ref{sec:2pVal}, where we will see the same issue emerging again in a more realistic context.

\subsection{Sensitivity to new physics}\label{sec:sens}

We conclude the discussion of the univariate example by testing its sensitivity to putative new physics effects. We consider three New Physics (NP) scenarios that foresee, respectively, the presence of a resonant bump in the tail of the $x$ distribution, a non-resonant enhancement and a resonant peak in the bulk of the distribution. Following Ref.~\cite{DAgnolo:2018cun}, we consider
\begin{itemize}[leftmargin=*,labelindent=22pt]
\item[{$\mathbf{{\rm{\bf{NP}}}_1}$:}]
a peak in the tail of the exponential Reference distribution, modeled by a Gaussian 
\beq\label{NP1}
n(x|{\rm{NP}}_{1;\nui})=\ndRef{x}+{\rm{N}}_1\frac{1}{\sqrt{2\pi}\sigma}e^{-\frac{(x-\bar{x}_1)^2}{2\sigma^2}}\,,
\eeq
with $\bar{x}_1=6.4$, $\sigma=0.16$ and ${\rm{N}}_1=10$.
\item[{$\mathbf{{\rm{\bf{NP}}}_2}$:}] a non resonant effect in the tail of the Reference distribution
\beq\label{NP2}
n(x|{\rm{NP}}_{2;\nui})=\ndRef{x}+{\rm{N}}_2 \frac{x^2}2\, e^{-x}\,,
\eeq
with ${\rm{N}}_2=180$.
\item[{$\mathbf{{\rm{\bf{NP}}}_3}$:}] a peak in the bulk, again modeled by a Gaussian shape  
\beq\label{NP3}
n(x|{\rm{NP}}_{3;\nui})=\ndRef{x}+{\rm{N}}_3 \frac{1}{\sqrt{2\pi}\sigma}e^{-\frac{(x-\bar{x}_3)^2}{2\sigma^2}} \,,
\eeq
with $\bar{x}_3=1.6$, $\sigma=0.16$ and ${\rm{N}}_3=90$.
\end{itemize}
All our putative new physics scenarios give a positive contribution to the Reference distribution. As such, they can be interpreted as an additional ``signal'' component in the distribution of the data, on top of the ``background'' Reference distribution. This is obviously not necessary for our method, which can equally well be sensitive to new physics effects that interfere quantum-mechanically with the Reference Model producing a non-additive contribution. Also notice that we decided not to include nuisance parameters in the new physics term, which is thus assumed to be perfectly known. Also this assumption is not crucial for the sensitivity since a modeling of the signal is not required in our method. Nuisance parameters related to the signal come at play whenever one wants to interpret the outcome of the method as a bound on the theoretical parameters of a specific scenario.

\begin{figure}[t]
\begin{center}
\begin{minipage}[c]{0.34\textwidth}
\centering
  \includegraphics[height=6.7cm]{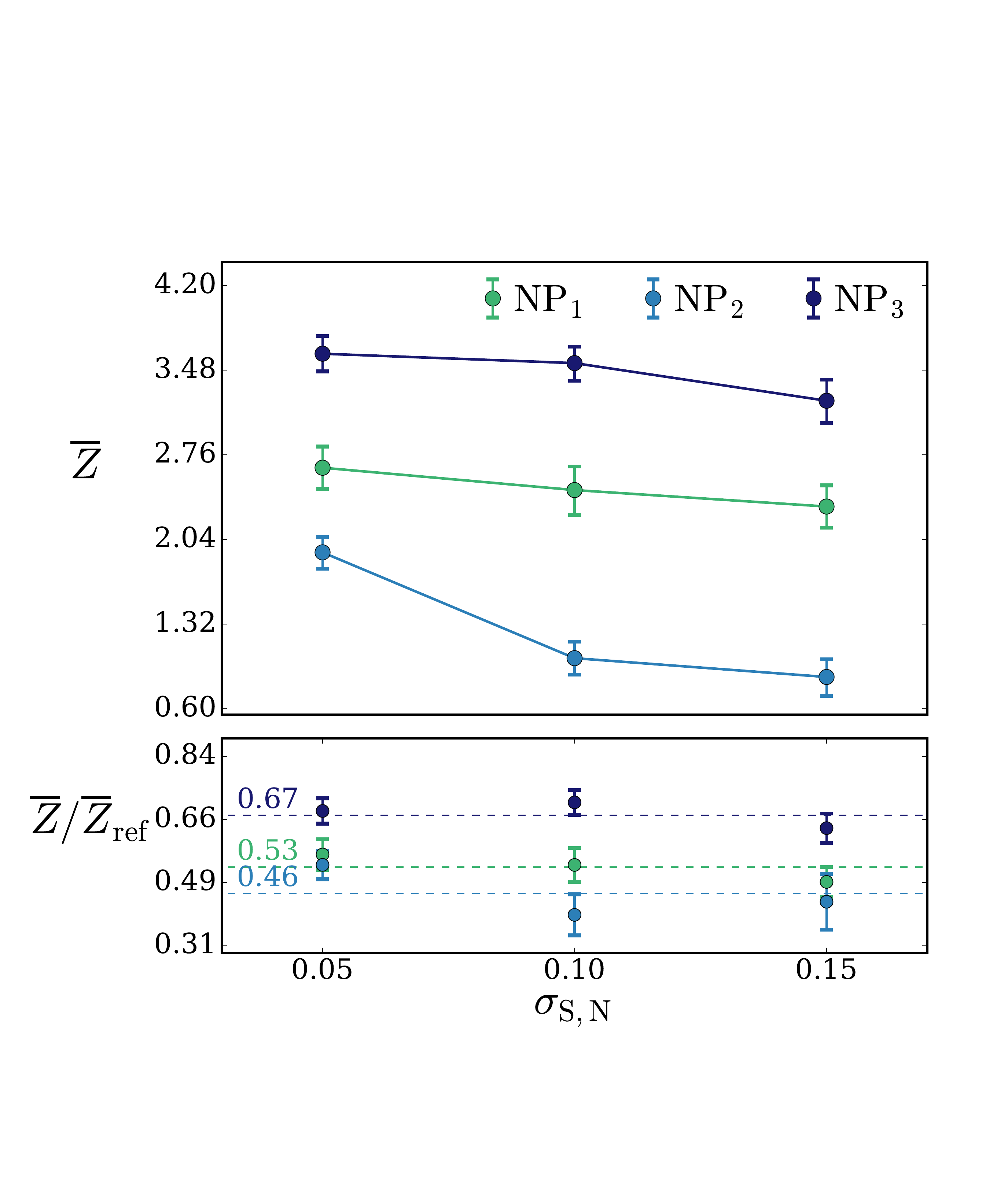}
  \subcaption{}
\end{minipage}\hfill
\begin{minipage}[c]{0.639\textwidth}
\centering
  \includegraphics[height=6.7cm]{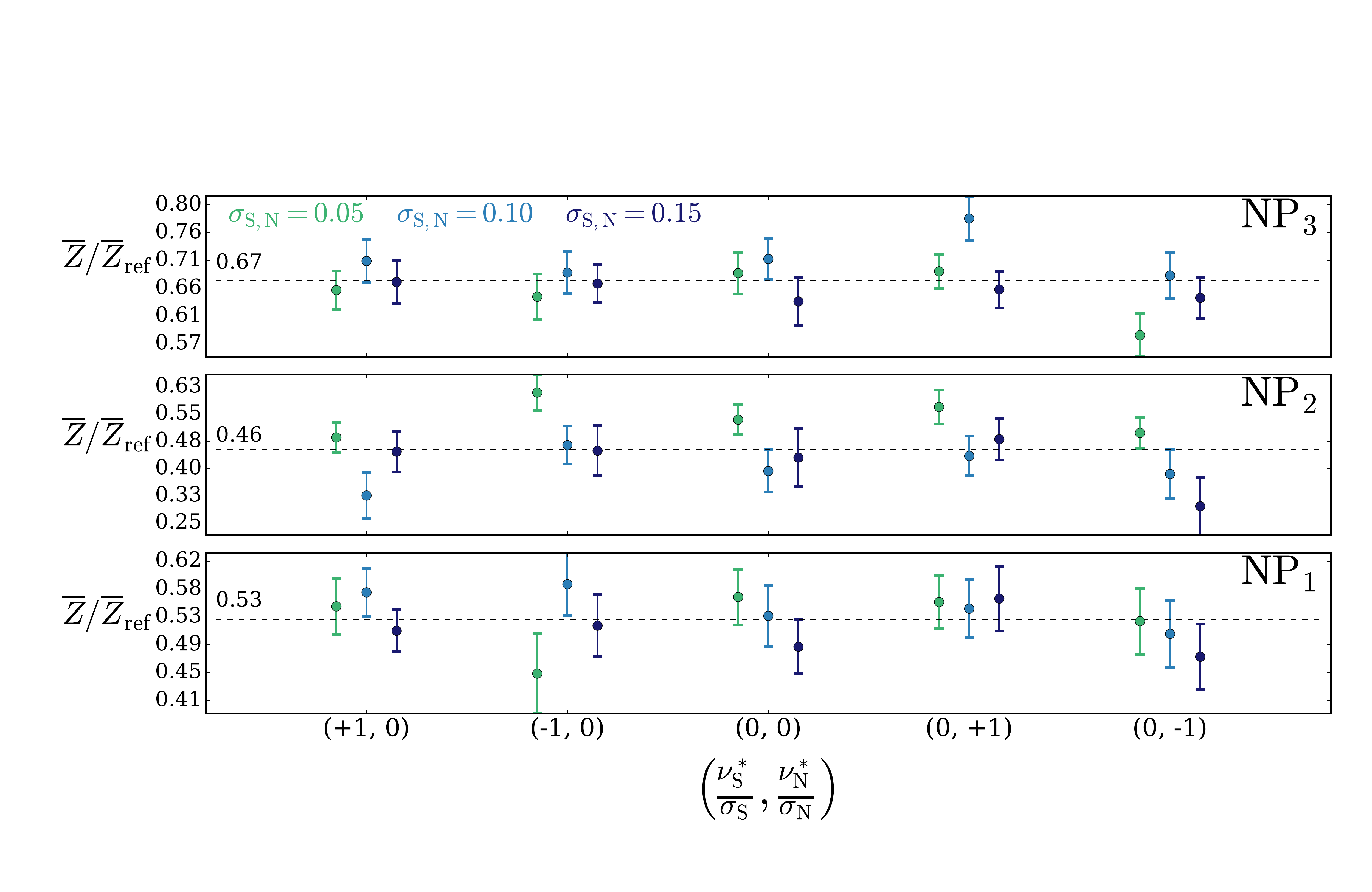}
  \subcaption{}
\end{minipage}

\caption{The median $Z$-score (${\overline{Z}}$) obtained with our model-independent strategy, compared to the median reference $Z$-score (${\overline{Z}}_{\rm{ref}}$) of a model-dependent search (see the main text) optimized for each of the three new physics scenarios in eq.s~(\ref{NP1}--\ref{NP3}). The left panel (a) shows the dependence of ${\overline{Z}}$ on the nuisance parameter uncertainties, and the mild dependence of the ratio ${\overline{Z}}/{\overline{Z}}_{\rm{ref}}$. The left panel is obtained with NP toys generated with central-value nuisance parameters $\nu_{{\rm\textsc{n}},{\rm\textsc{s}}}^*=0$. The right panel (b) displays ${\overline{Z}}/{\overline{Z}}_{\rm{ref}}$ under multiple assumptions are for the nuisance parameters uncertainties ($\sigma_{{\rm\textsc{n}},{\rm\textsc{s}}}=5,\,10,\,15\%$) and for the true values ($\nu_{{\rm\textsc{n}},{\rm\textsc{s}}}^*=0,\,\pm\sigma_{{\rm\textsc{n}},{\rm\textsc{s}}}$) of the nuisance parameters. The error bars quantify the statistical uncertainties (on $100$ toys) in the determination of the median.}
\label{1D_results}
\end{center}
\end{figure}

We quantify the potential of our strategy to detect departures from the Reference Model, if one of the three NP$_{1,2,3}$ models is present in the data, in terms of the median $Z$-score ${\overline{Z}}$ obtained by running our algorithm on toy datasets generated according to the $n(x|{\rm{NP}})$ distribution. For each NP-hypothesis toy we repeat the exact same operations we described in Section~\ref{sec:t} to obtain the test statistic $t$, in the exact same configuration (architecture, weight clipping, etc.) we used in Section~\ref{sec:validation} for validation on the Reference-hypothesis toy datasets. The linear-order reconstruction of $\log\,{{\widehat{r}}}$ is employed for the modeling of the nuisance parameters effect. We saw in Section~\ref{sec:validation} that this modeling is sufficiently accurate if we limit our analysis to the regime $\sigma_{{\rm\textsc{n}},{\rm\textsc{s}}}\leq15\%$. The value of $t$ on each NP toy is compared with the $\chi^2_{13}$ distribution and converted to a $p$-value by exploiting the asymptotic formulae we verified Section~\ref{sec:validation}. For each NP$_{1,2,3}$ new physics scenario, the median $p$-value is computed using $100$ NP toy datasets, obtaining ${\overline{Z}} = \Phi^{-1}(1-p)$, with $\Phi$ the cumulative of the Standard Gaussian. The results are reported in Figure~\ref{1D_results} under multiple assumptions ($\sigma_{{\rm\textsc{n}},{\rm\textsc{s}}}=5,\,10,\,15\%$) for the nuisance parameters standard deviations and for different choices ($\nu_{{\rm\textsc{n}},{\rm\textsc{s}}}^*=0,\,\pm\sigma_{{\rm\textsc{n}},{\rm\textsc{s}}}$) of the true values of the nuisance parameters that underly (through the $\RH$ component of $n(x|{\rm{NP}})$) the generation of the NP toys.

The figure also reports a ``reference'' median $Z$-score ${\overline{Z}}_{\rm{ref}}$, that quantifies the sensitivity of a model-dependent data analysis strategy targeted and optimized for the detection of each individual NP hypothesis. A model-dependent search is necessarily more powerful than a model-independent one for the detection of the NP signal it is designed for. Correspondingly, ${\overline{Z}}_{\rm{ref}}$ must be significantly larger than ${\overline{Z}}$ by consistency and the two quantities should not be compared directly. As in Ref.s~\cite{DAgnolo:2018cun,DAgnolo:2019vbw}, we use ${\overline{Z}}_{\rm{ref}}$ to quantify how ``difficult'' or ``easy'' the NP$_{1,2,3}$ signals are to detect in absolute terms, and we report the ratio ${\overline{Z}}/{\overline{Z}}_{\rm{ref}}<1$ as a measure of the degradation in sensitivity of our model-independent strategy relative to dedicated searches.

As a ``reference'' model-dependent search strategy we consider a hypothesis test based on the profile likelihood ratio, and more precisely on the test statistic ``$q_0$'' for the discovery of positive signals defined in Ref.~\cite{Cowan:2010js}. Namely, we extend the NP hypothesis by a ``signal strength'' parameter $\mu\geq0$ that rescales ${\rm{N}}_{i}\to \mu\,{\rm{N}}_{i}$ (for $i=1,2,3$) in eq.s~(\ref{NP1}--\ref{NP3}). Denoting as $\widehat\mu$ the value of the signal strength parameter that maximizes the likelihood of the NP hypothesis, and $\widehat\nui$ the maximum in the nuisances' space, we define
\beq
q_0=-2\,\log\frac{\max\limits_{\nui}\Lik({\rm{R}}_{\nui} |\data,\auxdata)}{\Lik({\rm{NP}}_{i;\widehat\nui;\widehat\mu} |\data,\auxdata)}\,,
\eeq
if $\widehat\mu>0$, and we set $q_0=0$ otherwise. In the equation, $\Lik$ denotes the extended likelihood constructed as in Section~\ref{sec:found}, exploiting the analytic knowledge of the new physics distributions provided by eq.s~(\ref{NP1}--\ref{NP3}). The ``numerator'' hypothesis ${\rm{R}}_{\nui}$ coincides by construction with the NP hypothesis at $\mu=0$. The distribution of $q_0$ under the Reference (numerator) hypothesis is known in the asymptotic limit. We can thus associate a $p$-value to the value of $q_0$ that is obtained on each NP toy data set. The median $p$-value over the toys provide the median $Z$-score~\cite{Cowan:2010js}
\beq
{\overline{Z}}_{\rm{ref}}={\rm{median}}\left[\sqrt{q_0}\,\right]\,.
\eeq

The physical interpretation of the results on the left panel of Figure~\ref{1D_results} is quite straightforward. The sensitivity to the resonant new physics scenarios NP$_{1,3}$ is not affected by the presence of nuisances, because the nuisance parameters we are considering can not produce deformations of the Reference distribution that mimic a resonant peak. On the contrary, the scale nuisance parameter can mimic non-resonant new physics and indeed the sensitivity to NP$_{2}$ considerably deteriorates as $\sigma_{{\rm\textsc{n}},{\rm\textsc{s}}}$ increases. The same behavior is observed for the model-dependent ${\overline{Z}}_{\rm{ref}}$, as well as for the sensitivity ${\bar{Z}}$ of our model-independent strategy. Indeed, we see that the ${\overline{Z}}/{\overline{Z}}_{\rm{ref}}$ ratio is quite stable under the variation of $\sigma_{{\rm\textsc{n}},{\rm\textsc{s}}}$. This confirms the existence of a direct correlation, as in previous studies~\cite{DAgnolo:2018cun,DAgnolo:2019vbw}, between the sensitivity of our model-dependent strategy and the ``absolute degree of detectability'' of the new physics scenario, as quantified by the sensitivity of a model-dependent search. A further confirmation of this correlation is provided by the right panel of the figure. 

Before concluding this section, it is interesting to consider a fourth scenario for new physics, which does not manifest itself in the variable of interest ``$x$'', but rather in the auxiliary measurements that constrain the nuisance parameters. As discussed in Section~\ref{sec:npcr}, our strategy is not necessarily blind to this type of effects. Consider a situation where the estimator for the scale nuisance parameter, $\widehat\nu_{\rm\textsc{s}}(\auxdata)$, is biased due to new physics by an amount $\Delta\nu_{\rm\textsc{s}}=5\,\sigma_{\rm\textsc{s}}$. Since we do not know about this bias, our auxiliary likelihood remains the one in eq.~(\ref{1dnui}), but $\widehat\nu_{\rm\textsc{s}}(\auxdata)$ in reality is not distributed around the true $\nu_{{\rm\textsc{s}}}^*$, but around $\nu_{{\rm\textsc{s}}}^*+\Delta\nu_{\rm\textsc{s}}$. In order to generate toy experiments that describe this scenario, one has to take $\nu_{{\rm\textsc{s}}}^*+\Delta\nu_{\rm\textsc{s}}$ as the central-value for the generation of the toy $\widehat\nu_{\rm\textsc{s}}$ values while using the true $\nu_{{\rm\textsc{s}}}^*$ for the generation of the $x$ toy datasets. The mismatch, on average, between $\widehat\nu_{\rm\textsc{s}}$ and the value of $\nu_{\rm\textsc{s}}$ that truly underlies the $x$ variable distribution can lead to the detection of new physics as explained in Section~\ref{sec:npcr}. For $\sigma_{\rm\textsc{s}}=15\%$ we find sensitivities
\begin{center}
\def\arraystretch{1.5}
\begin{tabular}{c|c|c|c|c|c}
$(\frac{\nu_{{\rm\textsc{s}}}^*}{\sigma_{\rm\textsc{s}}},\frac{\nu_{{\rm\textsc{n}}}^*}{\sigma_{\rm\textsc{n}}})$ & $(0, 0)$ &	$(+1, 0)$ & $(0, +1)$ & $(-1, 0)$& $(0, -1)$\\
	\hline
			${\overline{Z}}$	&  $2.87^{+0.16}_{-0.15}$	&  $3.53^{+0.12}_{-0.11}$	 &  $3.04^{+0.14}_{-0.14}$ &  $3.22^{+0.14}_{-0.14}$ &  $3.31^{+0.14}_{-0.14}$		\\
\end{tabular}
\end{center}

\section{Two-body final state}\label{sec:two}
In the previous section we described the practical implementation of our strategy and its validation in a very simple univariate toy problem. We now turn to a slightly more complex setup, which is inspired by the realistic problem of model-independent new physics searches in two-body final states at the LHC (see Ref.~\cite{DAgnolo:2019vbw}). While not yet a complete LHC analysis, the setup that we study in the present section is at a similar scale of complexity, and it poses novel challenges with respect to the univariate problem. We will show how to deal with them, aiming at providing the reader with useful indications on how to handle the various technical aspects that might show up in realistic physics analysis contexts.

A two-body final state can be characterized in terms of the five kinematical features $p_{T,1(2)}$, $\eta_{1(2)}$ and $\Delta\phi_{12}=\phi_1-\phi_2$, with $p_T$, $\eta$ and $\phi$ the transverse momentum, the pseudorapidity and the azimuthal angle of the individual particles.\footnote{Two additional variables such as the total transverse momentum and the pseudorapidity of the two-particles system could be included in order to enhance the sensitivity of the analysis to the production of the two particles in association with hard objects.} The particles are $p_T$-ordered, namely $p_{T,1}>p_{T,2}$. Data are supposed to be selected by requiring the two particles to have same flavor and opposite sign, but this information is not retained at this stage. We do not specify sharply the nature of the final state objects. In the typical cases we have in mind, these are either muons, electrons or $\tau$ leptons reconstructed by the detector. On the other hand, the same construction could be applied to objects with similar resolution, e.g., trading electrons for photons or taus for jets. The kinematical distributions would be quite different in the different cases, however we do not expect these differences to impact the technical viability of our strategy, which we aim at demonstrating. The total cross-section of the process would be also different. However we can compensate this adjusting the assumed integrated luminosity of the dataset, making the total number of expected events $\NRefCV$ roughly equal in the various cases. Therefore, for our purpose the only relevant difference between muons, electrons and $\tau$ final states resides in the increasingly large systematic uncertainties that affect the corresponding SM predictions. Since larger uncertainties are more difficult to handle, as outlined in the previous section, it is instructive to investigate these three scenarios.

Owing to the previous discussion, we ignore the difference in the distributions of the different final states and we model all of them as opposite-sign muons. Namely, the central-value Reference distribution $\ndRefCV{x}$ is the same in all cases, and it corresponds to the SM simulation of $pp\to\mu^+\mu^-+X$ at the $13$~TeV LHC obtained with \texttt{MadGraph5}~\cite{Alwall:2014hca} at LO, with extra jets matching and using \texttt{Pythia6}~\cite{Sjostrand:2006za} and \texttt{Delphes3}~\cite{deFavereau:2013fsa} for parton showering and detector simulation. 
The data samples we employ for the analysis are the ones described in Ref.~\cite{DAgnolo:2019vbw} and can be downloaded from Zenodo~\cite{grosso_gaia_2021_4442665}. We consider two Gaussian nuisance parameters $\nu_{\rm\textsc{n}}$ and $\nu_{\rm\textsc{s}}$ describing, as in the previous section, the uncertainty on the event yield normalization and on the scale factor in the measurement of the transverse momenta. We adopt a simple modeling of the normalization uncertainties by a global (phase-space-independent) factor with standard deviation $\sigma_{\rm\textsc{n}}=2.5\%$, corresponding to the uncertainty of the luminosity measurement. Since the normalization nuisance parameter can be incorporated analytically in the likelihood, as we have discussed, it is essentially trivial to deal with it. 

The scale factor, on the other hand, affects the input variable distributions in a non-trivial manner. Furthermore, the uncertainty in its determination widely depends on the nature of the particle. We consider three representative scenarios, having in mind the specific case of CMS\footnote{Our assumptions loosely apply also to the case of the ATLAS detector.}:
\begin{itemize}
\item muon-like: for the CMS experiment, the uncertainty on the muon momentum scale is very small due to the combined information of the inner tracker and the dedicated muon detectors. Based on Ref.~\cite{CMS:2018rym}, we set the uncertainty to a typical value $\sigma_{\textsc{s}}^{\rm{(b)}}=5\times10^{-4}$ for central muons with $|\eta|<2.1$ (barrel region) and $\sigma_{\textsc{s}}^{\rm{(e)}}=15\times10^{-4}$ for $|\eta|\ge2.1$ (endcaps region). Here and in the following cases we ignore the dependence of the uncertainty on the particle transverse momentum for simplicity, but a generalization in this direction is straightforward.
\item electron-like: the momentum reconstruction for electrons is instead based on the combination of the inner tracker information and the energy deposit in the electromagnetic calorimeter. The LHC pileup makes the trajectory reconstruction harder while for the energy reconstruction from the calorimeter information one has to consider the energy loss through bremsstrahlung in the detector material before the calorimeter is reached. The resulting uncertainty is then typically~\cite{CMS:2015xaf} an order of magnitude worse than the one affecting the muons. We here consider an error of $\sigma_{\textsc{s}}^{\rm{(b)}}=3\times10^{-3}$ and $\sigma_{\textsc{s}}^{\rm{(e)}}=9\times10^{-3}$.
\item $\tau$-like: tau leptons decay in the CMS detector and their 4-momenta has to be reconstructed starting from the decay products; the information of all sub-detectors is combined to reconstruct all the particles produced in the collision events in the so called ParticleFlow algorithm \cite{CMS:2017yfk}. For hadronically decaying taus the energy scale uncertainty was found to be always better than $3\%$; here we simply assume an error on the $\tau$-lepton momentum reconstruction of $3\times10^{-2}$ for both the barrel and the endcaps regions, independently of the magnitude of the momentum~\cite{CMS:2011eio}. 
\end{itemize}
In all cases, we treat the effects on the barrel and endcaps regions as fully correlated and we employ a single nuisance parameter $\nu_{\rm\textsc{s}}$ to describe both. Specifically, $\nu_{\textsc{s}}$ is the scale uncertainty in the barrel, with standard  deviation  $\sigma_{\textsc{s}}\equiv \sigma_{\textsc{s}}^{\rm{(b)}}$.

The Monte Carlo samples for non-central values ($\nu_{\rm\textsc{s}}\neq0$) of the scaling nuisance parameters, needed for the implementation and the validation of our strategy, are obtained by reprocessing the di-muon dataset with the transformation $p_{T,1(2)}^{\rm (b,e)}\to {\rm exp}\left({\nu_{\rm\textsc{s}}\sigma_{\rm\textsc{s}}^{\rm{(b,e)}}/\sigma_{\rm\textsc{s}}^{\rm{(b)}}}\right)p_{T,1(2)}^{\rm (b,e)}$, which acts differently on the barrel and endcaps regions. 
 After the transverse momenta rescaling, we apply acceptance cuts $p_{T,2(1)}>20$~GeV, as well as a lower threshold on the di-body invariant mass of $M_{12}>100$~GeV, in order to exclude the resonant peak associated with the $Z$ boson production. Indeed, if included the $Z$ peak would dominate the composition of the data sample by several orders of magnitude, and our analysis would effectively turn into a search for new physics at the $Z$-pole. We thus exclude the $Z$ peak for a more broad exploration of the two-body phase space. The invariant mass cut will have to be raised to $120$~GeV in the \mbox{$\tau$-like} scenario. As we will discuss, this is because $Z$-pole events contamination of the signal-region enhances the effect of scale uncertainties to a non-manageable level at low invariant mass. A similar analysis could also be repeated below the Z mass, as done by the CMS and the LHCb experiments, exploiting real-time analysis techniques \cite{CMS:2019buh, LHCb:2017trq}. We do not discuss this case here.

In what follows we describe the implementation of our model-independent search strategy on a dataset whose integrated luminosity corresponds to $\NRefCV=8\,700$ expected events in the signal region defined by the acceptance and the $100$~GeV invariant mass cut. In the case of opposite-sign muons, this number of events would correspond to an integrated luminosity of around $0.35$~fb$^{-1}$. The expected event yield in the non-central Reference hypothesis, $\NRefnu$, is computed with the same integrated luminosity, duly taking into account the normalization nuisance factor $e^{\nu_{\rm\textsc{n}}}$, and the effect of the scale nuisance $\nu_{\rm\textsc{s}}$ on the selection cuts efficiency. A higher integrated luminosity, of $1.1$~fb$^{-1}$, is considered in the $\tau$-like scenario in order to maintain $\NRefCV$ as large as (specifically, $\NRefCV=8\,400$) in the other scenarios compensating for the higher invariant mass cut.

Finally, in all scenarios we apply an upper cut $p_{T,1(2)}<1$~TeV. The phase space region excluded by this cut is populated, for the luminosity we are considering, with a probability as low as $10^{-5}$ in the Reference model. Therefore it has essentially no impact on the analysis and on its sensitivity to new physics, also in light of the fact that the mere observation of a few events in the region excluded by the cut would constitute a discovery. On the other hand, it is technically important to set some upper cut (though extremely mild, as in this case) in order to strictly avoid the presence in toy datasets of high-$p_T$ outliers, falling in a region that is too rare to be populated even in the Reference sample.\footnote{In a real-life situation, the value of this upper threshold would be set just above the highest $p_T$ value observed in data.} Indeed, we will see that our strategy would overreact to such outliers, similarly to what we discussed in Section~\ref{sec:validation} in the univariate example. 

\subsection{Model selection}\label{sec:MS5f}

The first step in our strategy implementation is the selection of a suitable neural network model ``$f(x;\w)$'', and of its weight-clipping regularization parameter, for the BSM network (see Figure~\ref{schematic2}). The principles underlying the selection, and its technical implementation, are described in detail in Section~\ref{sec3:model} for the univariate example. However the choice of the weight clipping parameter turns out to be more delicate for the multivariate analysis under examination. We believe that this is due to the enhanced sensitivity to the statistical fluctuations of the training sample, which in turn stems from two reasons. First, the sparsity of data in more dimensions unavoidably favors overfitting, to be mitigated with a more aggressive weight clipping. Second, in the current study we will employ a Reference sample size that is only $5$ times larger than $\NRefCV$, namely ${{\rm{N}}_{\cal{R}}}=5\,\NRefCV\simeq 40\,000$, to be compared with ${{\rm{N}}_{\cal{R}}}=100\,\NRefCV$ in the univariate case. This choice, which obviously enhances the statistical fluctuations of the Reference sample, was made in order to validate our strategy in a realistic context where an extremely abundant Reference sample might (possibly because of the resources needed to run the full detector simulation) not be available.\footnote{As a side remark, we acknowledge the importance of a reliable fast simulation to make it feasible to generate very large reference datasets. To this purpose, it would be crucial to explore the use of analysis-specific deep-learning based data augmentation techniques (as in Ref.~\cite{Chen:2020uds}), in conjunction with the speed up of event generators \cite{Hagiwara:2013oka}.} 

In the same spirit, the results of the present section are obtained (if not specified otherwise) using a single Monte Carlo sample of $3.6$ million unweighted events in total, generated with mild acceptance requirements. Each toy dataset was obtained by random sampling around (up to Poisson fluctuations) $200\,000$ events in the original sample. After the events are selected according to these requirements, the desired average number $\NRefCV$ of toy events is found. The Reference dataset employed for the training of each toy experiment was obtained by sampling $1$ million events from the original data, out of the remaining $3.4$ million. This way of proceeding is different from the one we adopted in the univariate example, where each toy and the corresponding Reference sample were generated independently. Clearly, this procedure dictated by the constraints of our limited computational power, is not ideal as it introduces unwanted correlation among the toys. Since we sample with probability $2\times10^{5}/3.6\times10^6=1/18$, we can still reasonably regard the different toys as independent if we generate around $100$ of them (but not more). The Reference samples are instead quite correlated because we extract $1$ million points out of $3.4$ million only. However there is no conceptual need for Reference samples being uncorrelated across toys. Indeed, we described the conceptual role played by the Reference sample, in Section~\ref{sec:mlml}, under the implicit assumption that only one such sample is available for the training of all the toys. The only condition on the Reference sample is $\Nreference/\NRefCV\gg1$. We are assuming here that $\Nreference/\NRefCV =5$ suffices. This assumption has been validated by verifying the stability of the training outcome of individual toys under re-sampling of the Reference sample. Further cross-checks of this and other aspects, including the approximate independence of the toys, have been performed using a second independent $3.6$ million points sample. In addition, the results of the present section concerning the tuning of the weight clipping and the hyperparameters optimization have been reproduced using this second sample.

In light of the items discussed above, it is important to study model selection in detail for the two-body final state problem outlining the differences with the univariate case results presented in Section~\ref{sec3:model}. This is the purpose of the present section.

\begin{figure}[t]
\begin{center}
\includegraphics[width=16cm]{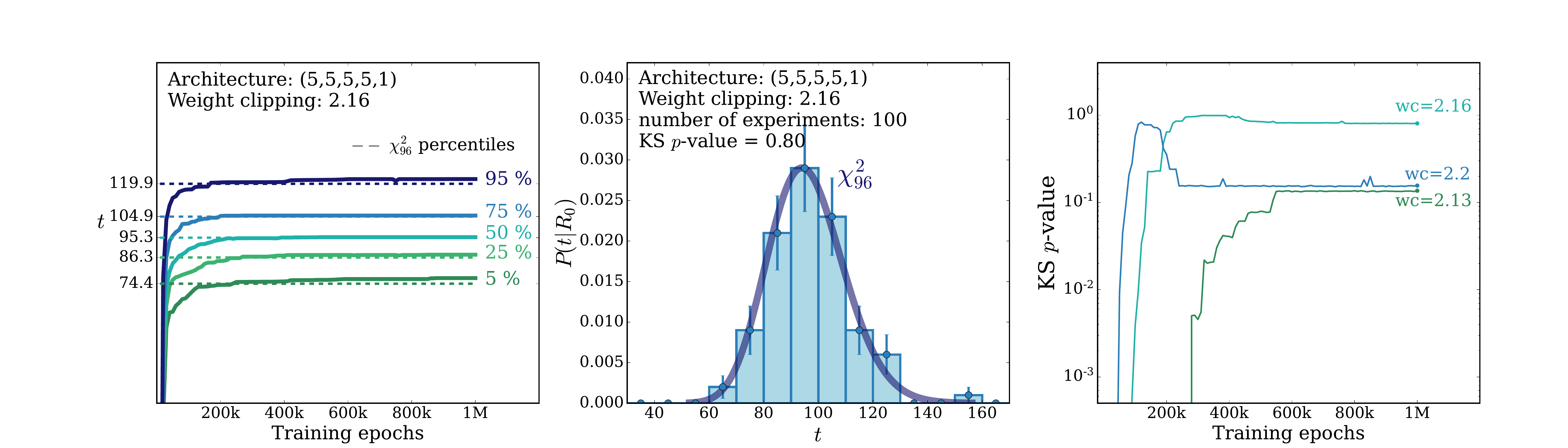}
\caption{Left panel: percentiles of the empirical ${\overline{t}}$ distribution for the $(5,5,5,5,1)$ network, with $100$~toys, as a function of the number of training epochs for the optimal value ($2.16$) of the weight clipping parameter. Middle: the distribution after $1$ million epochs. Right: the evolution during training of the KS $p$-value for different values of the weight clipping.}
\label{WCtuning_5D_3a}
\end{center}
\end{figure}

In a previous study~\cite{DAgnolo:2019vbw} of the same dataset we found that a $(5,5,5,5,1)$ network with $3$ hidden layers of $5$ nodes each (for a total of $96$ degrees of freedom) returns a distribution for the test statistic $\overline{t}$ which is well compatible with the target $\chi^2_{96}$ distribution, for an appropriate choice of the weight clipping parameter.\footnote{In this section we employ the concepts, terminology and notations introduced in Section~\ref{sec3:model}. In particular, $\overline{t}$ is the test statistic in the absence of nuisances, defined by eq.~(\ref{tbar}).} The weight clipping selection is performed with the algorithm described in Section~\ref{sec3:model}, which iteratively reduces the window of potentially viable values of the weight clipping parameter. The last step of the selection process, where the window is already as small as the $[2.1,\,2.2]$ interval, is illustrated in Figure~\ref{WCtuning_5D_3a}. A comparison with Figure~\ref{WCtuning_1D} and Table~\ref{tab_1D_wclip_optimization} immediately reveals a number of differences between the univariate and the multivariate case. First of all, the empirical ${\overline{t}}$ distribution is much more sensitive to the weight clipping. Values of the weight clipping that differ from the optimal one (of $2.16$) at the second digit produce distributions that are appreciably different from the target $\chi^2_{96}$, while in the univariate case good compatibility with the $\chi^2_{13}$ was observed in a quite wide range of weight clipping. Moreover, the stabilization of the distributions with a reasonable degree of compatibility is observed only after $500\,000$ training epochs or more, while $100\,000$ epochs were sufficient in the univariate case. For the problem at hand, such large number of epochs requires a few hours CPU time.\footnote{The training time required for a given architecture clearly depends on the problem. In particular it is proportional to the number of training points which in turn, keeping the ratio ${{\rm{N}}_{\cal{R}}}/\NRefCV=5$ fixed, scales with the number of expected events.}

No further studies were made in Ref.~\cite{DAgnolo:2019vbw} on the choice of the model architecture. On one hand, this is justified by the fact that identifying one single $\chi^2$-compatible configuration is sufficient for the applicability of our strategy. On the other hand, of the many configurations that potentially satisfy this requirement one should select the most complex model, because more expressive networks have more potential to fit putative BSM effects, enhancing the sensitivity of the search. There is not a unique notion of complexity for neural network models. Complexity can, for instance, be enhanced by increasing the number of hidden layers or the number of nodes per layer or, alternatively, by introducing more sophisticated activation functions and connection maps. It is hard to reduce such concepts to a unique scalar metric. One simple way to proceed would be to count the number of trainable parameters, but this would not discriminate between models with different architectures. In our study we restrict our attention to fully connected feedforward neural networks, with the same number of nodes at each layer. Different architectures are thus characterized by two parameters, namely the number of hidden layers and the number of nodes per layer, i.e. the depth and the width of the network. In what follows we explore this two-dimensional architectures space in slices of depth, trying to identify the maximum number of nodes that, for fixed number of layers, can be made compatible with the target $\chi^2$ distribution for an appropriate choice of the weight clipping parameter. 

The conceptual criteria for model selection discussed above must be combined with practical considerations, taking into account the available computational resources that limit the complexity of the model we can concretely handle. With ``computational resources'' we refer both to the memory required to store the model and its gradients during training, and to the training time needed to get a stable solution. For models with a good level of compatibility with the target $\chi^2$ distribution, we sharply define a solution as ``stable'' by requiring the KS $p$-value not to vary more than $10\%$ for at least $100\,000$ epochs. The memory is not a limiting factor. It does not exceed around $1$~GB even for the most complex models we have considered. The training time is instead considerable, because of the large number of epochs that is typically required. For the present study we consider a neural network model ``manageable'' when a stable training (on a single toy dataset) takes less than $6$~hours CPU time. This threshold takes into account the need of repeating training on many toys (we use $100$~toys to establish $\chi^2$-compatibility), of performing a scan on the value of the weight clipping parameter that ensures compatibility, and of exploring different architectures. One should notice that our procedure offers parallelization opportunities by running toy experiments in parallel. Because of this, and having at hand a large-size cluster of CPUs (CERN \texttt{lxplus} cluster) and a handful of GPUs, we found it convenient to run in parallel many time-consuming toys on CPUs as opposed of running a few fast toys on GPUs. 

\begin{figure}[t]
\begin{center}
\includegraphics[width=16cm]{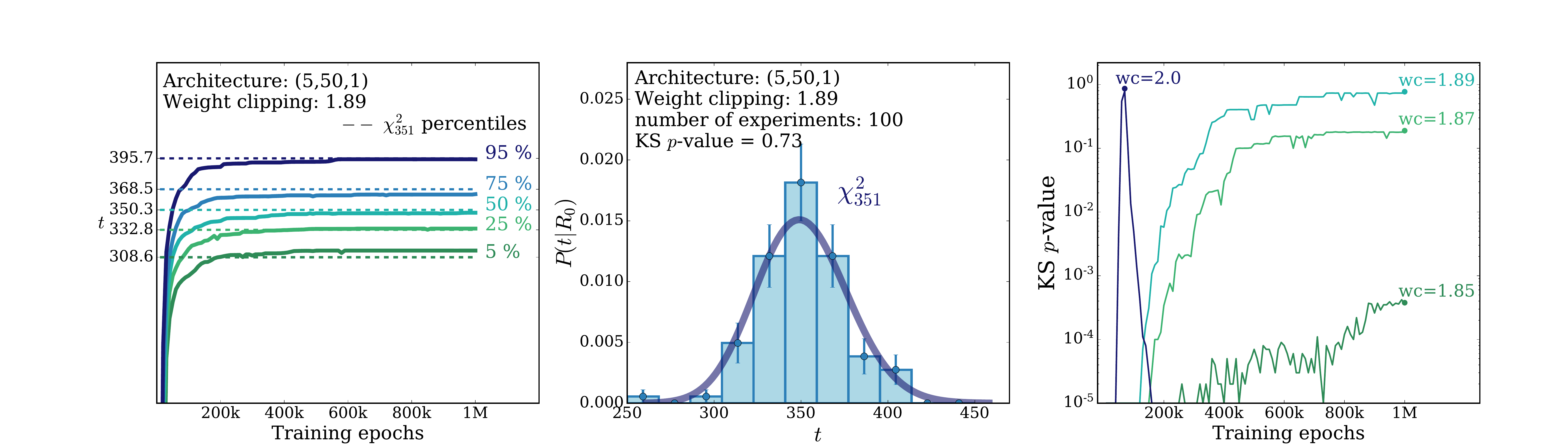}
\caption{Same as Figure~\ref{WCtuning_5D_3a}, but for the $(5,50,1)$ architecture.}
\label{WCtuning_5D_1a}
\includegraphics[width=16cm]{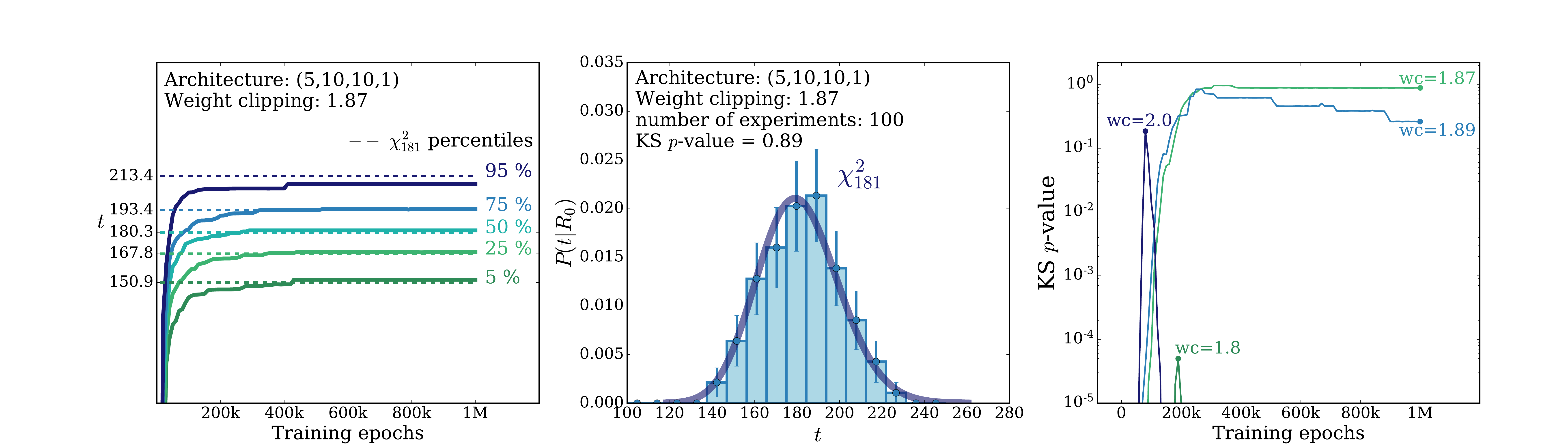}
\caption{Same as Figure~\ref{WCtuning_5D_3a}, but for the $(5,10,10,1)$ architecture.}
\label{WCtuning_5D_2a}
\end{center}
\end{figure}

Based on the above considerations, we identified the $(5,50,1)$ network as the most complex viable model among those with a single hidden layer. The last step of the weight clipping selection process is illustrated in Figure~\ref{WCtuning_5D_1a}. The observed behaviour is similar to the one of Figure~\ref{WCtuning_5D_3a} in terms of the sensitivity to the weight clipping and of the number of epochs required for training. The $(5,50,1)$ network has many more parameters ($351$ versus $96$) than the $(5,5,5,5,1)$ one, but all concentrated in one layer. These two aspects combined make the training time somewhat longer, but still within the boundary of $6$~hours CPU time that defines our computational threshold. Increasing the number of neurons of the network would further increase the training time, therefore the $(5,50,1)$ model is selected among the one-layer architectures. Among the architectures with two hidden layers, we selected by similar considerations (see Figure~\ref{WCtuning_5D_2a}) the $(5,10,10,1)$ network. 

\begin{table}[t]
\begin{center}
\begin{tabular}{c|c|c|c|c|c}
	\# of layers  & latent size 	&dof		& weight clipping	& training epochs	& KS $p$-value\\
	\hline
	\multirow{2}{*}{1}& 5			& 36 		& 1				& 100k		&0.33		\\
				   & {\bf{50}}		& 351	& 1.89			& 1M			& 0.90		\\
	\hline
	\multirow{3}{*}{2} & 6			& 85		& 1.8				& 200k		& 0.81		\\
				   & 7			& 106	& 1.84			& 200k		& 1.00		\\
				   & {\bf{10}}		& 181	& 1.87			& 1M			& 0.99		\\
	\hline
	\multirow{3}{*}{3}& {\bf{5}}			& 96		& 2.16			& 1M		& 0.89		\\
				   & 10		& 291	& ---				& ---		& ---		\\
\end{tabular}
\caption{Summary of the weight clipping tuning results for the architectures considered in this section.}\label{tab:5D_BSM_selection}
\end{center}
\vspace{-0.4cm}
\end{table}%

\begin{figure}[t]
\begin{center}
\includegraphics[width=16cm]{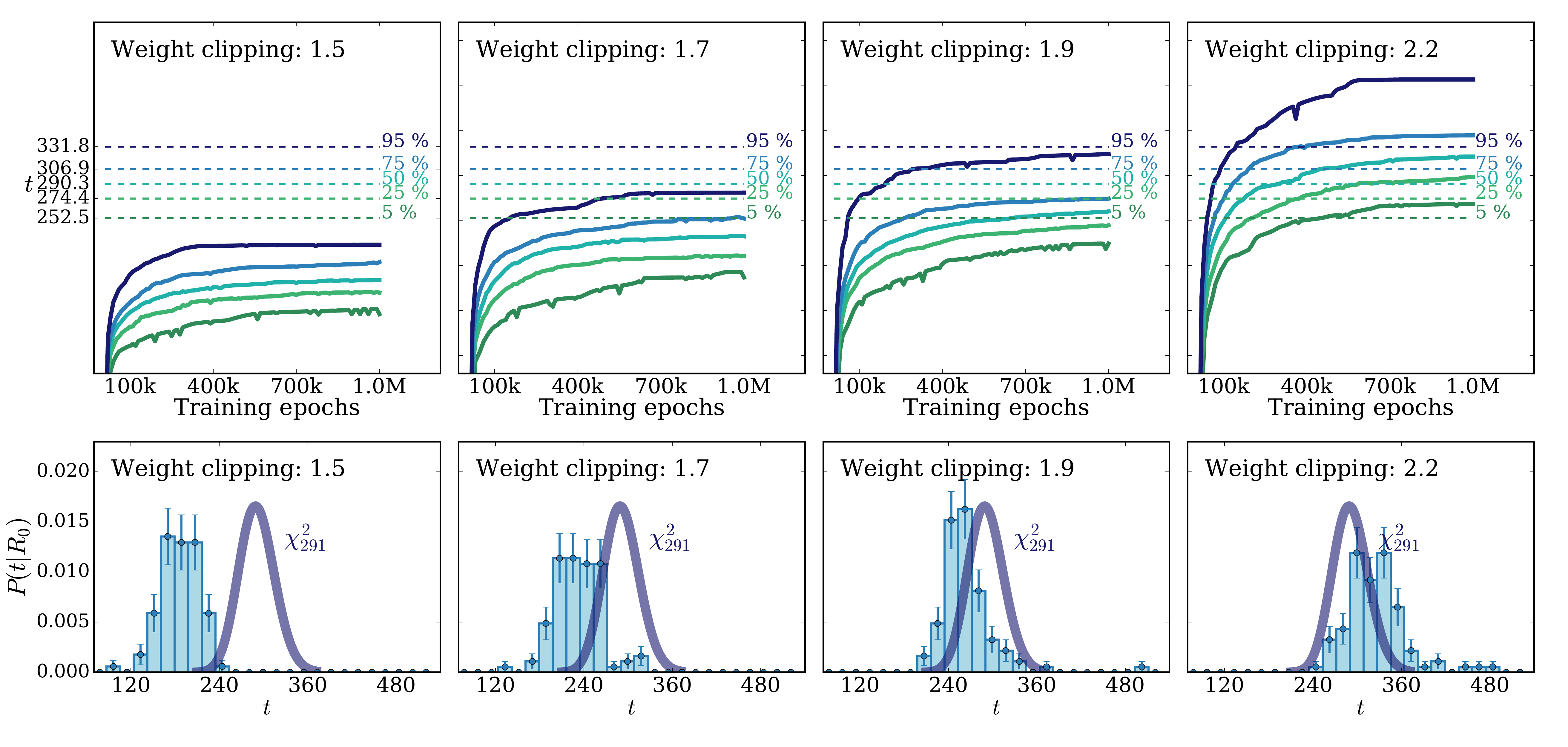}
\caption{The percentiles of the empirical $\bar t$ distribution as a function of the training epochs (top row) and the distribution of the empirical $\bar t$ distribution after 1M training epochs (bottom row) for the $(5, 10, 10, 10, 1)$ network at different values of the weight clipping.}
\label{WCtuning_5D_4}
\vspace{-0.4cm}
\end{center}
\end{figure}

We also tested other architectures, with the results summarized in Table~\ref{tab:5D_BSM_selection}. For networks with $1$ ($2$) hidden layers and less than $50$ ($10$) neurons, we could easily tune the weight clipping parameter obtaining a good level of compatibility with the target $\chi^2$. The number of epochs that safely ensures convergence, reported in the table, decreases with the network size as expected, and training becomes computationally less demanding. Networks with more neurons are beyond our computational threshold as previously explained. A $3$ layers network with $10$ neurons was also considered, but the weight clipping tuning could not be achieved, because of the behaviour displayed in Figure~\ref{WCtuning_5D_4}. If the weight clipping is small, training is stable but the $\overline{t}$ distribution strongly undershoots the target $\chi^2$. By raising the weight clipping the distribution moves to the right, but it is not stable even after one million epochs. More training time would be needed to establish if, for instance, the configuration with weight clipping equal to $1.9$ will eventually converge to the target $\chi^2$. Since this goes beyond our computational threshold, the $(5,10,10,10,1)$ network has to be discarded. We thus retained the $(5,5,5,5,1)$ network, in the $3$-layers class. We did not consider networks with four or more layers because we expect, in light of these results, that for these networks we would be obliged to use less than $5$ (the number of features) neurons in the hidden layers, entailing dimensionality reduction. In summary, the only architectures to be considered for further studies are $(5, 50, 1)$, $(5, 10, 10, 1)$ and $(5, 5, 5, 5, 1)$. We will refer to them as Model~$1$, ~$2$ and ~$3$ respectively.

\subsection{Learning nuisances and validation}\label{sec:2pVal}

\begin{figure}[t]
\centering
   \includegraphics[width=1.12\linewidth]{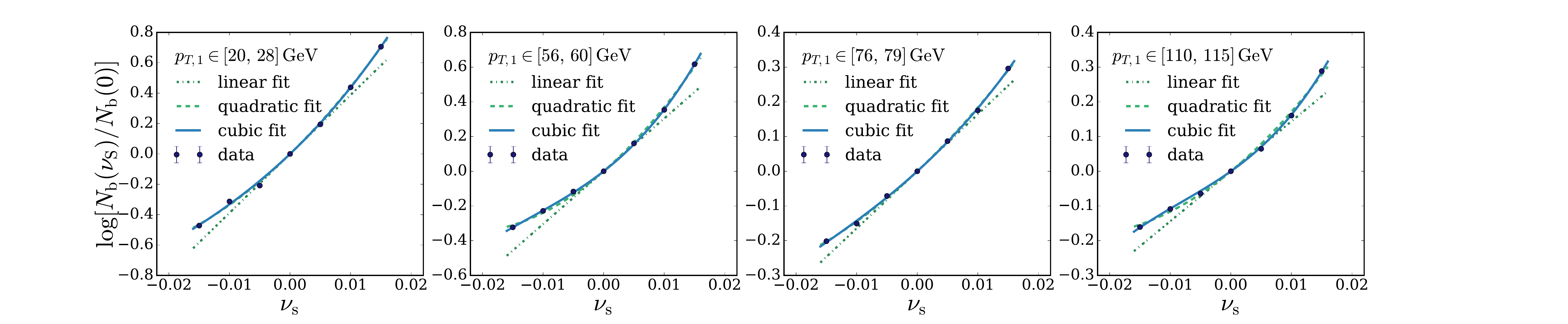}\hspace{0.4cm}
\caption{The dependence on $\nu_{{\rm\textsc{s}}}$ of $\log{N_{\rm{b}}(\nu_{{\rm\textsc{s}}})}/{N_{\rm{b}}(0)}$ in selected bins of the transverse momentum distribution. The dots represent the true value of the log-ratio. The linear, quadratic fits are performed using a subset of the true values points within $\pm 0.01$. The quartic one also considers points at $\pm 0.015$.}\label{pt_dep_nus}
\end{figure}

Our next task is to model the effect of nuisance parameters on the distribution log-ratio $\log r(x, \nui)$. This is a rather straightforward application of the methodology of Section~\ref{sec:nuplearn}, only slightly more computationally demanding than the one presented in Section~\ref{sec:lm} for the univariate problem. The normalization nuisance ${\nu_{\rm\textsc{n}}}$ contributes linearly to the log-ratio, we thus incorporate it analytically in the reconstructed $\log\,{{\widehat{r}}(x;\nui)}$, as in eq.~(\ref{lrapp}). The effect of the scale nuisance ${\nu_{\rm\textsc{s}}}$ is reconstructed locally in the five-dimensional space of features by means of two neural networks $\widehat\delta_{1,2}(x)$ that parametrize the Taylor expansion of the log-ratio up to quadratic order, again as in eq.~(\ref{lrapp}). The $\nu_{{\rm\textsc{s}},i}$ values used for training were selected by studying the effect of the scale uncertainty nuisance on the features distribution, like in Figure~\ref{pt_dep_nus}. The figure shows the dependence on $\nu_{{\rm\textsc{s}}}$ of the expected number of events in selected bins of the transverse momentum of the leading lepton ($p_{T,1}$). 
 The scale uncertainty in the endcaps region has been taken $3$ times the one in the barrel, as appropriate for the muon and electron scenarios defined at the beginning of this section. The result is expressed as a function of the scale in the barrel, $\nu_{{\rm\textsc{s}}}$. The uncertainty $\sigma_{\rm\textsc{s}}$ will be set to $5\times10^{-4}$ and to $3\times10^{-3}$ in the muon and electron scenarios, respectively. 
 We see that the dependence is quadratic to a good approximation in the interval $\nu_{{\rm\textsc{s}}}\in[-0.02,0.02]$, which comfortably covers the range that is relevant for the electron scenario up to more than $3$ sigma (and even more for the muon-like one). Training points $\nu_{{\rm\textsc{s}},i}=\{\pm1.5\times 10^{-3},\,\pm 1.5\times 10^{-2}\}$ are selected as a reasonable choice which exposes the $\widehat\delta$ networks both to the linear and to the quadratic component of the likelihood log-ratio. The validity of this choice was confirmed by also inspecting the nuisance dependence of other kinematical variables.

\begin{figure}[t]
\begin{minipage}[c]{0.5\textwidth}
\centering
   \includegraphics[width=0.998\linewidth]{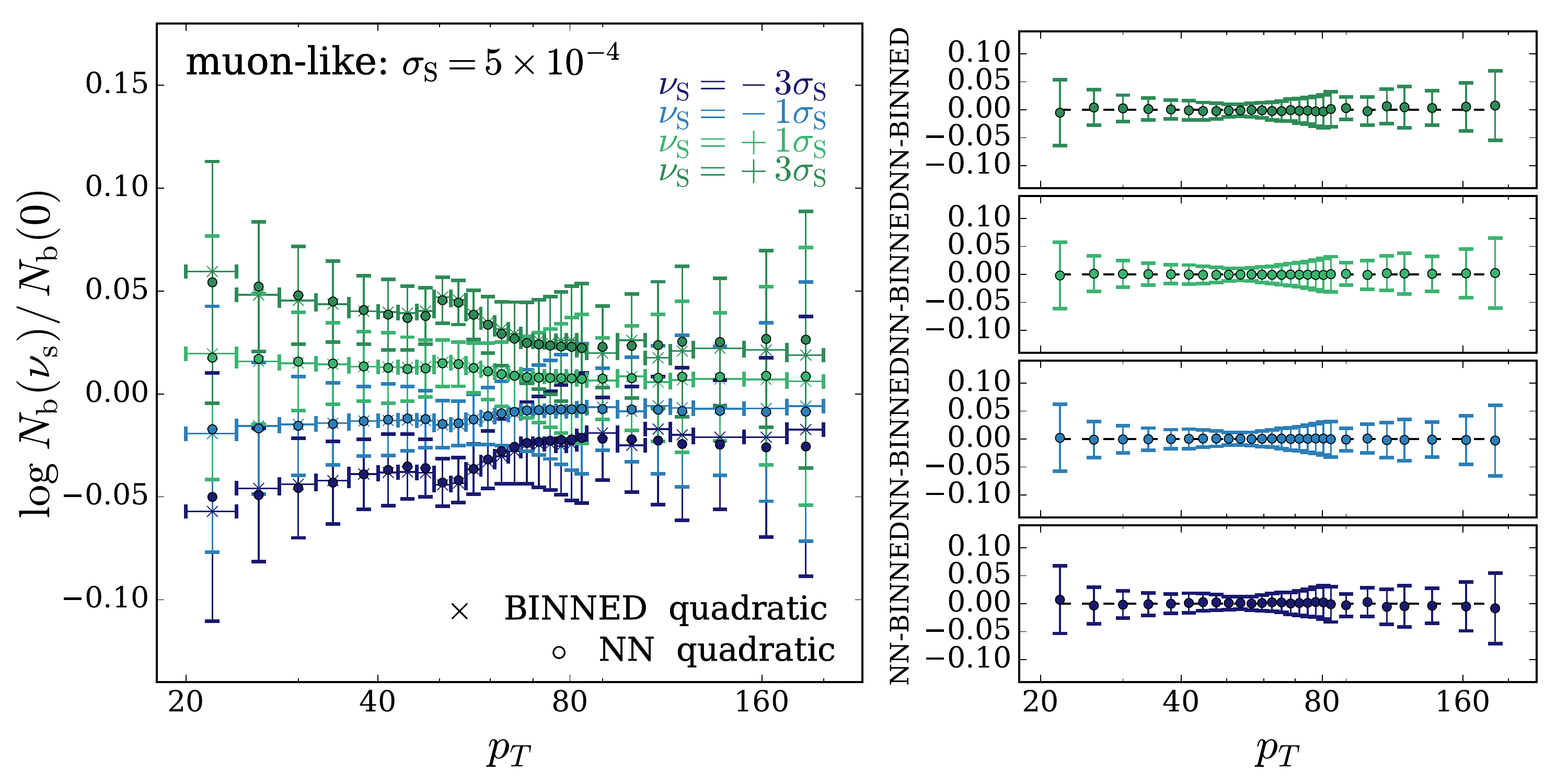}
\end{minipage}\hfill
\begin{minipage}[c]{0.5\textwidth}
\centering
   \includegraphics[width=0.998\linewidth]{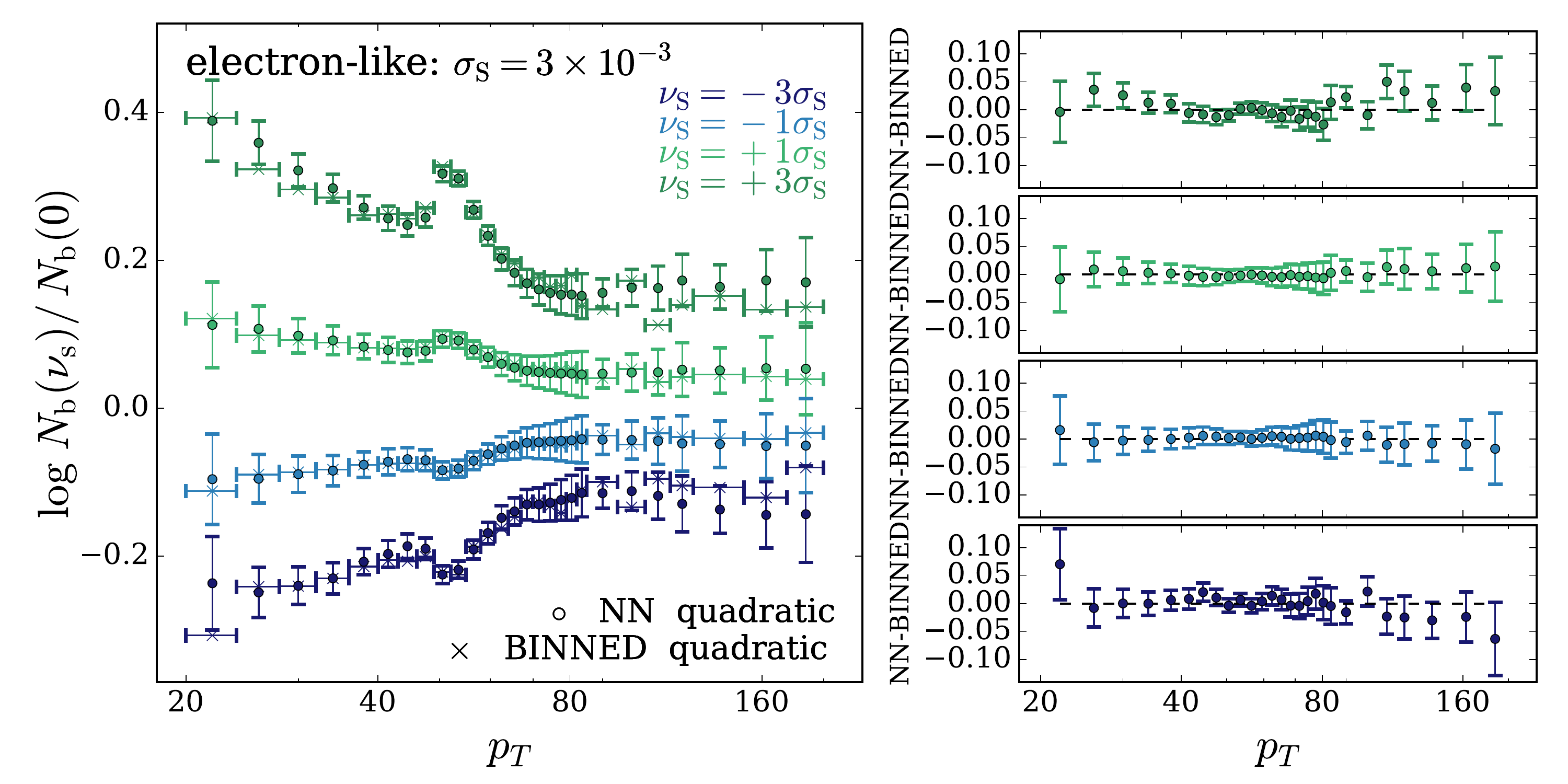}
\end{minipage}
\caption{The reconstructed distribution log-ratio (dots) for different values of $\nu_{{\rm\textsc{s}}}$, compared with the quadratic binned approximation. The two panels cover the ranges of $\nu_{{\rm\textsc{s}}}$ that are relevant for the muon- and electron like scenarios respectively.}
\label{5D_learning_nu_1}
\end{figure}

Five hidden layers with $10$ neurons (and ReLU activation functions) was identified as a viable architecture for the $\widehat\delta$ networks. The training samples ${{\rm{S}}_{0}}(\nui_i)$ were obtained using half of the original $3.6$ million sample. After the selection requirements are applied, they consist of around $80\,000$ events for each value of $\nu_i$. The ${{\rm{S}}_{1}}$ sample, with central-value nuisance, was provided by the remaining $1.8$ million events, weighted by a factor of $4$ in order to compensate for the presence of the four ${{\rm{S}}_{0}}(\nui_i)$ non-central-value samples. For training we applied an early stopping criterion based on the quality of the log-ratio reconstruction achieved by the networks. The quality of the reconstruction was monitored by plots like the one in Figure~\ref{5D_learning_nu_1} and also by testing the capability of the $\widehat\delta$ networks to reabsorb the effect of non-central nuisances in the test statistic distribution. Good performances were obtained with $2\,000$ epochs. A mild overfitting was observed training longer. 

In order to test the accuracy of the log-ratio reconstruction, we use the reconstructed ${{\widehat{r}}(x;\nui)}$ to re-weight the Monte Carlo sample with central-value nuisances, and we compare the predictions for the binned distribution log-ratio (in $p_T$ bins), as obtained by this re-weighting, with those obtained using non-central-value samples. Figure~\ref{5D_learning_nu_1} shows good agreement, for $\nu_{{\rm\textsc{s}}}$ in the range that is relevant to cover the muon- and electron- like scenarios. 

\begin{figure}[t]
\begin{center}
\includegraphics[width=16.5cm]{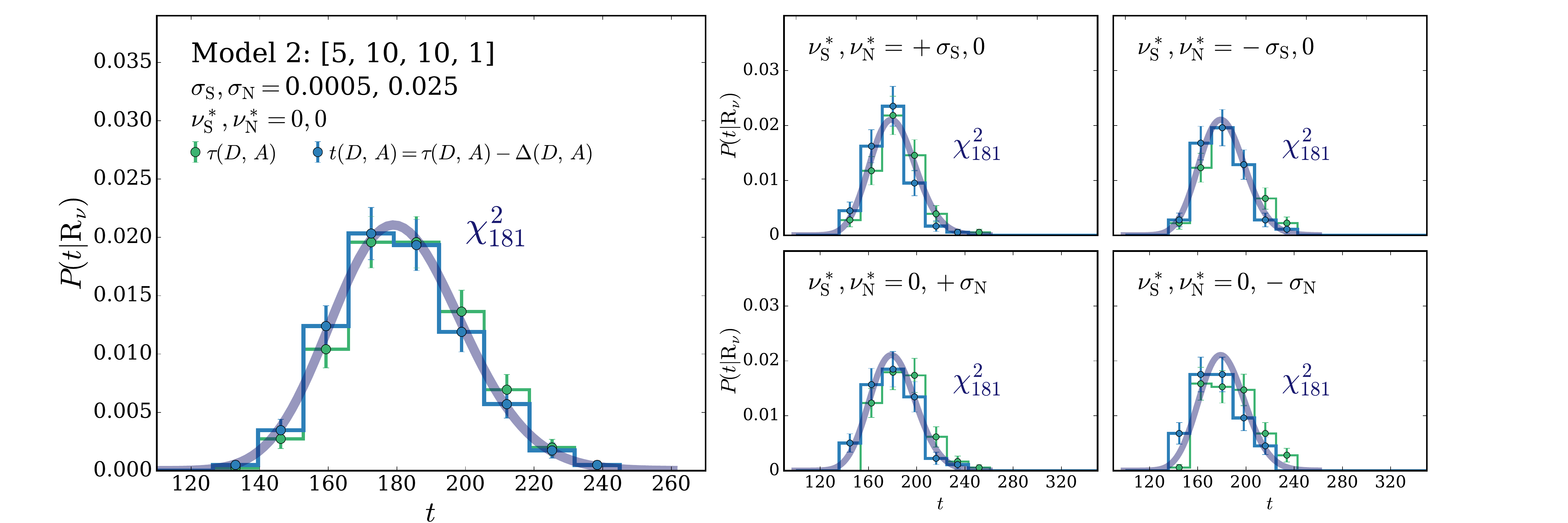}\\
\caption{The empirical distribution of $\tau$ (in green) and of $t$ (in blue) computed by $100$ toy experiments performed in the $\RH$ hypothesis at different points in the nuisance parameters space for the muon-like regime. The $\chi^2_{181}$ distribution is reported in blue in all the plots.}\label{5D_validation1}
\includegraphics[width=16.5cm]{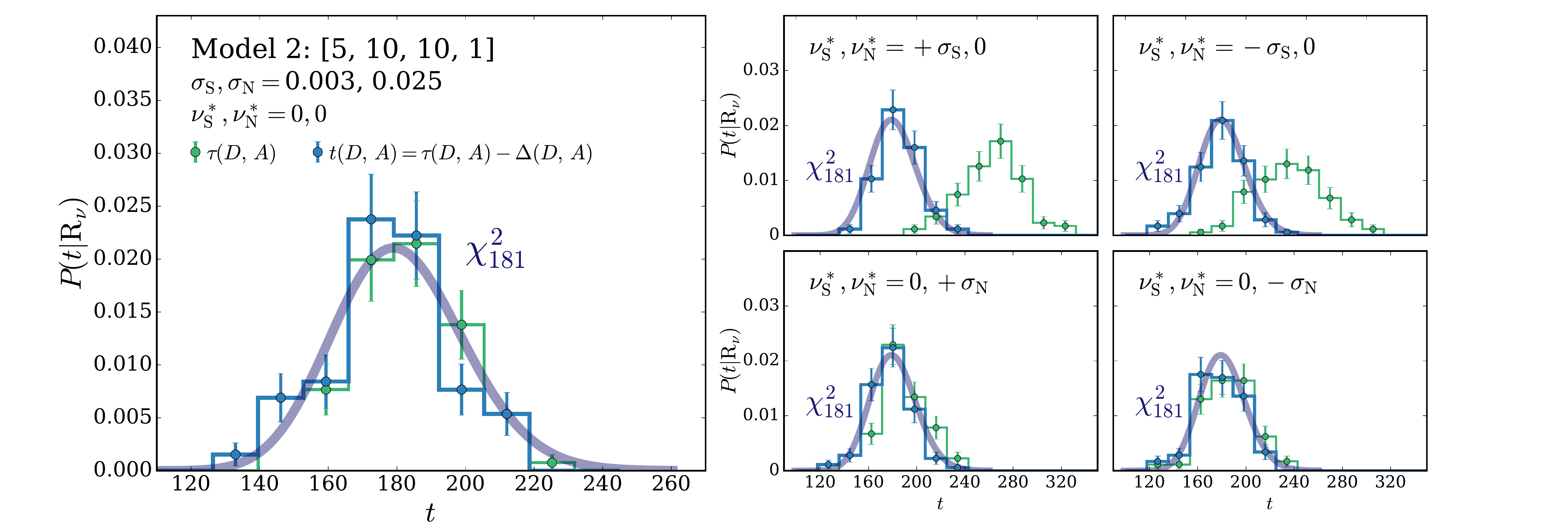}
\caption{Same as Figure~\ref{5D_validation1}, but for the electron-like regime.}
\label{5D_validation2}
\end{center}
\end{figure}

The most stringent cross-check of the quality of the log-ratio reconstruction is however provided by the final validation of the whole strategy, that consists in verifying the independence on the nuisance parameters of the distribution of the test statistic, $P(t|\RH)$. Indeed, as emphasized in previous sections (see in particular Section~\ref{sec:validation}), the emergence of a $\chi^2$ distribution for the test statistic $t=\tau-\Delta$, with the appropriate number of degrees of freedom, provides a highly non-trivial test of all aspects of the algorithm implementation, ranging from the selection of the BSM network hyperparameters (which affects $\tau$) to the accuracy of the log-ratio reconstruction (which affects both the $\tau$ and the $\Delta$ terms). In Figures~\ref{5D_validation1} and~\ref{5D_validation2} we display some of the validation plots that have been produced in order to verify the independence of the test statistic distribution on the true values ${\boldsymbol{\nu^*}}=(\nu_{\rm\textsc{n}}^*,\nu_{\rm\textsc{s}}^*)$ of the nuisance parameters. A summary of the results is provided in Table~\ref{tab:5D_validation}, covering the three neural network models (1, 2 and 3) selected in Section~\ref{sec:MS5f} for the BSM network, and in the electron-scenario for the scale uncertainty. The KS $p$-value is typically low in the ``w/o correction'' columns, showing that the presence of nuisances impacts the distribution of $\tau$ significantly. The asymptotic formula for the distribution of $t=\tau-\Delta$ is recovered by the inclusion of the $\Delta$ term, as shown by the higher $p$-values in the ``w/o correction'' columns. 

In summary, we have demonstrated the possibility to deal with a level of uncertainties that corresponds to the electron-like scenario, as defined at the beginning of Section~\ref{sec:two}. Trivially (since lower uncertainties are easier to manage), the same holds in the muon-like setup. The larger uncertainty that is foreseen in the $\tau$-like scenario is instead more difficult to manage, and deserves an extensive dedicated discussion, which is the subject of the following section.

\begin{table}[t]
\begin{center}
\scalebox{0.78}{
\begin{tabular}{c|cc|cc|cc|}
	\multirow{3}{*}{{$\displaystyle\left(\frac{\nu_{\rm\textsc{s}}^*}{\sigma_{\rm\textsc{s}}},\,\frac{\nu_{\rm\textsc{n}}^*}{\sigma_{\rm\textsc{n}}}\right)$}} 	& \multicolumn{2}{c|}{Model 1} 	& \multicolumn{2}{c|}{Model 2} 	& \multicolumn{2}{c|}{Model 3}\\	
													& \multicolumn{2}{c|}{KS $p$-value}		& \multicolumn{2}{c|}{KS $p$-value}		& \multicolumn{2}{c|}{KS $p$-value}\\
													& w/o correction 	& w/ correction		& w/o correction 	& w/ correction		& w/o correction 	& w/ correction	\\
	\hline
	(0, 0)												&  $0.59$		&  $0.86$  			&  $0.082$	&  $0.10$				&  $0.02$	&  $0.03$ \\
	(+1, 0)											&  $<10^{-5}$	&  $0.02$ 				&  $<10^{-5}$	&  $0.05$  			&  $<10^{-5}$		&  $0.18$\\
	(0, +1)											& $0.0002$	&  $0.58$ 				&  $0.002$	&  $0.18$				&  $0.11$			&  $0.13$\\
	(-1, 0)											&  $<10^{-5}$	&  $0.17$ 				&  $<10^{-5}$	&  $0.83$  			&  $<10^{-5}$		&  $0.20$\\
	(0, -1)											&  $0.24$		&  $0.71$ 				&  $0.09$		&  $0.24$  			&  $0.002$		&  $0.06$\\
\end{tabular}
}
\end{center}
\caption{Kolmogorov--Smirnov $p$-value for the compatibility of the $\tau$ (``w/o correction'' columns) and of the $t$ (``w/ correction'' columns) distributions with the target $\chi^2$ distribution for model 1, 2, 3 in the electron-like regime. The KS test is based $100$ toy experiments performed in the $\RH$ hypothesis at different points in the nuisance parameters space.}
\label{tab:5D_validation}
\end{table}

\newpage
\subsection[The $\tau$-like scenario]{The $\boldsymbol{\tau}$-like scenario}\label{sec:tlike}

\begin{figure}[t]
\centering
   \includegraphics[width=1.1\linewidth]{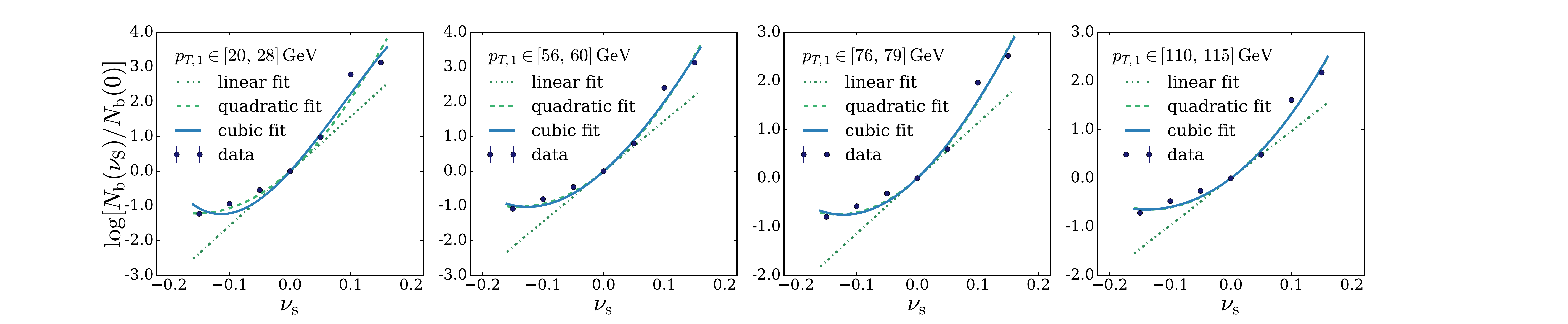}\\
   \caption{The dependence on $\nu_{{\rm\textsc{s}}}$ of $\log{N_{\rm{b}}(\nu_{{\rm\textsc{s}}})}/{N_{\rm{b}}(0)}$ in selected bins of the transverse momentum distribution for $M_{12}>100$~GeV. The dots represent the true value of the log-ratio. The linear and quadratic fits are performed using a subset of the true values points within $\pm 0.1$; the cubic one also considers two additional points at $\pm 0.15$.}\label{5D_learning_nu_2_a}
    \includegraphics[width=1.1\linewidth]{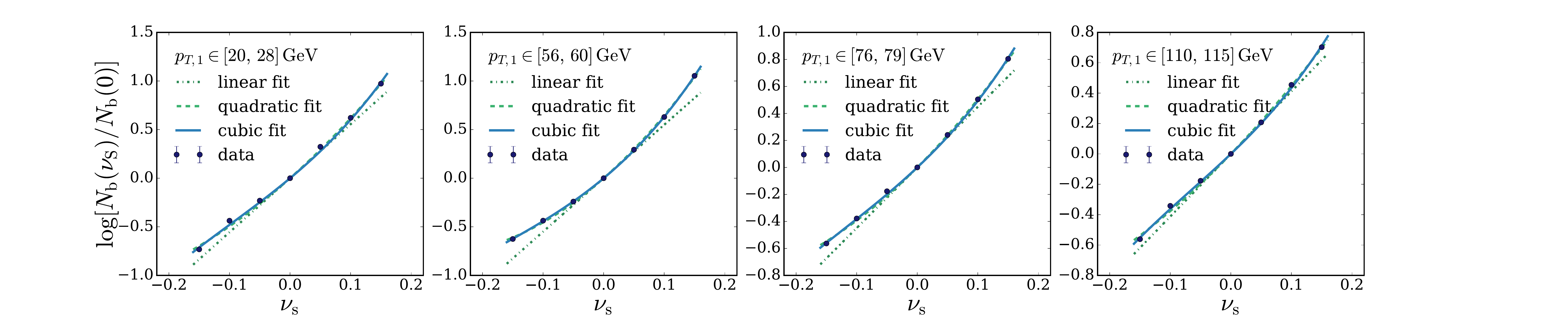}
\caption{Same as Figure~\ref{5D_learning_nu_2_a}, but for $M_{12}>120$~GeV (lower panel).}\label{5D_learning_nu_2_b}
\end{figure}

The first difficulty we encounter in the $\tau$-like scenario is the wild dependence of the distribution on the scale nuisance parameter, displayed in Figure~\ref{5D_learning_nu_2_a}. The effect is due to the migration of events from the $Z$-peak to the signal region defined by the invariant mass cut  $M_{12}>100$~GeV. Since the $Z$-peak events are overly abundant, even a small correction to the $Z$-peak rejection efficiency (of order $\sigma_{\rm\textsc{s}}=3\times10^{-2}$ in the $\tau$-like scenario) affects at order one the distribution in the signal region. Our current setup is only capable to deal with relatively small distortions, for which the Taylor expansion in eq.~(\ref{lrapp}) is justified. Therefore we do not even try to study the $\tau$-like scenario in the entire signal region $M_{12}>100$~GeV, but rather consider a harder cut $M_{12}>120$~GeV that mitigates the $Z$-peak migration effects. Figure~\ref{5D_learning_nu_2_b} shows that the effects of the nuisance are still sizable in this region, but moderate enough to justify the expansion in $\nu_{\rm\textsc{s}}$ up to the quadratic order. The harder invariant mass cut reduces the expected number of events by a factor of around $3$. We compensate by raising the luminosity as discussed at the beginning of this section, in order to maintain $\NRefCV=8\,400$ similar to the one of the muon- and electron-like setups. We also want to maintain a similar proportion between $\NRefCV$ and the total number of Monte Carlo events employed in the analyses. We must thus use three samples with $3.6$ million events (for a total of $10.8$ millions) each before cuts.

\begin{figure}[t]
\centering
   \includegraphics[width=0.95\linewidth]{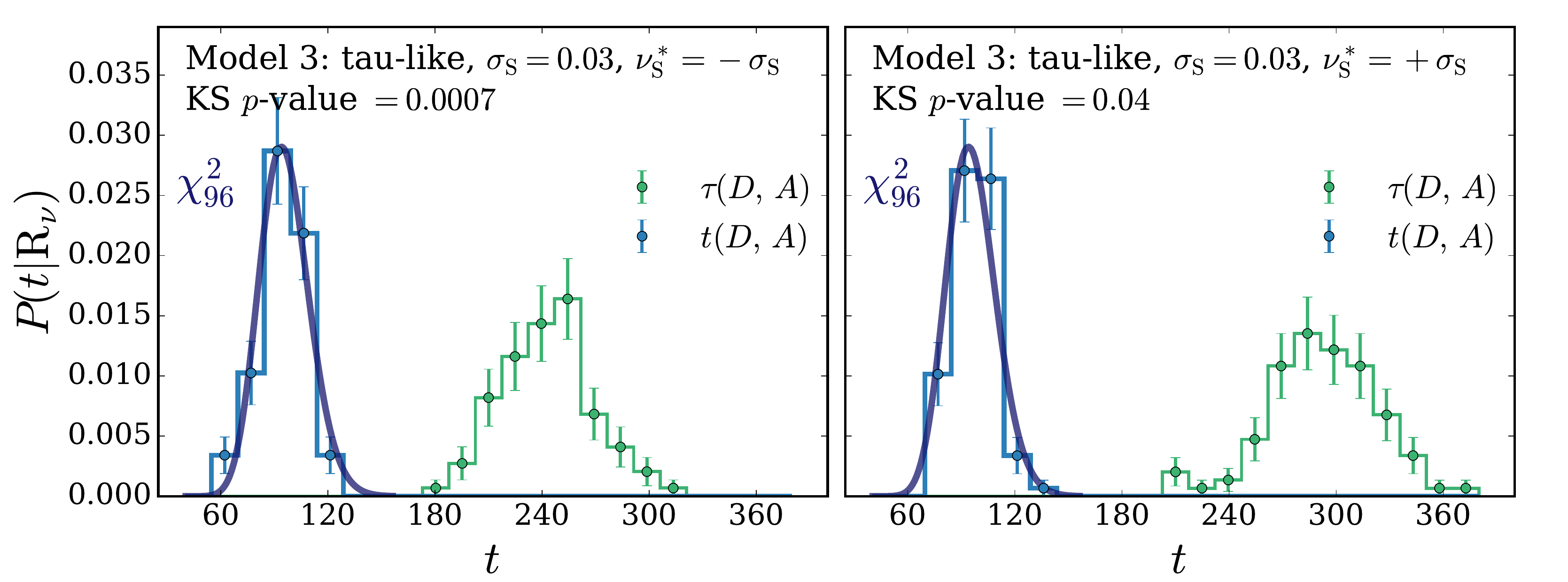}
   \caption{The empirical distribution of $\tau$ (in green) and of $t$ (in blue) computed by 100 toy experiments performed for Model~$3$ in the $\RH$ hypothesis at at $\nu_{\textsc{s}}=-1$ (left side) and $\nu_{\textsc{s}}=+1$ (right side) for the $\tau$-like regime before enriching the reference sample in the region of high transverse momentum.}\label{5D_taulike_model3_a}
    \includegraphics[width=0.95\linewidth]{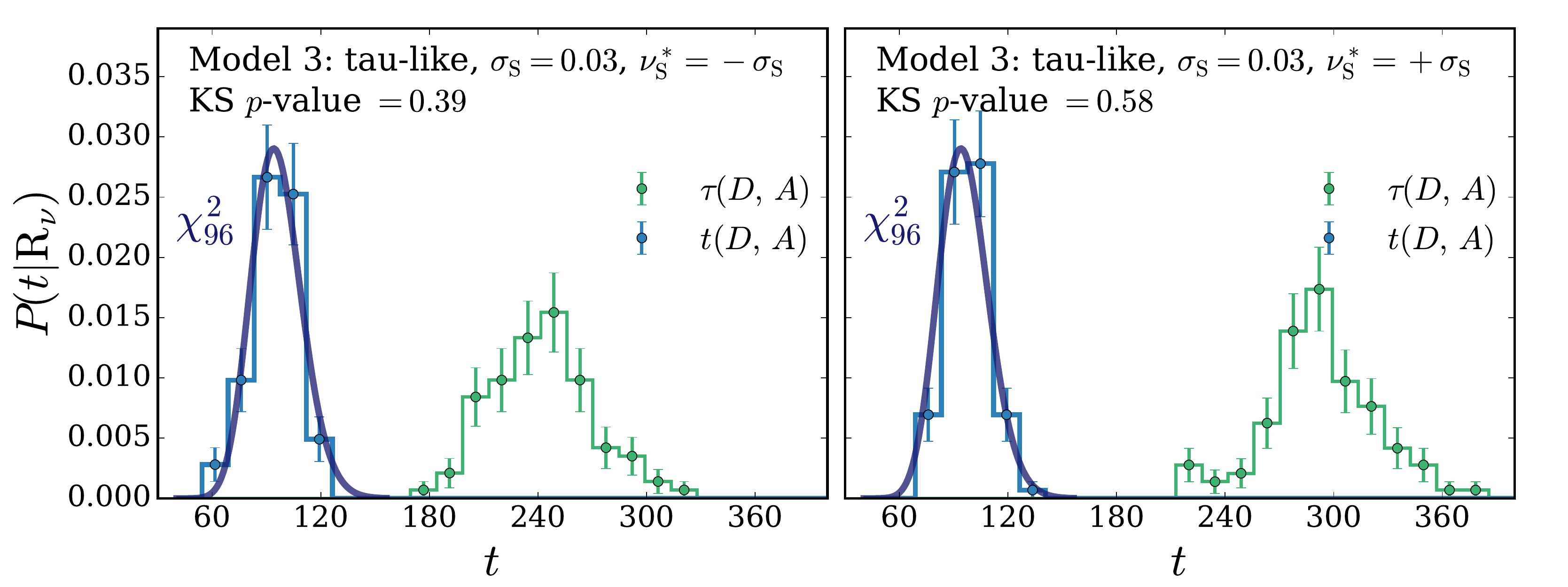}
\caption{Same as Figure~\ref{5D_taulike_model3_a}, but after enriching the reference sample in the region of high transverse momentum with $200'000$ additional events.}\label{5D_taulike_model3_b}
\end{figure}

It is straightforward to repeat in this new setup all the steps described in the previous section. In particular the three neural network architectures identified in Section~\ref{sec:MS5f} are still viable up to a mild retuning of the weight clipping parameter. However, validation is more delicate because of the stronger impact of systematics uncertainties on the distribution of $\tau$. As discussed in Section~\ref{af} and~verified in Section~\ref{sec:validation} in the univariate example, we expect that a higher accuracy is required in the computation of $\tau$ and of $\Delta$ in order to properly capture the cancellation that takes place in the test statistics $t=\tau-\Delta$. We observe that different levels of accuracy are required to validate the three neural network models, depending on the sensitivity of each model to the sparsity of input features. In particular, Model~$3$ (with $3$ hidden layers) turns out not to be particularly sensitive, and its validation does not pose any particular issue, even if the KS compatibility $p$-values for non-central nuisances (see Figure~\ref{5D_taulike_model3_a}) are somewhat lower than those we found in the previous section for the muon- and electron-like scenarios. For Model ~$1$ and~$2$, instead, the compatibility with the target $\chi^2$ is remarkably low, especially if $\nu_{{\rm\textsc{s}}}^*$ is positive. The exact same asymmetric behavior was found in Section~\ref{sec:validation} in the univariate example, and attributed (see Figure~\ref{1D_learning_nuS-0.6}) to the fact that positive scale variations push the data to the extreme tail of the Reference model distribution, which is not populated in the Reference sample. The same effect was found to be responsible for the behavior we observe in the present setup. Indeed we could check the presence of extreme outliers in the trained neural network output, localized in a transverse momentum region that is not populated in the Reference sample. 

Since the problem is due to lack of Reference data in the tail, a way out could be to add statistics to the Reference sample, which however is computationally costly. Certainly feasible with the computing power of a large experiment but beyond our capabilities. A more efficient solution is instead to enrich the Reference sample with a new Monte Carlo sample with a cut on the transverse momenta at generation-level. We then generate $200'000$ events with $200$~GeV generation-level cut on the minimal leading $p_T$ (plus basic acceptance cuts), and we further cut it at $250$~GeV on the reconstructed momenta. We add such events with appropriate weights to the original $10.8$ million sample, and we remove the original events with $p_T>250$~GeV. The so-obtained weighted sample is then employed to generate Reference samples, and the toy data by hit-or-miss unweighting. It is also used for the training of the $\widehat\delta$ networks, improving the quality of the distribution ratio reconstruction in the high-$p_T$ tail. 

\begin{figure}[t]
\begin{minipage}[c]{0.449\textwidth}
\centering
   \includegraphics[height=3.9cm]{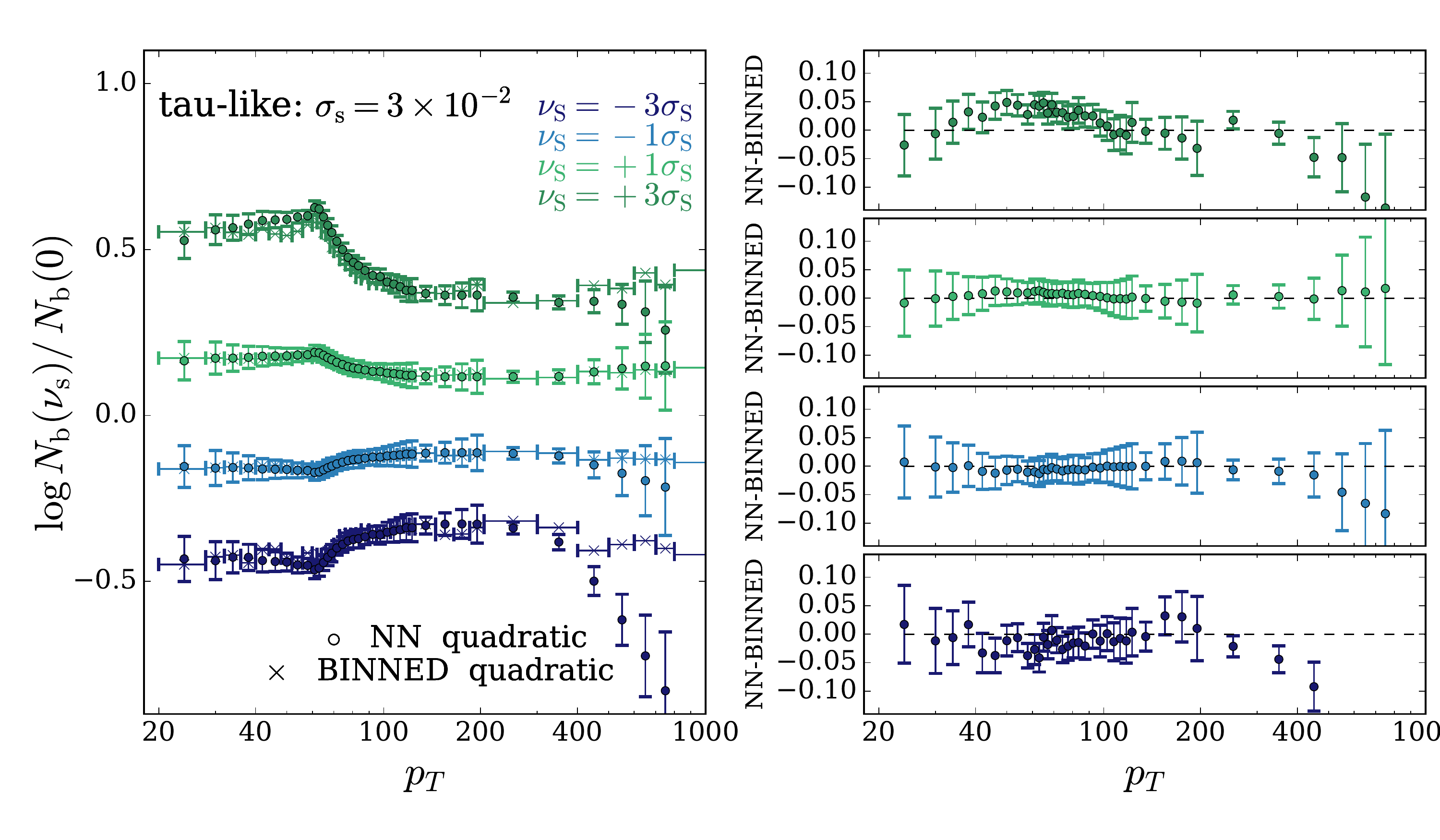}
   \subcaption{}
 \end{minipage}\hfill
 \begin{minipage}[c]{0.549\textwidth}
   \includegraphics[height=3.9cm]{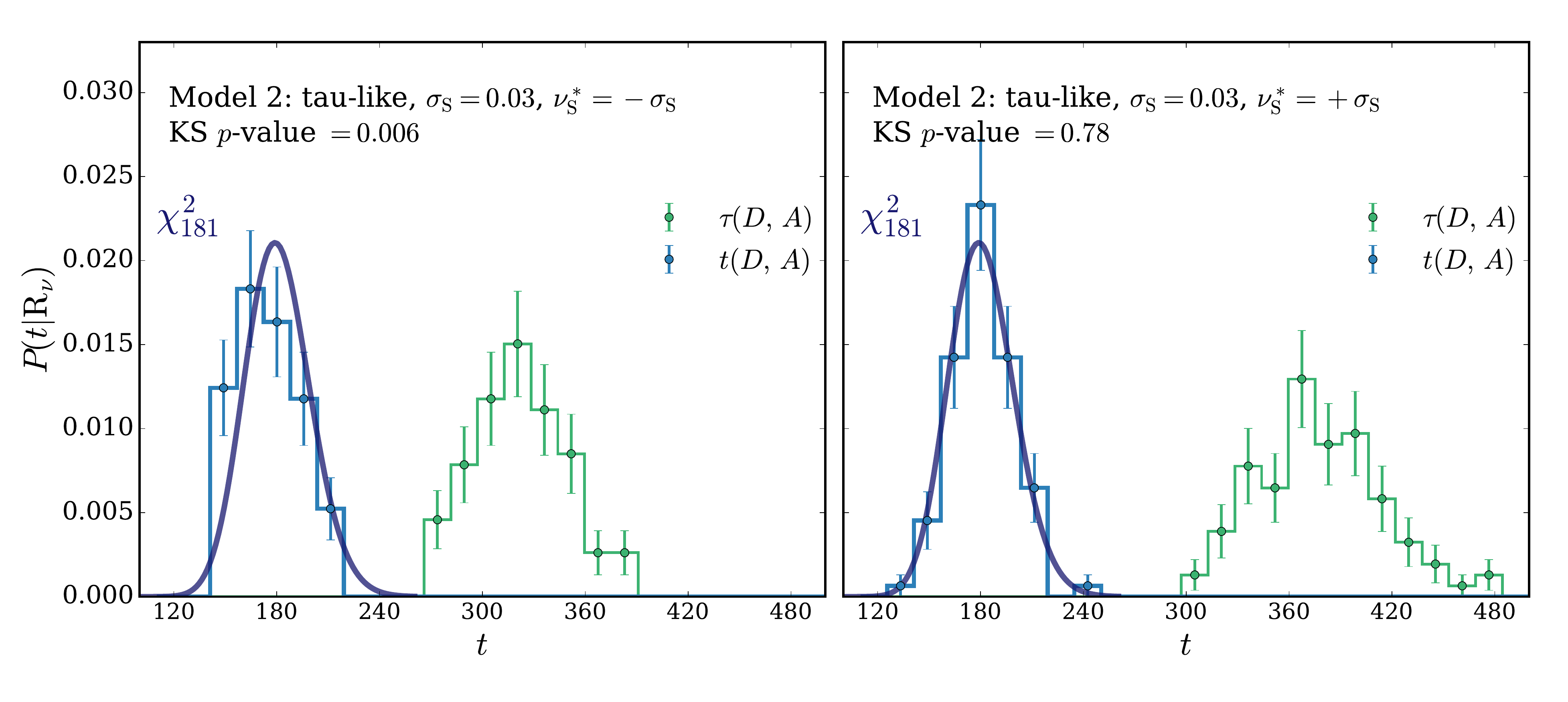}
   \subcaption{}
 \end{minipage}
   \caption{Left side: the reconstructed distribution log-ratio (dots) for different values of $\nu_{{\rm\textsc{s}}}$, compared with the quadratic binned approximation. Right side: the empirical distribution of $\tau$ (in green) and of $t$ (in blue) computed by 100 toy experiments performed for  Model~$2$ in the $\RH$ hypothesis at $\nu_{\textsc{s}}=-1$ (left side) and $\nu_{\textsc{s}}=+1$ (right side) for the $\tau$-like regime. Both plots have been obtained enriching the reference sample in the region of high transverse momentum with $400'000$ additional events.}\label{5D_taulike_model2}

\end{figure}

\begin{figure}[t]
\begin{minipage}[c]{0.449\textwidth}
\centering
    \includegraphics[height=3.9cm]{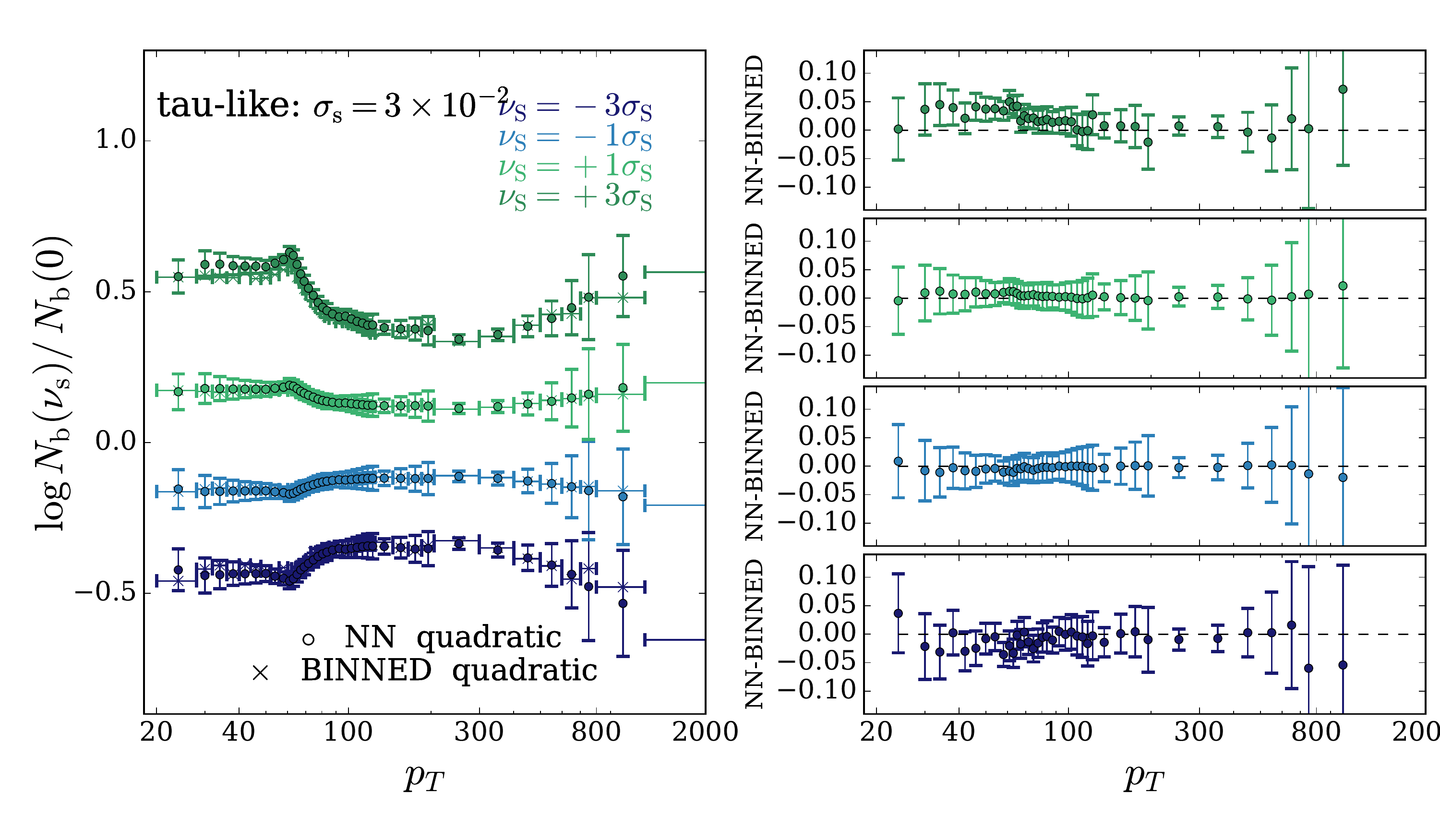}
    \subcaption{}
   \end{minipage}\hfill
   \begin{minipage}[c]{0.549\textwidth}
   \centering
    \includegraphics[height=3.9cm]{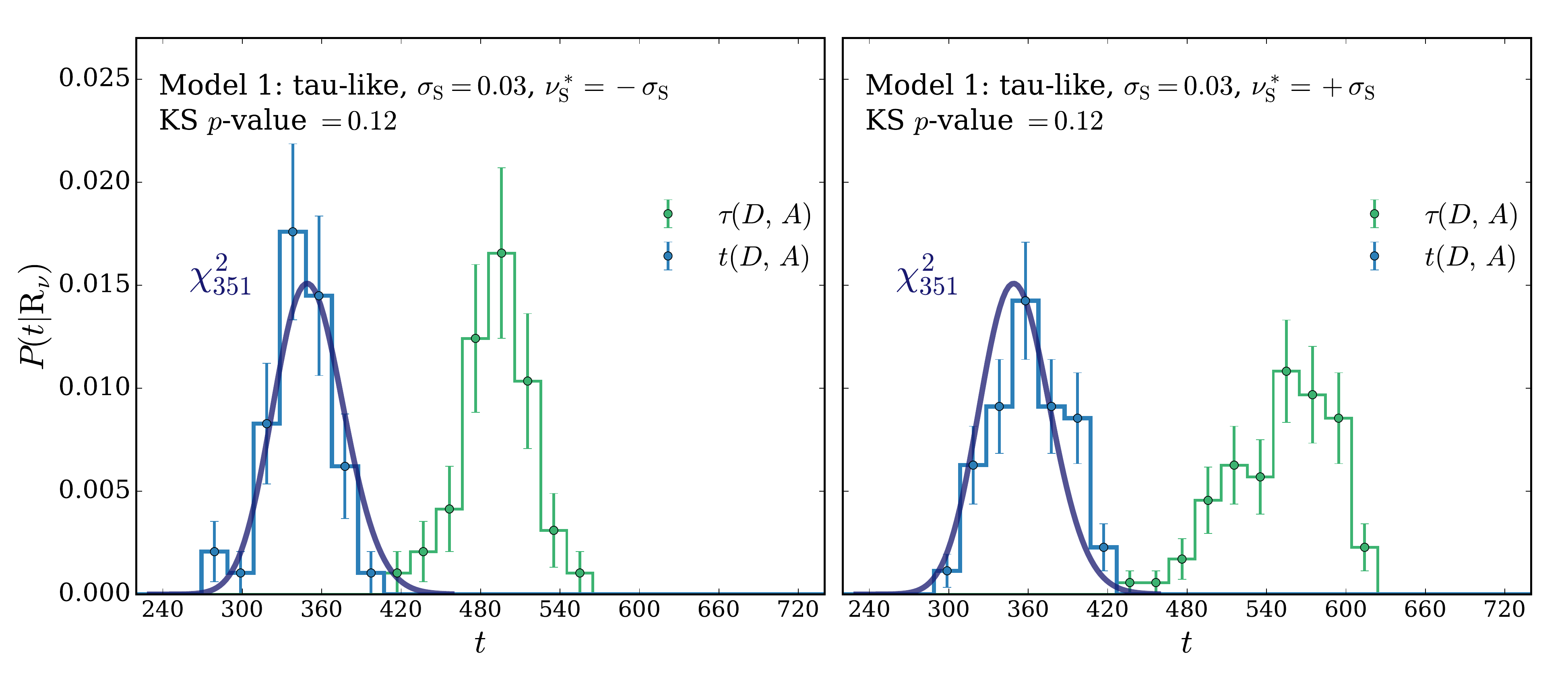}
    \subcaption{}
    \end{minipage}
    \caption{Same as Figure~\ref{5D_taulike_model2}, but for Model~$1$ and $600'000$ additional events in the region of high transverse momentum.}
\label{5D_taulike_model1}
\vspace{-0.2cm}
\end{figure}

The usage of the enriched sample allows us to validate Model~$3$ with higher KS $p$-value, as shown in Figure~\ref{5D_taulike_model3_b}. Furthermore it eliminates the outliers in the neural network output and drastically ameliorates the $\chi^2$-compatibility of Model~$1$ and ~$2$. On the other hand, a satisfactory validation of Model~$1$ and ~$2$ requires a further improvement of the sample. By increasing the number of events in the high $p_T$ tail from $200'000$ to $400'000$, good results are found for the validation of Model~$2$, shown in Figure~\ref{5D_taulike_model2}. Figure~\ref{5D_taulike_model1} shows that good compatibility can be obtained for Model~$1$ as well, but only with $600'000$ high-$p_T$ events. The improvement can be traced back to the more accurate reconstruction of the nuisance coefficient functions $\delta$, which can be monitored by comparing the left panels of the two figures.

 
\subsection{Sensitivity to new physics}
We conclude the section on the two-body final state experiments presenting some examples of the algorithm performances to detect New Physics in the data. For definiteness, we chose the Model~$3$ architecture to perform the sensitivity tests. We consider two new physics benchmark scenarios:~\footnote{The same benchmarks are employed in Ref.~\cite{DAgnolo:2019vbw}, to which the reader is referred for additional details.} 
\begin{itemize}
\item $Z'$ scenario:
a new vector boson with the same couplings to SM fermions as the SM $Z$ boson and mass of $300$~GeV;.
\item EFT scenario: a non-resonant effect due to a dimension-6 4-fermion interaction
\beq
\frac{c_W}{\Lambda}J_{L\mu}^aJ_{La}^{\mu}
\eeq
where $J_{La}^{\mu}$ is the $\rm{SU(2)}_L$ SM current, the energy scale $\Lambda$ is fixed at $1$ TeV and the Wilson coefficient $c_W$ determines the coupling strength.
\end{itemize}

Both benchmarks are studied in the three regimes of systematic uncertainties considered so far and the median observed $Z$-score ($\overline{Z}$) is compared with a median reference $Z$-score ($\overline{Z}_{\rm{ref}}$). As in Section~\ref{sec:sens}, the reference $Z$-score is define as a model dependent measure of the significance, performed by assuming that the specific new physics model is known a priori. As a first approximation, in both scenarios a model dependent analysis would select the two-body invariant mass as the variable of interest. We thus compute the test statistic in eq.~(\ref{tstat}) by binning the two-body invariant mass and studying the effects of the nuisance parameters and that of the signals in each bin. For the SM hypothesis the dependence on the momentum scale nuisance parameter $\nu_{\textsc{s}}$ is approximated by a quadratic polynomial, whereas for the $Z'$ signal we use a quartic one. We call ${\rm{N}}(S)$ the total number of expected $Z'$ events, and we introduce a global exponential factor to describe the normalization uncertainty. Namely, we parametrize the number of events expected in each bin as
\beq
\hat n_i^{(Z')}({\rm{N}}(S), \nu_{\textsc{s}}, \nu_{\textsc{n}}) = [ (a_{0i} +a_{1i}\nu_{\textsc{s}}+a_{2i}\nu_{\textsc{s}}^2) + {\rm{N}}(S)\,(b_{0i}+b_{1i}\nu_{\textsc{s}}+b_{2i}\nu_{\textsc{s}}^2+b_{3i}\nu_{\textsc{s}}^3+b_{4i}\nu_{\textsc{s}}^4)]\cdot e^{\nu_{\textsc{n}}}\,.
\eeq
For the EFT instead, the number of events in each bin depends quadratically on the Wilson coefficient $c_W$, while the dependence on $\nu_{\textsc{s}}$ on the New Physics term (i.e., on the linear and quadratic $c_W$ terms) can be safely ignored. Therefore, we have
\beq
\hat n_i^{(\rm{EFT})}(c_W, \nu_{\textsc{s}}, \nu_{\textsc{n}}) = (a_{0i}+a_{1i}^{\nu_{\textsc{s}}}\nu_{\textsc{s}}+a_{2i}^{\nu_{\textsc{s}}}\nu_{\textsc{s}}^2+a_{1i}^{c_W} c_W+a_{2i}^{c_W} c_W^2 )\cdot e^{\nu_{\textsc{n}}}\,.
\eeq
The numerical $a$ and $b$ coefficients in the above equations where determined by a fit to the Monte Carlo simulations in each bin.

Denoting collectively as ``$\mu$'' the signal strengths in the two scenarios, namely $\mu={\rm{N}}(S)$ or $\mu=c_W$, respectively, the binned log-likelihood reads (up to an irrelevant additive constant)
\beq
\log\Lik(\mu, \nu_{\textsc{s}}, \nu_{\textsc{n}}|\data,\auxdata)
= \sum\limits_{i\in\rm{bins}}n_i\,\log[\hat n_i(\mu, \nu_{\textsc{s}}, \nu_{\textsc{n}})] -\rm{N}(\mu, \nu_{\textsc{s}}, \nu_{\textsc{n}}) + \log\Lik({\boldsymbol0}|\auxdata)\,,
\eeq
where $n_i$ denotes the number of observed events in the $i$-th bin. The binned log-likelihood is then used to compute the test statistic
\beq
t_{\textrm{ref}}(\data,\auxdata)=2 \frac{\max\limits_{\mu,\nui} [\log\,\Lik(\mu, \nu_{\textsc{s}}, \nu_{\textsc{n}}|\data,\auxdata)]}{  \max\limits_{\nui} [\log\,\Lik(0, \nu_{\textsc{s}}, \nu_{\textsc{n}}|\,.\data,\auxdata)]}  \,.
\eeq 
The reference $Z$-score is finally obtained by throwing toy experiments in the new physics hypothesis and computing the $p$-value of the median of the empirical test statistic distribution. In the regimes considered for this work, the counts per bin are always greater than $4$. Therefore it is legitimate to assume the asymptotic behavior for the distribution of the test statistic under the null (SM) hypothesis to be valid, and compute the $p$-value with respect to a $\chi^2_1$. The asymptotic behavior has been verified by running the procedure on SM-distributed toys.

\begin{figure}[t]
\begin{minipage}[c]{0.49\textwidth}
\centering
\includegraphics[width=6.8cm]{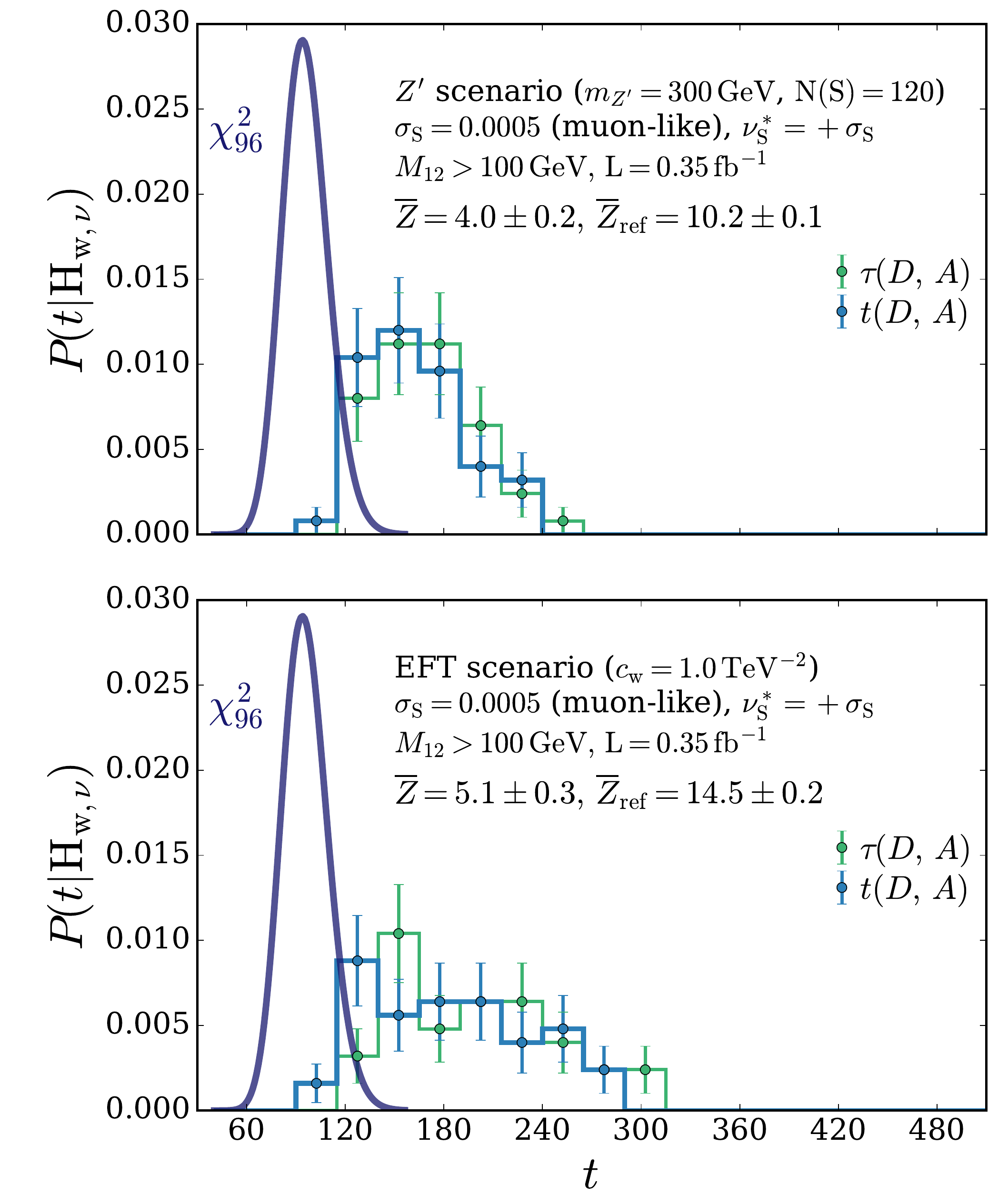}
\subcaption{}
\end{minipage}\hfill
\begin{minipage}[c]{0.49\textwidth}
\centering
\includegraphics[width=6.8cm]{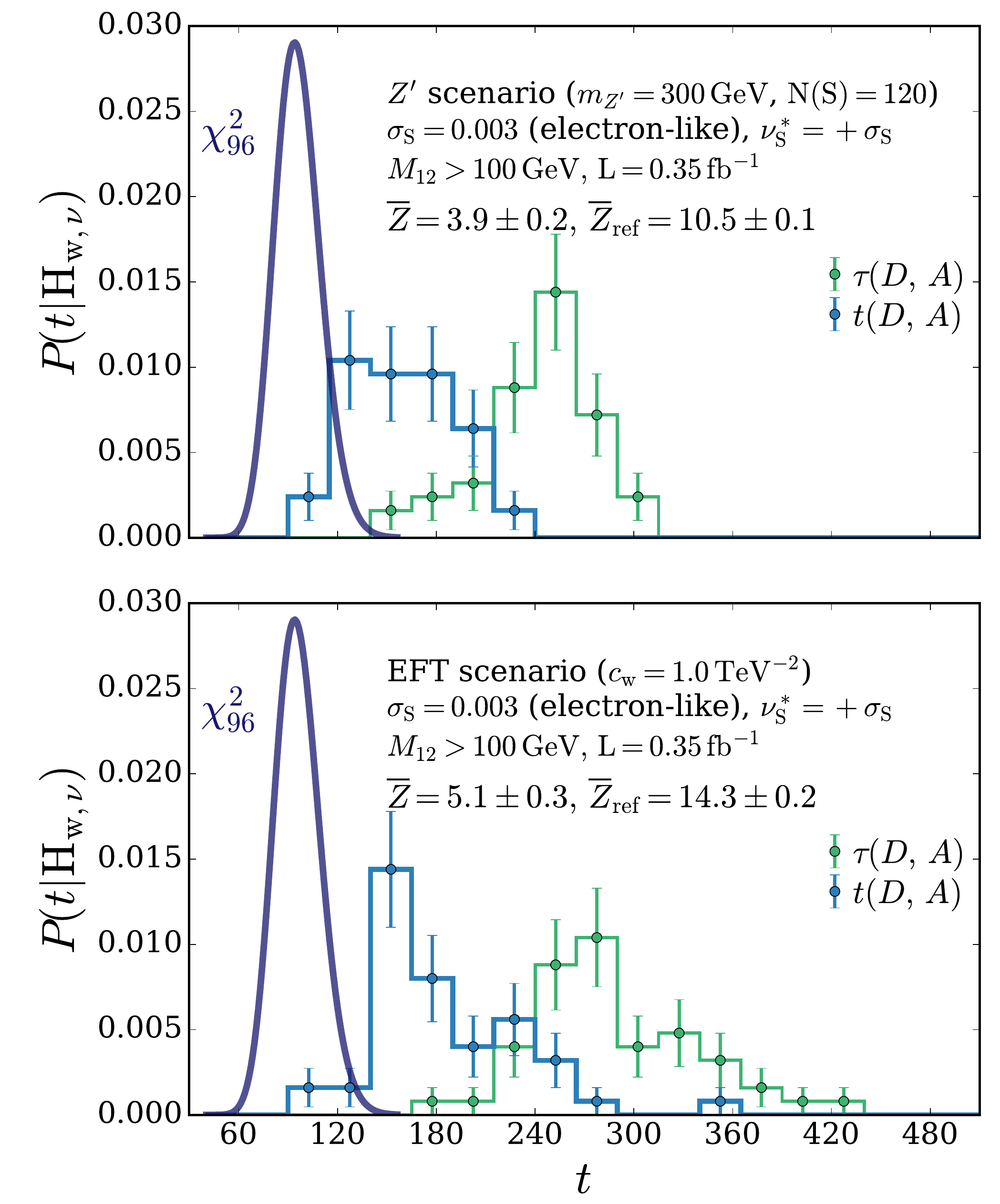}
\subcaption{}
\end{minipage}
\caption{Sensitivity to two New Physics scenarios in the muon-like (a) and electron-like (b) regimes. The upper panels show the sensitivity of the method to the presence of a $Z'$  ($m_{Z'}=300$~GeV, ${\rm{N}}(S)=120$) resonance in the two leptons invariant mass. The lower panels show the sensitivity of the method to a non resonant effect due to a dimension-$6$ $4$-fermions interaction (EFT scenario, $c_W=1.0\,\rm{TeV^{-2}}$). In all panels the true value of the scale nuisance parameter is assumed to be $1$ standard deviation above the central value.}\label{5D_sensitivity_1}
\vspace{-0.2cm}
\end{figure}
\begin{figure}[h]
\begin{minipage}[c]{0.49\textwidth}
\centering
\includegraphics[width=6.8cm]{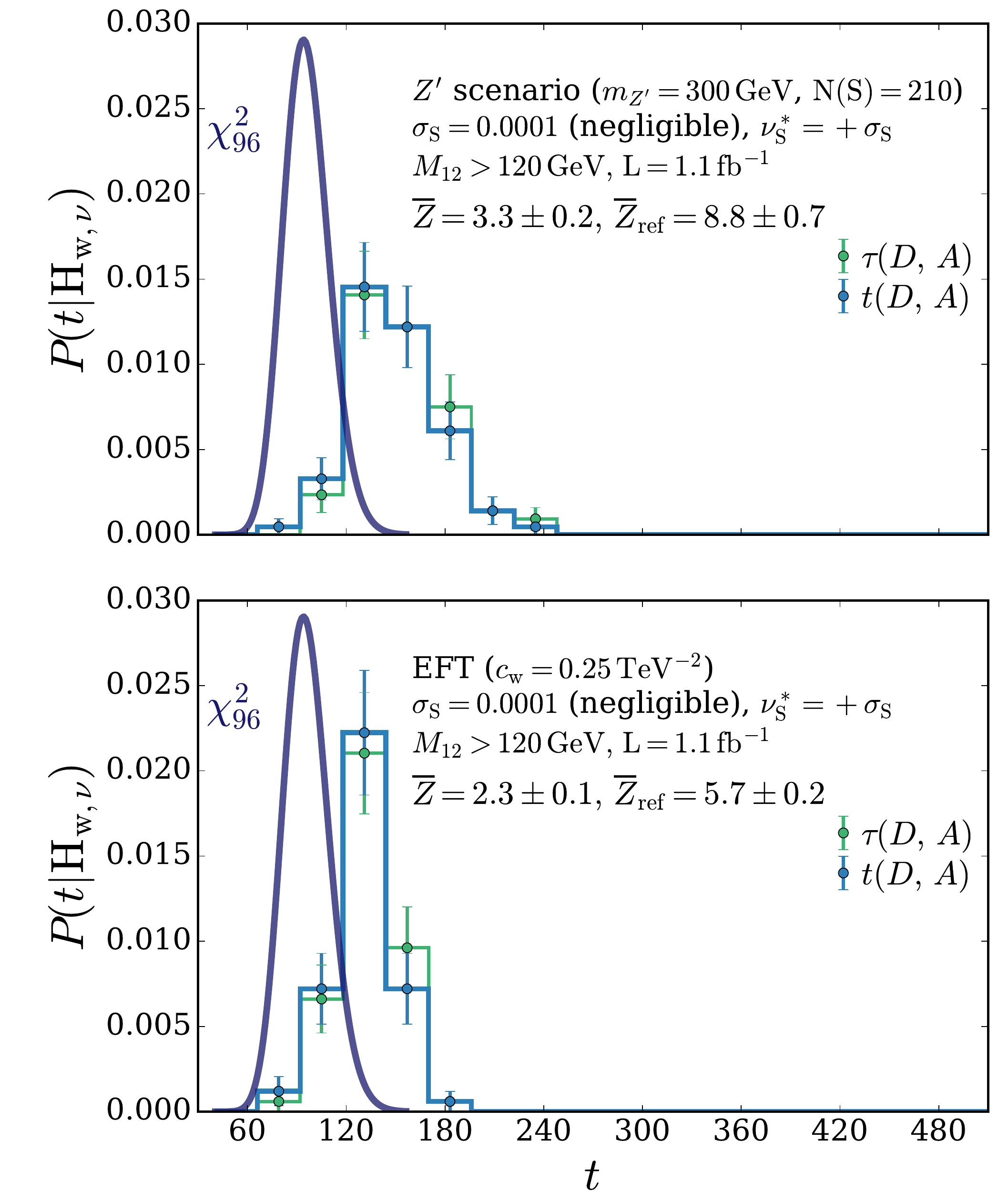}
\subcaption{}
\end{minipage}\hfill
\begin{minipage}[c]{0.49\textwidth}
\centering
\includegraphics[width=6.8cm]{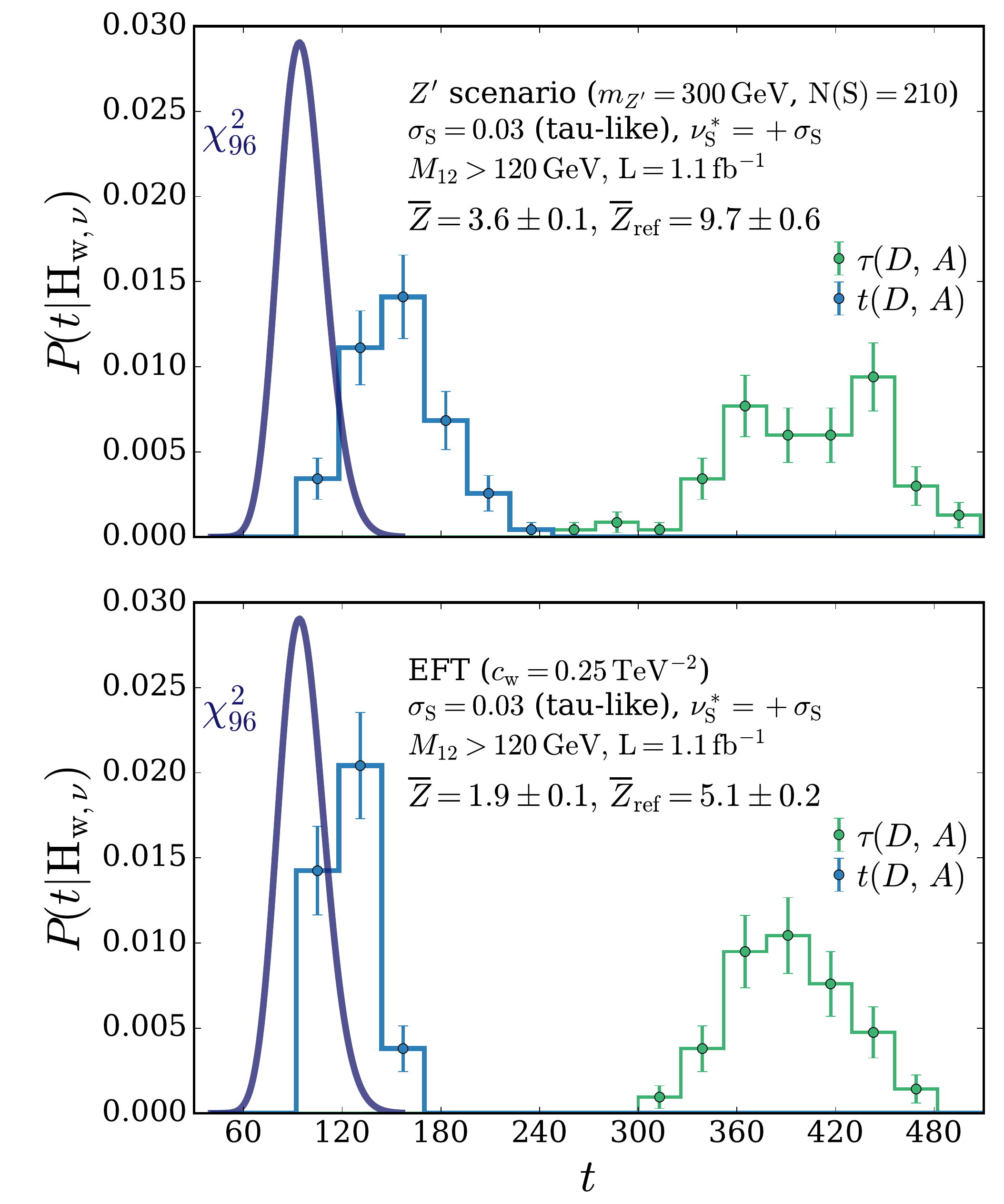}
\subcaption{}
\end{minipage}
\caption{Sensitivity to two New Physics scenarios in the case of negligible uncertainties (a) and $\tau$-like regime (b). The upper panels show the sensitivity of the method to the presence of a $Z'$  ($m_{Z'}=300$~GeV, ${\rm{N}}(S)=210$) resonance in the two leptons invariant mass. The lower panels show the sensitivity of the method to the EFT scenario, with $c_W=0.25\,\rm{TeV^{-2}}$). In all panels the true value of the scale nuisance parameter is assumed to be $1$ standard deviation above the central value.}\label{5D_sensitivity_2}
\vspace{-0.2cm}
\begin{center}
\includegraphics[width=13.8cm]{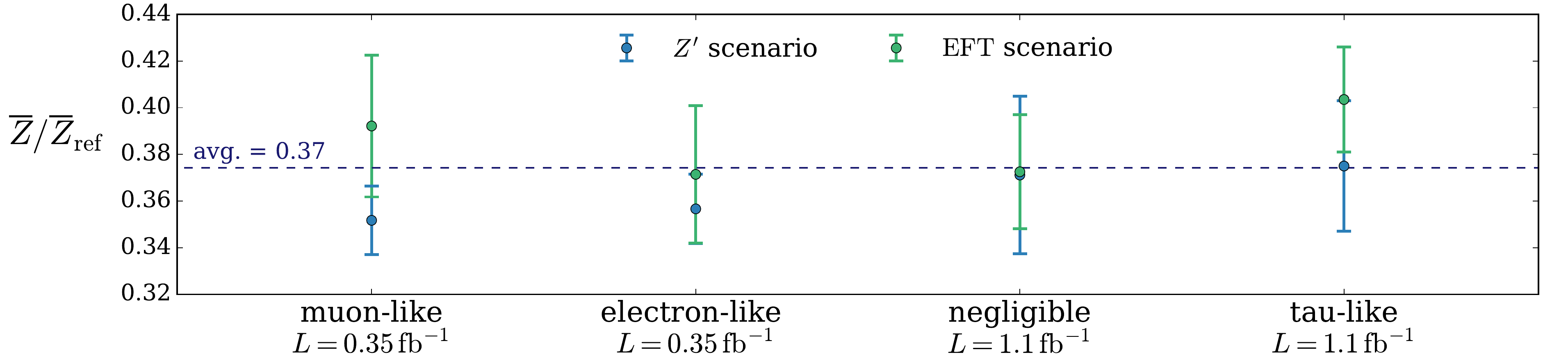}
\caption{Summary of the sensitivity of our method, relative to the sensitivity of dedicated model-dependent searches, to selected New Physics benchmark models. The relative performances depend neither on the New Physics model nor on the assumed scenario for systematic uncertainties.}\label{5D_sensitivity_summary}
\end{center}
\vspace{-0.6cm}
\end{figure}

Figure~\ref{5D_sensitivity_1} shows the algorithm performances in the muon-like and electron-like regimes. The setup is the one described at the beginning of this section, with an effective luminosity (set by assuming the cross-section of the di-muon process) of $0.35\,\rm{fb}^{-1}$ and a cut on the two-body invariant mass at $100$~GeV, which leads to approximately $\NRefCV=8\,400$ expected SM events in the search region. For the $Z'$ scenario we inject a number of signal events which is Poisson-distributed around the expected value ${\rm{N}}(S)=120$, which is around $1\%$ of $\NRefCV$. Whereas for the EFT scenario we generate a Monte Carlo sample with Wilson coefficient set to $1\,\rm{TeV}^{-2}$, which increases the total cross section only at the $2$ per mille level. Figure~\ref{5D_sensitivity_1} shows that muon- and electron-like systematics do not affect appreciably the sensitivity of our method, nor the sensitivity of the model-dependent analysis strategy that we take as reference. 

The results in the $\tau$-like regime are presented in Figure~\ref{5D_sensitivity_2}. As previously explained the effective luminosity is now set to $1.1\,\rm{fb}^{-1}$ and the cut on the two-body invariant mass is moved to $120$~GeV. Since the data integrated luminosity is now a factor $3$ larger than what used in the previous cases, the sensitivity to new physics improves making the previous benchmark models visible with overly high significance. In order to define realistically challenging benchmarks we thus reduce the $Z'$ cross-section such that ${\rm{N}}(S)=210<3\cdot 120$, while in the EFT scenario we lower the Wilson coefficient to $c_W=0.25\,\rm{TeV}^{-2}$. In order to asses the role of systematics, we compare the $\tau$-like setup to an idealized experiment where the uncertainties are negligible (specifically, $\sigma_{\textsc{s}}=\sigma_{\textsc{n}}=1\times10^{-4}$). We observe a slight degradation of the sensitivity due to the uncertainties, but only in the case of the EFT new physics scenario, as expected because the resonant $Z'$ signal can not be mimicked by systematics effects. 

We conclude that our strategy to deal with systematic uncertainties, on top of being robust against false positives as verified in the previous sections, maintains a remarkably high sensitivity to putative new physics effects. The observed mild sensitivity loss due to uncertainties, when present, is perfectly in line with the degradation of the model-dependent reference analysis performances, signally that the sensitivity lowers because the new physics signal is genuinely harder to see and not because of an intrinsic limitation of our model-independent method. Furthermore, the results of the present section confirm the weak dependence on the specific type of new physics, claimed in our previous works~\cite{DAgnolo:2018cun,DAgnolo:2019vbw}, of the ratio ${\overline{Z}}/{\overline{Z}}_{\rm{ref}}$. This is shown in Figure~\ref{5D_sensitivity_summary} by summarizing the performances we have obtained at different luminosities, systematic uncertainties regime and new physics scenarios. In all the experiments our reach is a factor $\sim2.7$ lower than the reference $Z$-score.
\newpage

\section{Conclusions and outlook}\label{sec:conc}
We have proposed and validated a strategy for model-independent new physics searches that duly takes into account the imperfect knowledge of the Reference model predictions. The methodology is robustly based on the canonical Maximum Likelihood ratio treatment of uncertainties as nuisance parameters for hypothesis testing, which emerges as a completely natural and conceptually straightforward extension of the basic framework we proposed and developed in Ref.s~\cite{DAgnolo:2018cun,DAgnolo:2019vbw}. Our findings open the door to real analysis applications, where a ``New-Physics-Learning'' Machine (NPLM) inspects the LHC data in search from departures from the Standard Model, with no bias on the nature and the origin of the putative discrepancy.

The detailed study of the method in real LHC analyses will be essential in order to identify possible implementation issues, which might require further developments of the NLPM strategy itself or methodological advances in related domains. 
Based on the studies performed in the present paper we can anticipate interesting directions for future developments:
\begin{enumerate}
\item The need of a statistically accurate enough (large or ``smart'') Reference sample. We have seen in the study of the two-body final state example how a limited Reference to Data ratio $\Nreference/\NRefCV =5$ (to be compared with $\Nreference/\NRefCV =100$ in the univariate problem) poses a number of technical difficulties, ranging from an enhanced sensitivity to the weight clipping parameter (see  Section~\ref{sec:MS5f}) to possible validation failures (see Sections~\ref{sec:validation} and~\ref{sec:tlike}) due to Data outliers in regions that are not populated in the Reference sample. Raising the Reference sample statistic is not the only way to address these issues. Our results in Section~\ref{sec:tlike} suggest that with a suitably weighted Reference one might obtain the same effect without increasing $\Nreference$, i.e. without impacting the training execution time.
\item The generation of Reference-distributed toys. Our strategies for model selection and validation heavily rely on the availability of toy datasets, namely sets of unweighted data that mimic the outcome of the real experiment under the Standard Model hypothesis. Generating a large set of toys requires, in the first place, a large enough sample of Standard Model data. The potential issue is, as for item~1, that such large sample might not be available or it might be computationally too demanding to be generated. Furthermore, if the Standard Model data are weighted, producing unweighted events with the hit-or-miss technique can be highly inefficient in the presence of large weights, and conceptually impossible if some of the weights are negative as it is the case for simulations at Next-to-Leading order.
\item Accurate learning of nuisance effects. We have seen in Sections~\ref{sec:validation} and~\ref{sec:tlike} that an accurate reconstruction of $\log{{{r}}(x;\nui)}$ is essential and that higher accuracy is needed for those nuisance parameter that impact the distribution of $\tau$ more considerably. On the other hand the accuracy could be limited by an insufficient statistical accuracy of the data used for training the $\widehat\delta$ networks. Moreover when the dependence on the nuisance parameters is not a small correction to the central-value distribution, such that it can not be Taylor-expanded in $\nui$, we expect that learning $\log{{{r}}(x;\nui)}$ might become more demanding. 
\item Training execution time. The time needed for training the ``BSM'' network is considerable, and entails (see Section~\ref{sec:MS5f}) a computational constraint on the maximal neural network complexity that we can handle. The time obviously increases with $\Nreference$, potentially posing an obstruction to the data statistics we can handle, at fixed $\Nreference/\NRefCV$, or to $\Nreference$ itself, which on the other hand we might need to take large as per item~1.
\end{enumerate}

It should be noted that items~1 and 3, as well as item~4, are not absolute obstructions to the applicability of the NPLM strategy. They rather limit the integrated luminosity of the data (i.e., $\NRefCV$) that our algorithm can handle. Indeed item~1 can be addressed by lowering $\NRefCV$, and item~3 as well because the impact of systematic uncertainties on the analysis is relatively smaller if the data statistics is lower. On the other hand an upper limit on $\NRefCV$ does not prevent us from employing the full data luminosity for the analysis. One could indeed split the data in several independent datasets, run NPLM on each and combine statistically the corresponding $p$-values. However, this necessarily entails a reduced sensitivity to new physics effects. 

We also see that most of the items listed above are not specific of the NPLM methodology. In particular the availability of sufficient samples of Standard Model data is a generic need of any LHC analysis, which will become more pressing with the high data statistics of the HL-LHC. Similarly, the generation of toy datasets is in principle a need for any un-binned analysis that can not rely on  asymptotic formulas. Finally, learning the effect of nuisance parameters is methodologically identical to (and directly relevant for) the regression on the distribution dependence on parameters of interest, which is being studied extensively for other applications such as inference on new physics parameters. Potential limitations related with the training time are instead obviously specific of the NPLM methods. It is not excluded that the training time could be substantially reduced by a better choice of the training algorithm or of its implementation, which is an aspect we did not investigate in great detail so far. A more radical solution is to trade neural networks with non-parametric Kernel models, which are radically faster to train~\cite{nplm:Falkon}.

In summary, NPLM emerges as a promising option for the development of a new kind of model-independent new physics searches. The extensive deployment of this type of analyses might play a vital role in experimental programs where, like at the LHC, increasingly rich experimental data are accompanied by an increasingly blurred theoretical guidance in their interpretation. Furthermore, designing NPLM analyses and addressing the corresponding challenges might trigger developments in event generation and in likelihood-free inference techniques, with broader implications on LHC physics.

\section*{Acknowledgements}
M.P. and G.G. are supported by the European Research Council (ERC) under the European Union's Horizon 2020 research and
innovation program (grant agreement n$^o$ 772369). A.W.~acknowledges support from the Swiss National Science Foundation under contract 200021-178999. 

\appendix

\section{Model-independent strategies}\label{app:MI}

In the categorisation of model-independent LHC searches, using either machine learning techniques or more traditional statistical methods, one should first of all keep ``Anomaly Detection'' strategies, like the ones in Ref.s~\cite{Blance:2019ibf,Knapp:2020dde,Cheng:2020dal,Roy:2019jae, Heimel:2018mkt, Cerri:2018anq, Farina:2018fyg, Aguilar-Saavedra:2017rzt}, distinct from the other methods. Anomaly Detection algorithms aim at detecting outliers in the data, namely at classifying events (i.e., instances of $x$ in $\data$) that are ``rare'' in the dataset. The great advantage of this strategy is that it relies marginally or does not rely at all on the availability of a Reference dataset because the notion of ``rarity'' can in principle be extracted exclusively from the observed data. The disadvantage is that its sensitivity is limited to those manifestations of new physics that take place in regions of the phase-space that are weakly populated in the Reference Model. Consider for illustration a univariate feature variable $x$ and a new physics model producing a peak in that variable. The model can be detected if the peak falls in the extreme tail of the Reference Model distribution, because the events falling in the tail will be identified as rare and thus selected by the algorithm. If instead the peak emerges in the bulk of the Reference Model distribution, the events originating from resonance production will not be selected because they are not rare in the dataset. The sensitivity to new physics in this case can only emerge from inspecting the observed dataset as a whole, rather than searching for individual anomalous events. Similar considerations apply to realistic multivariate problems where the presence of the resonance also affects the distribution of other variables. New physics will be detected only if the resonance effects are pronounced in a region that is rare in the Reference Model distribution of these variables. 

Furthermore, it should be kept in mind that the selection of ``rare'' events that is achieved by Anomaly Detection algorithms is only the first step of a search for new physics. The second one is to assess whether or not the observed number of selected rare events signals the presence of new physics. The first step can be achieved purely based on the observed data, but the second one requires comparing with the predictions of the Reference Model. 

Anomaly Detection methods are thus more limited in scope, both in terms of new physics targets and methodologically, since they constitute only one step of the new physics search. Therefore they should be kept distinct from the strict model-independent strategies as we defined them in Section~\ref{intro}. 

It should be noted that even the strict (in our definition) model-independent strategies ~\cite{DAgnolo:2018cun, DAgnolo:2019vbw, D0:2000vuh,Abbott:2001ke, D0:2000dnz,Abazov:2011ma,Aktas:2004pz,H1:2008aak, Aaltonen:2007dg, CDF:2008voc,CMS:2011fra, CMS:2008gya, CMS:2017yoc, CMS:2020zjg, ATLAS:2012qna, ATLAS:2014sxa, ATLAS:2017irs, Aaboud:2018ufy,Alwall:2008va} are not ``fully'' model-independent because they rely on the prior choice of the feature variables $x$ and possibly of the region $X$ ($x\in{X}$) of the phase-space where they are considered to be of interest for the analysis.  A fully model-independent test should ideally employ the entire ATLAS and CMS raw datasets (i.e., the collection of detector hits) before the reconstruction of high-level objects, which is however unfeasible particularly because the trigger selection would be still an implicit bias. On the other hand it is not hard to identify physically motivated features and search regions where new physics could emerge. For instance new heavy particles and short-distance interactions would generically show up in final states with high-level reconstructed SM particles or jets with high transverse momentum. New light particles might instead be found in jets and alter their inner structure, and/or produce anomalous tracks to be found in the analysis of detector data of even lower level, etc. By restricting to the corresponding datasets one could perform new physics searches that are still way more model-independent than the search for one postulated new physics model. Clearly if many such model-independent searches were actually performed one should in principle include an estimate of the look-elsewhere effect in the assessment of the statistical significance of a putative excess. However this effect would be present, and to a larger extent, also in an agnostic interpretation of the regular model-dependent LHC searches.

In order to proceed to a finer characterization of the strict model-independent methods it is important to clarify the role and the origin of the Reference dataset $\reference$, which is an essential element of these strategies. The availability of sufficiently accurate SM background predictions is obviously a potential concern for any LHC analysis, and in particular it is a concern for the deployment of model-independent methods. In order to emphasize this issue, it was proposed in Ref.~\cite{Nachman:2020lpy} to accompany the regular notion of (signal) model-independence with the one of ``background model independence'' and to treat the two notions on equal footing. 

As mentioned above, to perform a new physics search up to the quantification of an excess significance it is necessary to have a trustable Reference Model. The need of a Reference prediction is conceptually unavoidable in any strategy to search for ``new'' phenomena, as it provides the necessary notion of ``old'' phenomena.  In absence of a Reference Model one can only perform the first step of a new physics search, as discussed for Anomaly Detection methods. We can identify ``rare" events, but we can not say whether they are present in the dataset because of new or old physics. 

The need of a trustable Reference prediction is clearly not a new conclusion, it is a common requirement of any model-dependent or model-independent search ever performed. Our interpretation of the emphasis on ``background model independence'' given in~\cite{Nachman:2020lpy} is the quest for methods with a built-in data-driven estimate of the background. We do not consider this aspect relevant, for two reasons. First, because it very commonly happens in concrete LHC final states that a data-driven background estimate is not available, and Monte Carlo simulations need to be employed for at least one of the dominant components of the background. We thus need a method that can also employ Monte Carlo background simulation, rather than being limited to data-driven estimates at least for some of the background components. Second, because there are strategies, like ours, that indeed work both with first-principle and with data-driven background estimates. Therefore it is possible, and convenient, to keep the background estimate problem separate from the development of the search strategy itself.

The regular notion of (signal) model-independence is also subject to caveats as detailed above. But is still possible to classify and rank different methodologies by their ``degree of model-independence'' as we did for Anomaly Detection, by trying to figure out which type of new physics signals they might or might not be sensitive to. From this viewpoint one would rank {\textsc{BumpHunter}}~\cite{Choudalakis:2011qn} and similar strategies~\cite{Metodiev:2017vrx, Collins:2018epr, Nachman:2020lpy, Collins:2019jip, Andreassen:2020nkr, Benkendorfer:2020gek,Amram:2020ykb,Aad:2020cws}, which target resonant signals in a pre-specified variable, lower than methods with a broader target~\cite{DAgnolo:2018cun, DAgnolo:2019vbw, Kuusela:2011aa, DeSimone:2018efk, Chakravarti:2021svb, Matchev:2020wwx, D0:2000vuh, Abbott:2001ke, D0:2000dnz, Abazov:2011ma, Aktas:2004pz, H1:2008aak, Aaltonen:2007dg, CDF:2008voc,CMS:2011fra, CMS:2008gya, CMS:2017yoc, CMS:2020zjg, ATLAS:2012qna, ATLAS:2014sxa, ATLAS:2017irs, Aaboud:2018ufy,Alwall:2008va}. On the other hand, one should not employ these generic considerations to tell which one is the ``right'' strategy to pursue. That depends on aspects that are specific of the final state (features set and phase-space region) one is willing to explore, such as the availability (or not) of a trustable Reference sample and the actual perspectives of progress in the characterization of the data relative to more standard analysis techniques. This is why, ultimately, all these methods should be tried with data and their complementarities exploited to extract the most out of the LHC datasets.

The aim of this Appendix was to introduce some elements for the classification of model-independent strategies, not to provide an exhaustive overview of the field. Strategies like those in Ref.~\cite{Park:2020pak} and Ref.~\cite{Casa:2018avf} would require a more in-depth exposition as they do not fully fall in any of our categories. The approach in Ref.~\cite{Park:2020pak} is to train a multi-categories classifier to tag multiple SM processes and specific putative new physics signals, with the idea that if used on the data this classifier will be sensitive to new physics models not used in training, which can be verified with numerical experiments. In Ref.~\cite{Casa:2018avf}, one employs machine learning techniques to cluster the data into categories that correspond, in the Reference hypothesis, to the different components of the background. A new category can emerge in the presence of new physics. This approach is conceptually interesting as an Anomaly Detection (dubbed ``Collective Anomaly Detection'' in Ref.~\cite{chandola2009anomaly}) performed on the entire dataset rather than on individual events.

\bibliographystyle{JHEP}

\begin{thebibliography}{10}

\bibitem{DAgnolo:2018cun}
R.~T. D'Agnolo and A.~Wulzer, \emph{{Learning New Physics from a Machine}},
  \href{https://doi.org/10.1103/PhysRevD.99.015014}{\emph{Phys. Rev. D}
  {\bfseries 99} (2019) 015014}
  [\href{https://arxiv.org/abs/1806.02350}{{\ttfamily 1806.02350}}].

\bibitem{DAgnolo:2019vbw}
R.~T. D'Agnolo, G.~Grosso, M.~Pierini, A.~Wulzer and M.~Zanetti,
  \emph{{Learning multivariate new physics}},
  \href{https://doi.org/10.1140/epjc/s10052-021-08853-y}{\emph{Eur. Phys. J. C}
  {\bfseries 81} (2021) 89} [\href{https://arxiv.org/abs/1912.12155}{{\ttfamily
  1912.12155}}].

\bibitem{Zyla:2020zbs}
{\scshape Particle Data Group} collaboration, P.~Zyla et~al., \emph{{Review of
  Particle Physics}}, \href{https://doi.org/10.1093/ptep/ptaa104}{\emph{PTEP}
  {\bfseries 2020} (2020) 083C01}.

\bibitem{Neyman:1933wgr}
J.~Neyman and E.~S. Pearson, \emph{{On the Problem of the Most Efficient Tests
  of Statistical Hypotheses}},
  \href{https://doi.org/10.1098/rsta.1933.0009}{\emph{Phil. Trans. Roy. Soc.
  Lond. A} {\bfseries 231} (1933) 289}.

\bibitem{Wilks:1938dza}
S.~S. Wilks, \emph{{The Large-Sample Distribution of the Likelihood Ratio for
  Testing Composite Hypotheses}},
  \href{https://doi.org/10.1214/aoms/1177732360}{\emph{Annals Math. Statist.}
  {\bfseries 9} (1938) 60}.

\bibitem{Wald1943}
A.~Wald, \emph{{Tests of Statistical Hypotheses Concerning Several Parameters
  When the Number of Observations is Large}},
  \href{https://doi.org/10.2307/1990256}{\emph{Trans. Am. Math. Soc.}
  {\bfseries 54} (1943) 426}.

\bibitem{Cranmer:2015bka}
K.~Cranmer, J.~Pavez and G.~Louppe, \emph{{Approximating Likelihood Ratios with
  Calibrated Discriminative Classifiers}},
  \href{https://arxiv.org/abs/1506.02169}{{\ttfamily 1506.02169}}.

\bibitem{Baldi:2016fzo}
P.~Baldi, K.~Cranmer, T.~Faucett, P.~Sadowski and D.~Whiteson,
  \emph{{Parameterized neural networks for high-energy physics}},
  \href{https://doi.org/10.1140/epjc/s10052-016-4099-4}{\emph{Eur. Phys. J. C}
  {\bfseries 76} (2016) 235}
  [\href{https://arxiv.org/abs/1601.07913}{{\ttfamily 1601.07913}}].

\bibitem{Brehmer:2018hga}
J.~Brehmer, G.~Louppe, J.~Pavez and K.~Cranmer, \emph{{Mining gold from
  implicit models to improve likelihood-free inference}},
  \href{https://doi.org/10.1073/pnas.1915980117}{\emph{Proc. Nat. Acad. Sci.}
  {\bfseries 117} (2020) 5242}
  [\href{https://arxiv.org/abs/1805.12244}{{\ttfamily 1805.12244}}].

\bibitem{Brehmer:2019xox}
J.~Brehmer, F.~Kling, I.~Espejo and K.~Cranmer, \emph{{MadMiner: Machine
  learning-based inference for particle physics}},
  \href{https://doi.org/10.1007/s41781-020-0035-2}{\emph{Comput. Softw. Big
  Sci.} {\bfseries 4} (2020) 3}
  [\href{https://arxiv.org/abs/1907.10621}{{\ttfamily 1907.10621}}].

\bibitem{Chen:2020mev}
S.~Chen, A.~Glioti, G.~Panico and A.~Wulzer, \emph{{Parametrized classifiers
  for optimal EFT sensitivity}},
  \href{https://doi.org/10.1007/JHEP05(2021)247}{\emph{JHEP} {\bfseries 05}
  (2021) 247} [\href{https://arxiv.org/abs/2007.10356}{{\ttfamily
  2007.10356}}].

\bibitem{Chen:1}
S.~Chen, A.~Glioti, G.~Panico and A.~Wulzer, \emph{Boosted likelihood learning from event re-weighting}, {to appear (2021)}.

\bibitem{Chen:2}
S.~Chen, A.~Glioti, G.~Panico and A.~Wulzer, \emph{Learning systematic uncertainties}, {to appear (2021)}.

\bibitem{Cowan:2010js}
G.~Cowan, K.~Cranmer, E.~Gross and O.~Vitells, \emph{{Asymptotic formulae for
  likelihood-based tests of new physics}},
  \href{https://doi.org/10.1140/epjc/s10052-011-1554-0}{\emph{Eur. Phys. J. C}
  {\bfseries 71} (2011) 1554}
  [\href{https://arxiv.org/abs/1007.1727}{{\ttfamily 1007.1727}}].

\bibitem{tensorflow2015-whitepaper}
M.~Abadi, A.~Agarwal, P.~Barham, E.~Brevdo, Z.~Chen, C.~Citro et~al.,
  \emph{{TensorFlow}: Large-scale machine learning on heterogeneous systems},
  2015.

\bibitem{Grosso_New_Physics_Learning_2021}
G.~Grosso, \emph{{New Physics Learning Machine (NPLM): tools}},  11, 2021
[\href{https://github.com/GaiaGrosso/Learning_NP}{{\ttfamily GitHub}}].

\bibitem{Alwall:2014hca}
J.~Alwall, R.~Frederix, S.~Frixione, V.~Hirschi, F.~Maltoni, O.~Mattelaer
  et~al., \emph{{The automated computation of tree-level and next-to-leading
  order differential cross sections, and their matching to parton shower
  simulations}}, \href{https://doi.org/10.1007/JHEP07(2014)079}{\emph{JHEP}
  {\bfseries 07} (2014) 079} [\href{https://arxiv.org/abs/1405.0301}{{\ttfamily
  1405.0301}}].

\bibitem{Sjostrand:2006za}
T.~Sjostrand, S.~Mrenna and P.~Z. Skands, \emph{{PYTHIA 6.4 Physics and
  Manual}}, \href{https://doi.org/10.1088/1126-6708/2006/05/026}{\emph{JHEP}
  {\bfseries 05} (2006) 026}
  [\href{https://arxiv.org/abs/hep-ph/0603175}{{\ttfamily hep-ph/0603175}}].

\bibitem{deFavereau:2013fsa}
{\scshape DELPHES 3} collaboration, J.~de~Favereau, C.~Delaere, P.~Demin,
  A.~Giammanco, V.~Lema\^\i{}tre, A.~Mertens et~al., \emph{{DELPHES 3, A
  modular framework for fast simulation of a generic collider experiment}},
  \href{https://doi.org/10.1007/JHEP02(2014)057}{\emph{JHEP} {\bfseries 02}
  (2014) 057} [\href{https://arxiv.org/abs/1307.6346}{{\ttfamily 1307.6346}}].

\bibitem{grosso_gaia_2021_4442665}
G.~Grosso, R.~T. D'Agnolo, M.~Pierini, A.~Wulzer and M.~Zanetti, \emph{Nplm:
  Learning multivariate new physics},  Jan., 2021. [\href{https://doi.org/10.5281/zenodo.4442665}{{\ttfamily 10.5281/zenodo.4442665}}].

\bibitem{CMS:2018rym}
{\scshape CMS} collaboration, A.~M. Sirunyan et~al., \emph{{Performance of the
  CMS muon detector and muon reconstruction with proton-proton collisions at
  $\sqrt{s}=$ 13 TeV}},
  \href{https://doi.org/10.1088/1748-0221/13/06/P06015}{\emph{JINST} {\bfseries
  13} (2018) P06015} [\href{https://arxiv.org/abs/1804.04528}{{\ttfamily
  1804.04528}}].

\bibitem{CMS:2015xaf}
{\scshape CMS} collaboration, V.~Khachatryan et~al., \emph{{Performance of
  Electron Reconstruction and Selection with the CMS Detector in Proton-Proton
  Collisions at \ensuremath{\sqrt{}}s = 8 TeV}},
  \href{https://doi.org/10.1088/1748-0221/10/06/P06005}{\emph{JINST} {\bfseries
  10} (2015) P06005} [\href{https://arxiv.org/abs/1502.02701}{{\ttfamily
  1502.02701}}].

\bibitem{CMS:2017yfk}
{\scshape CMS} collaboration, A.~M. Sirunyan et~al., \emph{{Particle-flow
  reconstruction and global event description with the CMS detector}},
  \href{https://doi.org/10.1088/1748-0221/12/10/P10003}{\emph{JINST} {\bfseries
  12} (2017) P10003} [\href{https://arxiv.org/abs/1706.04965}{{\ttfamily
  1706.04965}}].

\bibitem{CMS:2011eio}
{\scshape CMS} collaboration, S.~Chatrchyan et~al., \emph{{Performance of
  tau-lepton reconstruction and identification in CMS}},
  \href{https://doi.org/10.1088/1748-0221/7/01/P01001}{\emph{JINST} {\bfseries
  7} (2012) P01001} [\href{https://arxiv.org/abs/1109.6034}{{\ttfamily
  1109.6034}}].

\bibitem{CMS:2019buh}
{\scshape CMS} collaboration, A.~M. Sirunyan et~al., \emph{{Search for a Narrow
  Resonance Lighter than 200 GeV Decaying to a Pair of Muons in Proton-Proton
  Collisions at $\sqrt{s} =$ TeV}},
  \href{https://doi.org/10.1103/PhysRevLett.124.131802}{\emph{Phys. Rev. Lett.}
  {\bfseries 124} (2020) 131802}
  [\href{https://arxiv.org/abs/1912.04776}{{\ttfamily 1912.04776}}].

\bibitem{LHCb:2017trq}
{\scshape LHC}b collaboration, R.~Aaij et~al., \emph{{Search for Dark Photons
  Produced in 13 TeV $pp$ Collisions}},
  \href{https://doi.org/10.1103/PhysRevLett.120.061801}{\emph{Phys. Rev. Lett.}
  {\bfseries 120} (2018) 061801}
  [\href{https://arxiv.org/abs/1710.02867}{{\ttfamily 1710.02867}}].

\bibitem{Chen:2020uds}
C.~Chen, O.~Cerri, T.~Q. Nguyen, J.-R. Vlimant and M.~Pierini, \emph{{Data
  Augmentation at the LHC through Analysis-specific Fast Simulation with Deep
  Learning}},  \href{https://arxiv.org/abs/2010.01835}{{\ttfamily 2010.01835}}.

\bibitem{Hagiwara:2013oka}
K.~Hagiwara, J.~Kanzaki, Q.~Li, N.~Okamura and T.~Stelzer, \emph{{Fast
  computation of MadGraph amplitudes on graphics processing unit (GPU)}},
  \href{https://doi.org/10.1140/epjc/s10052-013-2608-2}{\emph{Eur. Phys. J. C}
  {\bfseries 73} (2013) 2608}
  [\href{https://arxiv.org/abs/1305.0708}{{\ttfamily 1305.0708}}].

\bibitem{nplm:Falkon}
M.~Letizia, G.~Losapio, M.~Rando, G.~Grosso, A.~Wulzer, M.~Pierini et~al.,
  \emph{Learning new physics efficiently with kernel methods}, {to
  appear (2021)}.

\bibitem{Blance:2019ibf}
A.~Blance, M.~Spannowsky and P.~Waite, \emph{{Adversarially-trained
  autoencoders for robust unsupervised new physics searches}},
  \href{https://doi.org/10.1007/JHEP10(2019)047}{\emph{JHEP} {\bfseries 10}
  (2019) 047} [\href{https://arxiv.org/abs/1905.10384}{{\ttfamily
  1905.10384}}].

\bibitem{Knapp:2020dde}
O.~Knapp, O.~Cerri, G.~Dissertori, T.~Q. Nguyen, M.~Pierini and J.-R. Vlimant,
  \emph{{Adversarially Learned Anomaly Detection on CMS Open Data:
  re-discovering the top quark}},
  \href{https://doi.org/10.1140/epjp/s13360-021-01109-4}{\emph{Eur. Phys. J.
  Plus} {\bfseries 136} (2021) 236}
  [\href{https://arxiv.org/abs/2005.01598}{{\ttfamily 2005.01598}}].

\bibitem{Cheng:2020dal}
T.~Cheng, J.-F. Arguin, J.~Leissner-Martin, J.~Pilette and T.~Golling,
  \emph{{Variational Autoencoders for Anomalous Jet Tagging}},
  [\href{https://arxiv.org/abs/2007.01850}{{\ttfamily 2007.01850}}].

\bibitem{Roy:2019jae}
T.~S. Roy and A.~H. Vijay, \emph{{A robust anomaly finder based on
  autoencoders}},  [\href{https://arxiv.org/abs/1903.02032}{{\ttfamily
  1903.02032}}].

\bibitem{Heimel:2018mkt}
T.~Heimel, G.~Kasieczka, T.~Plehn and J.~M. Thompson, \emph{{QCD or What?}},
  \href{https://doi.org/10.21468/SciPostPhys.6.3.030}{\emph{SciPost Phys.}
  {\bfseries 6} (2019) 030} [\href{https://arxiv.org/abs/1808.08979}{{\ttfamily
  1808.08979}}].

\bibitem{Cerri:2018anq}
O.~Cerri, T.~Q. Nguyen, M.~Pierini, M.~Spiropulu and J.-R. Vlimant,
  \emph{{Variational Autoencoders for New Physics Mining at the Large Hadron
  Collider}}, \href{https://doi.org/10.1007/JHEP05(2019)036}{\emph{JHEP}
  {\bfseries 05} (2019) 036}
  [\href{https://arxiv.org/abs/1811.10276}{{\ttfamily 1811.10276}}].

\bibitem{Farina:2018fyg}
M.~Farina, Y.~Nakai and D.~Shih, \emph{{Searching for New Physics with Deep
  Autoencoders}},
  \href{https://doi.org/10.1103/PhysRevD.101.075021}{\emph{Phys. Rev. D}
  {\bfseries 101} (2020) 075021}
  [\href{https://arxiv.org/abs/1808.08992}{{\ttfamily 1808.08992}}].

\bibitem{Aguilar-Saavedra:2017rzt}
J.~A. Aguilar-Saavedra, J.~H. Collins and R.~K. Mishra, \emph{{A generic
  anti-QCD jet tagger}},
  \href{https://doi.org/10.1007/JHEP11(2017)163}{\emph{JHEP} {\bfseries 11}
  (2017) 163} [\href{https://arxiv.org/abs/1709.01087}{{\ttfamily
  1709.01087}}].

\bibitem{D0:2000vuh}
{\scshape D0} collaboration, B.~Abbott et~al., \emph{{Search for new physics in
  e\ensuremath{\mu}X data at D\O{} using SLEUTH: A quasi-model-independent
  search strategy for new physics}},
  \href{https://doi.org/10.1103/PhysRevD.62.092004}{\emph{Phys. Rev. D}
  {\bfseries 62} (2000) 092004}
  [\href{https://arxiv.org/abs/hep-ex/0006011}{{\ttfamily hep-ex/0006011}}].

\bibitem{Abbott:2001ke}
{\scshape D0} collaboration, B.~Abbott et~al., \emph{{A quasi-model-independent
  search for new high $p_T$ physics at D\textbackslash{}O}},
  \href{https://doi.org/10.1103/PhysRevLett.86.3712}{\emph{Phys. Rev. Lett.}
  {\bfseries 86} (2001) 3712}
  [\href{https://arxiv.org/abs/hep-ex/0011071}{{\ttfamily hep-ex/0011071}}].

\bibitem{D0:2000dnz}
{\scshape D0} collaboration, V.~M. Abazov et~al., \emph{{A Quasi model
  independent search for new physics at large transverse momentum}},
  \href{https://doi.org/10.1103/PhysRevD.64.012004}{\emph{Phys. Rev. D}
  {\bfseries 64} (2001) 012004}
  [\href{https://arxiv.org/abs/hep-ex/0011067}{{\ttfamily hep-ex/0011067}}].

\bibitem{Abazov:2011ma}
{\scshape D0} collaboration, V.~M. Abazov et~al., \emph{{Model independent
  search for new phenomena in $p \bar{p}$ collisions at $\sqrt{s}=1.96$ TeV}},
  \href{https://doi.org/10.1103/PhysRevD.85.092015}{\emph{Phys. Rev. D}
  {\bfseries 85} (2012) 092015}
  [\href{https://arxiv.org/abs/1108.5362}{{\ttfamily 1108.5362}}].

\bibitem{Aktas:2004pz}
{\scshape H1} collaboration, A.~Aktas et~al., \emph{{A General search for new
  phenomena in ep scattering at HERA}},
  \href{https://doi.org/10.1016/j.physletb.2004.09.057}{\emph{Phys. Lett. B}
  {\bfseries 602} (2004) 14}
  [\href{https://arxiv.org/abs/hep-ex/0408044}{{\ttfamily hep-ex/0408044}}].

\bibitem{H1:2008aak}
{\scshape H1} collaboration, F.~D. Aaron et~al., \emph{{A General Search for
  New Phenomena at HERA}},
  \href{https://doi.org/10.1016/j.physletb.2009.03.034}{\emph{Phys. Lett. B}
  {\bfseries 674} (2009) 257}
  [\href{https://arxiv.org/abs/0901.0507}{{\ttfamily 0901.0507}}].

\bibitem{Aaltonen:2007dg}
{\scshape CDF} collaboration, T.~Aaltonen et~al., \emph{{Model-Independent and
  Quasi-Model-Independent Search for New Physics at CDF}},
  \href{https://doi.org/10.1103/PhysRevD.78.012002}{\emph{Phys. Rev. D}
  {\bfseries 78} (2008) 012002}
  [\href{https://arxiv.org/abs/0712.1311}{{\ttfamily 0712.1311}}].

\bibitem{CDF:2008voc}
{\scshape CDF} collaboration, T.~Aaltonen et~al., \emph{{Global Search for New
  Physics with 2.0 fb$^{-1}$ at CDF}},
  \href{https://doi.org/10.1103/PhysRevD.79.011101}{\emph{Phys. Rev. D}
  {\bfseries 79} (2009) 011101}
  [\href{https://arxiv.org/abs/0809.3781}{{\ttfamily 0809.3781}}].

\bibitem{CMS:2011fra}
{\scshape CMS} collaboration, \emph{{Model Unspecific Search for New Physics in
  pp Collisions at $\sqrt(s) = 7$ TeV}}, [\href{https://cds.cern.ch/record/1360173}{{\ttfamily CMS-PAS-EXO-10-021}}].

\bibitem{CMS:2008gya}
{\scshape CMS} collaboration, \emph{{MUSIC -- An Automated Scan for Deviations
  between Data and Monte Carlo Simulation}}, [\href{https://cds.cern.ch/record/1152572}{{\ttfamily CMS-PAS-EXO-08-005}}].

\bibitem{CMS:2017yoc}
{\scshape CMS} collaboration, \emph{{MUSiC, a Model Unspecific Search for New
  Physics, in pp Collisions at $\sqrt{s}=8\,\mathrm{TeV}$}}, [\href{https://cds.cern.ch/record/2256653}{{\ttfamily CMS-PAS-EXO-14-016}}].

\bibitem{CMS:2020zjg}
{\scshape CMS} collaboration, A.~M. Sirunyan et~al., \emph{{MUSiC: a model
  unspecific search for new physics in proton-proton collisions at $\sqrt{s} =
  $ 13 TeV}},  [\href{https://arxiv.org/abs/2010.02984}{{\ttfamily 2010.02984}}].

\bibitem{ATLAS:2012qna}
{\scshape ATLAS} collaboration, \emph{{A general search for new phenomena with
  the ATLAS detector in pp collisions at $\sqrt(s)=7$ TeV.}}, .

\bibitem{ATLAS:2014sxa}
{\scshape ATLAS} collaboration, \emph{{A general search for new phenomena with
  the ATLAS detector in pp collisions at $\sqrt{s}=8$ TeV}}, .

\bibitem{ATLAS:2017irs}
{\scshape ATLAS} collaboration, \emph{{A model independent general search for
  new phenomena with the ATLAS detector at $\sqrt{s} = 13\;\textrm{TeV}$}}, .

\bibitem{Aaboud:2018ufy}
{\scshape ATLAS} collaboration, M.~Aaboud et~al., \emph{{A strategy for a
  general search for new phenomena using data-derived signal regions and its
  application within the ATLAS experiment}},
  \href{https://doi.org/10.1140/epjc/s10052-019-6540-y}{\emph{Eur. Phys. J. C}
  {\bfseries 79} (2019) 120}
  [\href{https://arxiv.org/abs/1807.07447}{{\ttfamily 1807.07447}}].

\bibitem{Alwall:2008va}
J.~Alwall, M.-P. Le, M.~Lisanti and J.~G. Wacker, \emph{{Model-Independent Jets
  plus Missing Energy Searches}},
  \href{https://doi.org/10.1103/PhysRevD.79.015005}{\emph{Phys. Rev. D}
  {\bfseries 79} (2009) 015005}
  [\href{https://arxiv.org/abs/0809.3264}{{\ttfamily 0809.3264}}].

\bibitem{Nachman:2020lpy}
B.~Nachman and D.~Shih, \emph{{Anomaly Detection with Density Estimation}},
  \href{https://doi.org/10.1103/PhysRevD.101.075042}{\emph{Phys. Rev. D}
  {\bfseries 101} (2020) 075042}
  [\href{https://arxiv.org/abs/2001.04990}{{\ttfamily 2001.04990}}].

\bibitem{Choudalakis:2011qn}
G.~Choudalakis, \emph{{On hypothesis testing, trials factor, hypertests and the
  BumpHunter}},  in \emph{{PHYSTAT 2011}}, 1, 2011,
  [\href{https://arxiv.org/abs/1101.0390}{{\ttfamily 1101.0390}}].

\bibitem{Metodiev:2017vrx}
E.~M. Metodiev, B.~Nachman and J.~Thaler, \emph{{Classification without labels:
  Learning from mixed samples in high energy physics}},
  \href{https://doi.org/10.1007/JHEP10(2017)174}{\emph{JHEP} {\bfseries 10}
  (2017) 174} [\href{https://arxiv.org/abs/1708.02949}{{\ttfamily
  1708.02949}}].

\bibitem{Collins:2018epr}
J.~H. Collins, K.~Howe and B.~Nachman, \emph{{Anomaly Detection for Resonant
  New Physics with Machine Learning}},
  \href{https://doi.org/10.1103/PhysRevLett.121.241803}{\emph{Phys. Rev. Lett.}
  {\bfseries 121} (2018) 241803}
  [\href{https://arxiv.org/abs/1805.02664}{{\ttfamily 1805.02664}}].

\bibitem{Collins:2019jip}
J.~H. Collins, K.~Howe and B.~Nachman, \emph{{Extending the search for new
  resonances with machine learning}},
  \href{https://doi.org/10.1103/PhysRevD.99.014038}{\emph{Phys. Rev. D}
  {\bfseries 99} (2019) 014038}
  [\href{https://arxiv.org/abs/1902.02634}{{\ttfamily 1902.02634}}].

\bibitem{Andreassen:2020nkr}
A.~Andreassen, B.~Nachman and D.~Shih, \emph{{Simulation Assisted
  Likelihood-free Anomaly Detection}},
  \href{https://doi.org/10.1103/PhysRevD.101.095004}{\emph{Phys. Rev. D}
  {\bfseries 101} (2020) 095004}
  [\href{https://arxiv.org/abs/2001.05001}{{\ttfamily 2001.05001}}].

\bibitem{Benkendorfer:2020gek}
K.~Benkendorfer, L.~L. Pottier and B.~Nachman, \emph{{Simulation-Assisted
  Decorrelation for Resonant Anomaly Detection}},
  [\href{https://arxiv.org/abs/2009.02205}{{\ttfamily 2009.02205}}].

\bibitem{Amram:2020ykb}
O.~Amram and C.~M. Suarez, \emph{{Tag N\textquoteright{} Train: a technique to
  train improved classifiers on unlabeled data}},
  \href{https://doi.org/10.1007/JHEP01(2021)153}{\emph{JHEP} {\bfseries 01}
  (2021) 153} [\href{https://arxiv.org/abs/2002.12376}{{\ttfamily
  2002.12376}}].

\bibitem{Aad:2020cws}
{\scshape ATLAS} collaboration, G.~Aad et~al., \emph{{Dijet resonance search
  with weak supervision using $\sqrt{s}=13$ TeV $pp$ collisions in the ATLAS
  detector}}, \href{https://doi.org/10.1103/PhysRevLett.125.131801}{\emph{Phys.
  Rev. Lett.} {\bfseries 125} (2020) 131801}
  [\href{https://arxiv.org/abs/2005.02983}{{\ttfamily 2005.02983}}].

\bibitem{Kuusela:2011aa}
M.~Kuusela, T.~Vatanen, E.~Malmi, T.~Raiko, T.~Aaltonen and Y.~Nagai,
  \emph{{Semi-Supervised Anomaly Detection - Towards Model-Independent Searches
  of New Physics}},
  \href{https://doi.org/10.1088/1742-6596/368/1/012032}{\emph{J. Phys. Conf.
  Ser.} {\bfseries 368} (2012) 012032}
  [\href{https://arxiv.org/abs/1112.3329}{{\ttfamily 1112.3329}}].

\bibitem{DeSimone:2018efk}
A.~De~Simone and T.~Jacques, \emph{{Guiding New Physics Searches with
  Unsupervised Learning}},
  \href{https://doi.org/10.1140/epjc/s10052-019-6787-3}{\emph{Eur. Phys. J. C}
  {\bfseries 79} (2019) 289}
  [\href{https://arxiv.org/abs/1807.06038}{{\ttfamily 1807.06038}}].

\bibitem{Chakravarti:2021svb}
P.~Chakravarti, M.~Kuusela, J.~Lei and L.~Wasserman, \emph{{Model-Independent
  Detection of New Physics Signals Using Interpretable Semi-Supervised
  Classifier Tests}},  \href{https://arxiv.org/abs/2102.07679}{{\ttfamily
  2102.07679}}.

\bibitem{Matchev:2020wwx}
K.~T. Matchev, P.~Shyamsundar and J.~Smolinsky, \emph{{A quantum algorithm for
  model independent searches for new physics}},
  [\href{https://arxiv.org/abs/2003.02181}{{\ttfamily 2003.02181}}].

\bibitem{Park:2020pak}
S.~E. Park, D.~Rankin, S.-M. Udrescu, M.~Yunus and P.~Harris, \emph{{Quasi
  Anomalous Knowledge: Searching for new physics with embedded knowledge}},
  \href{https://doi.org/10.1007/JHEP06(2021)030}{\emph{JHEP} {\bfseries 21}
  (2020) 030} [\href{https://arxiv.org/abs/2011.03550}{{\ttfamily
  2011.03550}}].

\bibitem{Casa:2018avf}
A.~Casa and G.~Menardi, \emph{{Nonparametric semisupervised classification for
  signal detection in high energy physics}},
  \href{https://arxiv.org/abs/1809.02977}{{\ttfamily 1809.02977}}.

\bibitem{chandola2009anomaly}
V.~Chandola, A.~Banerjee and V.~Kumar, \emph{Anomaly detection: A survey},
  \href{https://doi.org/10.1145/1541880.1541882}{\emph{ACM computing surveys
  (CSUR)} {\bfseries 41} (2009) 1}.

\end{thebibliography}

\providecommand{\href}[2]{#2}\begingroup\raggedright\endgroup

\end{document}